\documentclass[twocolumn]{aastex63}
\usepackage{hyperref,graphics,graphicx,amsmath,amssymb}
\usepackage{multirow}
\usepackage{soul}
\usepackage{epstopdf}

\graphicspath{{./}{figures/}}

\submitjournal{ApJ}
\shorttitle{The kpc-scale outflow of MaNGA 1-166919}
\shortauthors{al Yazeedi et al.}

\newcommand{\ii}{\,{\sc ii}}
\newcommand{\iii}{\,{\sc iii}}
\newcommand{\Ha}{H$\alpha$}

\newcommand{\kms}{km~s$^{-1}$}

\begin{document}
\title{The impact of low luminosity AGN on their host galaxies: A radio and optical investigation of the kpc-scale outflow in MaNGA 1$-$166919}

\correspondingauthor{A. al Yazeedi}
\email{aaa1007@nyu.edu}

\author[0000-0003-2577-6799]{Aisha al Yazeedi}
\affiliation{New York University Abu Dhabi, PO Box 129188, Abu Dhabi, UAE}
\affiliation{Center for Astro, Particle, and Planetary Physics, NYU Abu Dhabi, PO Box 129188, Abu Dhabi, UAE}

\author[0000-0002-6425-6879]{Ivan Yu. Katkov}
\affiliation{New York University Abu Dhabi, PO Box 129188, Abu Dhabi, UAE}
\affiliation{Center for Astro, Particle, and Planetary Physics, NYU Abu Dhabi, PO Box 129188, Abu Dhabi, UAE}
\affiliation{Sternberg Astronomical Institute, Lomonosov Moscow State University, Universitetskij pr., 13,  Moscow, 119234, Russia\\}

\author[0000-0003-4679-1058]{Joseph D. Gelfand}
\affiliation{New York University Abu Dhabi, PO Box 129188, Abu Dhabi, UAE}
\affiliation{Center for Astro, Particle, and Planetary Physics, NYU Abu Dhabi, PO Box 129188, Abu Dhabi, UAE}
\affiliation{Center for Cosmology and Particle Physics, New York University, 726 Broadway, room 958, New York, NY 10003}

\author[0000-0003-2212-6045]{Dominika Wylezalek}
\affiliation{Astronomisches Rechen-Institut, Zentrum f\"{u}r Astronomie der Universit\"{a}t Heidelberg, M\"{o}nchhofstr. 12-14, 69120 Heidelberg, Germany}

\author[0000-0001-6100-6869]{Nadia L. Zakamska}
\affiliation{Department of Physics \& Astronomy, Johns Hopkins University, Bloomberg Center, 3400 N. Charles St., Baltimore, MD 21218, USA}

\author[0000-0003-3762-7344]{Weizhe Liu}
\affiliation{Department of Astronomy, University of Maryland, College Park, MD 20742, USA}

\begin{abstract}
One way an Active Galactic Nucleus (AGN) influences the evolution of their host galaxy is by generating a large-scale (kpc-scale) outflow.  The content, energetics, and impact of such outflows depend on the properties of both the AGN and host galaxy, and understanding the relationship between them requires measuring the properties of all three.  In this paper, we do so by analyzing recent radio and optical integral field unit (IFU) spectroscopic observations of MaNGA 1-166919.  Our results indicate that the bi-conical outflow in this galaxy is powered by a low-luminosity, low-Eddington ratio AGN ejecting material that drives $\sim100-200~{\rm km~s}^{-1}$ shocks into the surrounding interstellar medium (ISM) -- producing the hot, ionized gas and relativistic particles associated with the observed outflow.  The energetics of the relativistic and ionized gas material produced at this shock are comparable, and both the mass outflow and kinetic power of the ionized gas in this outflow are higher than other AGN with similar bolometric luminosities.  Lastly, while the host galaxy's total star formation rate is comparable to that of other star-forming galaxies with a similar stellar mass, there is evidence that the outflow both suppresses and enhances star formation in its immediate surroundings.

\end{abstract}


\section{Introduction} \label{sec:intro}

The observed correlation between the properties of a galaxy and its supermassive black hole (SMBH) suggests the evolution of the two are related (e.g., \citealt{kormendy13}).  In current models for galaxy evolution, an important component of this relationship are kpc-scale outflows powered by accretion onto the SMBH, resulting in an Active Galactic Nucleus (AGN; e.g., see \citealt{king15} for a recent review). There believed to be two different classes of outflows: ``winds'', produced by the radiation emitted by the accreting material (e.g., \citealt{king15}) and ``jets'', highly collimated streams of relativistic particles.  Observations of a X-ray binaries and low-luminosity AGN suggest a possible connection between the type of outflow and mode of accretion onto the black hole (e.g., \citealt{kording06}), with 
(e.g., \citealt{HeckmanBest2014}):
\begin{itemize}
    \item winds primarily resulting from ``radiative-mode'' accretion, where the inflowing material is primarily constrained to a geometrically thin, optically thick accretion disk that extends all the way to the innermost stable circular orbit (ISCO) of the super-massive black hole, while
    \item jets are often produced by ``jet-mode'' accretion, in which the thin accretion disk does not reach the ISCO, but instead converted into a geometrically thick structure near the event horizon.
\end{itemize}
In both cases, the interaction between these outflows and surrounding medium generates shocks (e.g., \citealt{faucher12}) that can propagate through, and affect the properties of the entire galaxy (e.g., \citealt{nelson19}). In most models, the primary roles of these outflow is to suppress future star formation in these galaxies -- though whether they do so is uncertain (e.g., \citealt{bae17}).

Such outflows are believed to have multiple constituents, such as hot ionized gas produced at the shock, neutral and molecular material entrained in the flow, (e.g., \citealt{oosterloo17, richings18, Hall19, murthy19} and references therein), and cosmic rays -- highly relativistic particles accelerated at the shock.  Recent results suggest that, under certain conditions (e.g., $M_\star \gtrsim10^{10}~M_\odot$ galaxies; \citealt{hopkins20a}), 
the pressure of the resultant cosmic rays can actually play an important role in driving massive amounts of material from a galaxy (e.g., \citealt{mao18, hopkins20b}).  An important way of studying these particles is to measure the morphological and spectral properties of the radio synchrotron emission resulting from the interaction between cosmic rays and magnetic fields (e.g., \citealt{Zakamska14, Alexandroff16, Hwang18}).

\begin{figure}
    \centering
    \includegraphics[width=0.45\textwidth]{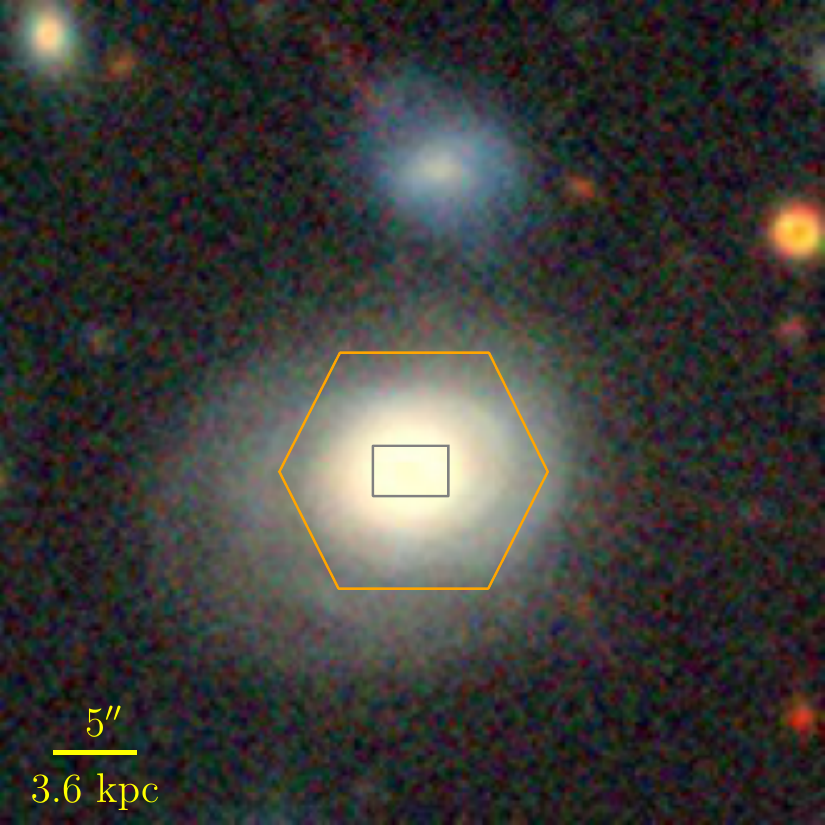}
    \caption{Composite RGB image of the Blob Source extracted from the DESI Legacy Imaging Surveys (\citet{Dey2019_legacysurveys}, \href{http://legacysurvey.org/viewer?ra=146.7093&dec=43.4238&layer=dr8&zoom=15}{legacysurvey.org}). MaNGA field of view is shown in orange. Gray box corresponds to the GMOS field of view.}
    \label{fig:desi_img}
\end{figure}

\begin{figure}
    \centering
    \includegraphics[width=0.45\textwidth]{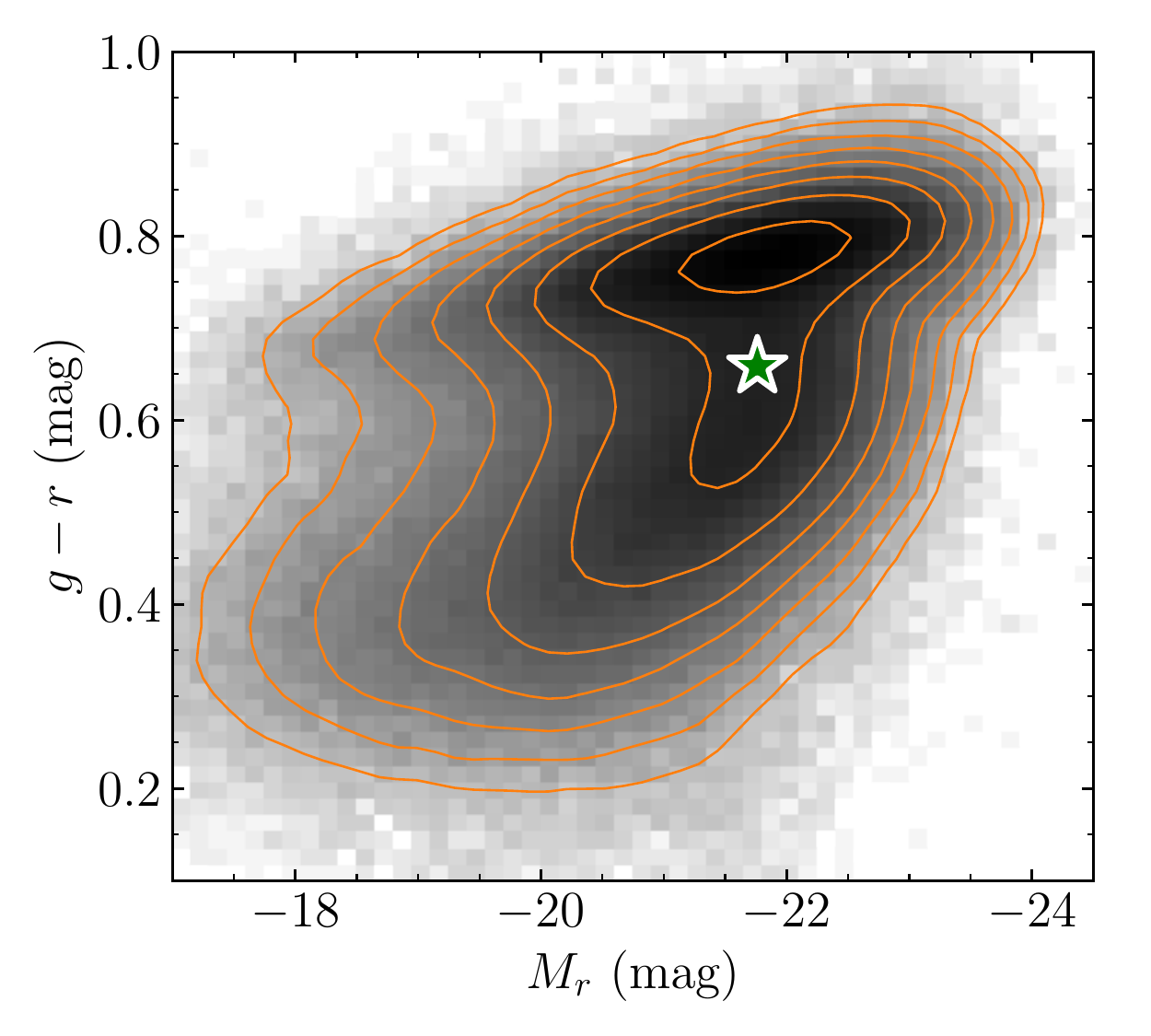}
    \caption{Location of studied galaxy (green symbol) in the color-magnitude diagram. Grey distribution is shown as reference based on the $k$-corrected photometry taken from the \href{http://rcsed.sai.msu.ru/}{Reference Catalog of galaxy Spectral Energy Distributions} (RCSED; \citealt{Chilingarian2017rcsed}).}
    \label{fig:cmd}
\end{figure}

In this paper we present a detailed study of the radio and optical emission of MaNGA 1$-$166919 (Figure \ref{fig:desi_img}), a fairly nearby ($z\sim0.07$, Table~\ref{tab:galprop}) galaxy whose optical colors suggest it lies within the ``green valley'' (see Figure \ref{fig:cmd}).  Such galaxies are believed to be transitioning from the ``blue'' (star-forming) cloud to the ``red''   (quiescent) sequence, possibly as a result of a large scale outflow removing and/or reheating the gas needed to form additional generations of stars.  A previous study of this galaxy by \citet{Wylezalek2017}, showed that it indeed hosts such an outflow.  As demonstrated below, analyzing the multi-properties of the outflow and host galaxies provides important insight into how the outflow is produced by the central AGN and how it interacts with the surrounding galaxy.

In \S\ref{analysis_radio}, we discuss our analysis of recent radio observations of this galaxy, presenting our measurement its radio morphology \S\ref{sec:radio_img} and spectrum \S\ref{sec:radio_spec}.  In \S\ref{analysis_optical}, we discuss the analysis of recent MaNGA (\S\ref{sec:manga}) and GMOS (\S\ref{sec:gmos}) IFU observations of this source.  In \S\ref{sec:outflow}, we present our measurements of the relativistic material (\S\ref{sec:outflow_cr}), kinematics (\S\ref{sec:outflow_kin}), and ionized gas (\S\ref{sec:outflow_thermal}) in this outflow.  In \S\ref{sec:agn}, we discuss the relationship between this outflow and the AGN in this galaxy, while in \S\ref{sec:hostgal} we discuss the interaction between this outflow and the surrounding medium.  In \S\ref{sec:summary} we summarize our results and their implications.  Throughout the paper we use a luminosity distance $D_L=330$~Mpc, an angular-size distance $D_A=287$~Mpc and cosmology corrected scale 1.397~kpc~arcsec$^{-1}$ according to the \href{https://ned.ipac.caltech.edu}{NASA/IPAC Extragalactic Database} (NED), assuming a flat $\Lambda$CDM cosmology $H_0=70$~km~s$^{-1}$ Mpc$^{-1}$, $\Omega_\mathrm{m}=0.279$ and $\Omega_\mathrm{\Lambda}=0.721$.

\begin{deluxetable}{lc}[tb]
\tablecaption{General properties of the studied galaxy. \label{tab:galprop}}
\tablecolumns{2}
\tablewidth{0pt}
\tablehead{
\colhead{Parameter} & \colhead{Studied galaxy}}
\startdata
\multirow{2}{*}{R.A. Dec. [J2000]} & 09:46:50.18  +43:25:25.8 \\
 & 146.709110	43.423861 \\
\multirow{3}{*}{IDs} & SDSS J094650.17+432525.8 \\
                     & WISEA~J094650.18+432525.8 \\
                     & LEDA~2220412 \\ 
 MaNGA-ID & \href{https://dr15.sdss.org/marvin/galaxy/1-166919}{1-166919} \\
 Plate-IFU & 8459-3702 \\
 \hline
 Redshift & 0.07221 \\
 Luminosity distance $D_L$ & 330 Mpc \\
 Angular-size distance $D_A$ & 287 Mpc \\
 Scale & 1.39 kpc arcsec$^{-1}$ \\
 Galactic $A_V$\tablenotemark{a} & 0.0478 mag \\ 
 $R_{\rm eff}$ ($r$-band) \tablenotemark{b} & 4.0\arcsec \\
 M$_\star$\tablenotemark{b} & $6.1\times10^{10}$ $M_\odot$ \\
 $g^{\prime} - r^{\prime}$\tablenotemark{b} & 0.658 \\ 
\enddata
\tablenotetext{a}{Galactic extinctions $A_V$ are taken from \citet{Schlegel1998} maps.}
\tablenotetext{b}{These parameters are from \href{https://www.sdss.org/dr13/manga/manga-target-selection/nsa/}{NASA-Sloan Atlas} as provided by  \href{https://dr15.sdss.org/marvin/galaxy/1-166919/}{Marvin} \citep{Cherinka2019AJ....158...74C}.}
\end{deluxetable}

\begin{table*}[tbh]
\begin{center}
\caption{JVLA Observations of MaNGA 1-166919}
\label{tab:radio_blob_obs}
\begin{tabular}{cccc}
\hline
\hline
 Project & \multicolumn{3}{c}{VLA/16B-082 (AG984)} \\
 Observation Date & 2016 Nov 13 & 2016 Oct 3 & 2017 Dec 30 \\
 Band (Frequency) & L (1-2 GHz) & S (2-4~GHz) & C (4-8~GHz) \\
 Configuration & A & A & B \\
 Time on Source & 16m24s & 21m52s & 24m40s \\
 Thermal RMS\tablenotemark{a} & $17~\frac{\rm \mu Jy}{\rm beam}$\tablenotemark{b} & $8.0~\frac{\rm \mu Jy}{\rm beam}$\tablenotemark{c} & $4.5~\frac{\rm \mu Jy}{\rm beam}$\tablenotemark{d}\\
Field of View\tablenotemark{e} & 30$^\prime$ & $15^\prime$ & $7\farcm5$ \\
L.A.S.\tablenotemark{f} & $18\arcsec$ & $9\arcsec$ & $14\farcs5$ \\
Number of Spectral Windows & 16 & 16 & 32 \\
Number of Spectral Channels/Window & 64 & 64 & 64 \\ 
Width of Spectral Channels & 1~MHz & 2~MHz & 2~MHz \\ 
\hline
\hline
\end{tabular}
\end{center}
\tablenotetext{a}{The image RMS due to thermal noise, calculated using the \href{https://obs.vla.nrao.edu/ect/}{VLA Exposure Calculator} assuming ``natural'' weighting.}
\tablenotetext{b}{Calculated assuming a bandwidth of 0.6~GHz to account for radio frequency interference (RFI) in this band.}
\tablenotetext{c}{Calculated assuming a bandwidth of 1.5~GHz to account for RFI in this band.}
\tablenotetext{d}{Calculated assuming a bandwidth of 3.35~GHz to account for RFI in this band.}
\tablenotetext{e}{Full-Width Half Power (FWHP) of primary beam.}
\tablenotetext{f}{Largest Angular Scale (LAS) from Table 3.1.1 \href{https://science.nrao.edu/facilities/vla/docs/manuals/oss/performance/resolution}{JVLA Resolution Webpage}, divided by 2 to account for the short on source time of these observations.}
\end{table*}

\section{Jansky Very Large Array Observations}
\label{analysis_radio}

To better measure the properties of the radio emission of MaNGA 1-166919, we analyzed the data collected in the three Jansky Very Large Array (JVLA) observations of this galaxy listed in Table \ref{tab:radio_blob_obs}.  For each observation, the raw ASDM files were converted into a measurement set (MS) using the {\tt importevla} task included in Common Astronomy Software Application ({\sc casa}; \citealt{casa}) version 5.1.2-4, and was calibrated using the VLA CASA Calibration Pipeline 5.1.2.  The delays, bandpass, and flux density scale were calibrated using short observations of 3C286 (J1331+3030), while the gains were calibrated using observations of quasar J0920+446 (B3 0917+441).  The calibrated data were then imaged using the {\sc casa} task {\tt tclean} using natural weighting to maximize the sensitivity (at the expense of angular resolution).  The large fractional bandwidth of these datasets results in substantial differences in the primary and synthesized beams and the intrinsic source flux across the band, which can create artifacts in the resultant images.  To mitigate these effects, we deconvolved the image using a two-term, multi-frequency synthesis (MTMFS) algorithm \citep{rau11}. During the deconvolution process, the residual maps were smoothed on scales on 0, 4, and 20 pixels to better identify sources of different angular sizes.  Furthermore, at L- and S-bands sufficient flux was detected in the field to use the {\sc casa} task {\tt gaincal} to recalculate the phase calibration assuming the intensity model generated from this imaging, with the new gain table applied to the data using the {\sc casa} {\tt appylcal} task.  Lastly, before any further analysis, the resultant total intensity and spectral index maps were corrected for primary beam attenuation using the {\sc casa} task {\tt widebandpbcor}, which accounts for changes in the primary beam across the large fractional bandwidth of these datasets.  The properties of the resultant images are listed in Table \ref{tab:radio_img_prop}, and this process resulted in images with background comparable to the thermal noise limit.

\begin{table*}[tbh]
\begin{center}
\caption{Properties of wideband images derived from the JVLA observations listed in Table \ref{tab:radio_blob_obs}.}
\label{tab:radio_img_prop}
\begin{tabular}{cccc}
    \hline
    \hline
    Band (Frequency) & L (1-2~GHz) & S (2-4~GHz) & C (4-8~GHz) \\
    Pixel Size & $0\farcs4\times0\farcs4$ & $0\farcs2\times0\farcs2$ & $0\farcs3\times0\farcs3$ \\
    Image Size [pixels] & $6400\times6400$ & $6400\times6400$ & $2500\times2500$ \\
    Self-calibration & 1~iter & 1~iter & None \\ 
    Beam & $1\farcs4 \times 1\farcs3$ & $0\farcs9\times0\farcs7$ & $1\farcs8\times1\farcs3$ \\
    Image RMS\tablenotemark{a} $\left[\frac{\rm \mu Jy}{\rm beam} \right]$  & $\approx18$ & $\approx8.5$ & $\approx5.5$ \\
    Dynamic Range\tablenotemark{b} & $\sim10^{3.2}$ & $\sim10^{3.2}$ & $\sim185$\\
    \hline
    \hline
\end{tabular}
\end{center}
\tablenotetext{a}{RMS is the ``root mean squared'' of the flux density within a source free region near MaNGA 1$-$166919.}
\tablenotetext{b}{Dynamic Range is the ratio of the peak flux density to the RMS around the brightest sources in the field.}
\end{table*}

In \S\ref{sec:radio_img}, we present our analysis of the images produced from the calibrated data, while in \S\ref{sec:radio_spec} we present our measurements of the spectrum of this galaxy's radio emission.

\subsection{Radio Morphology}
\label{sec:radio_img}

\begin{figure*}
\begin{center}
    \includegraphics[width=0.31\textwidth]{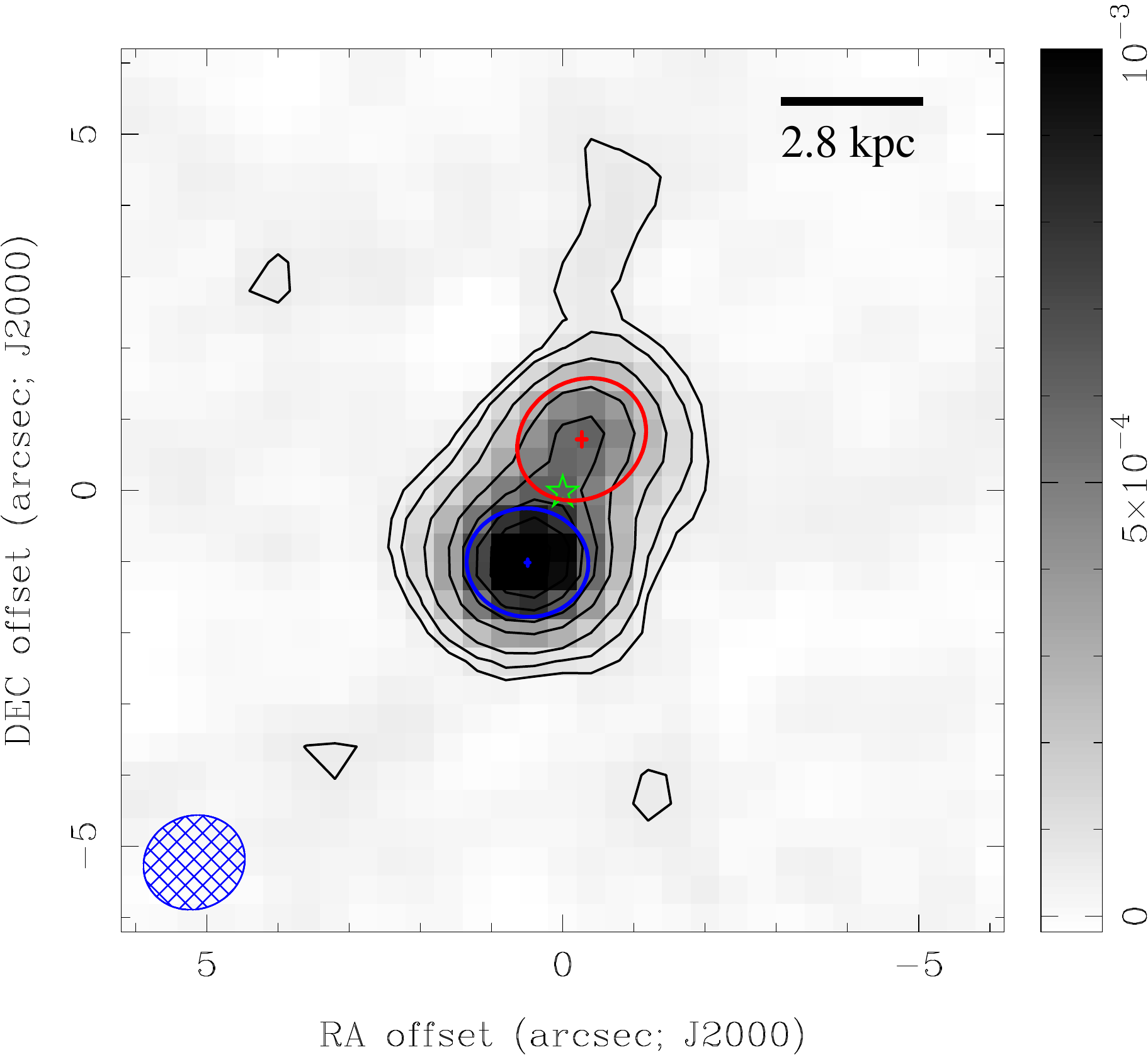}
    \includegraphics[width=0.31\textwidth]{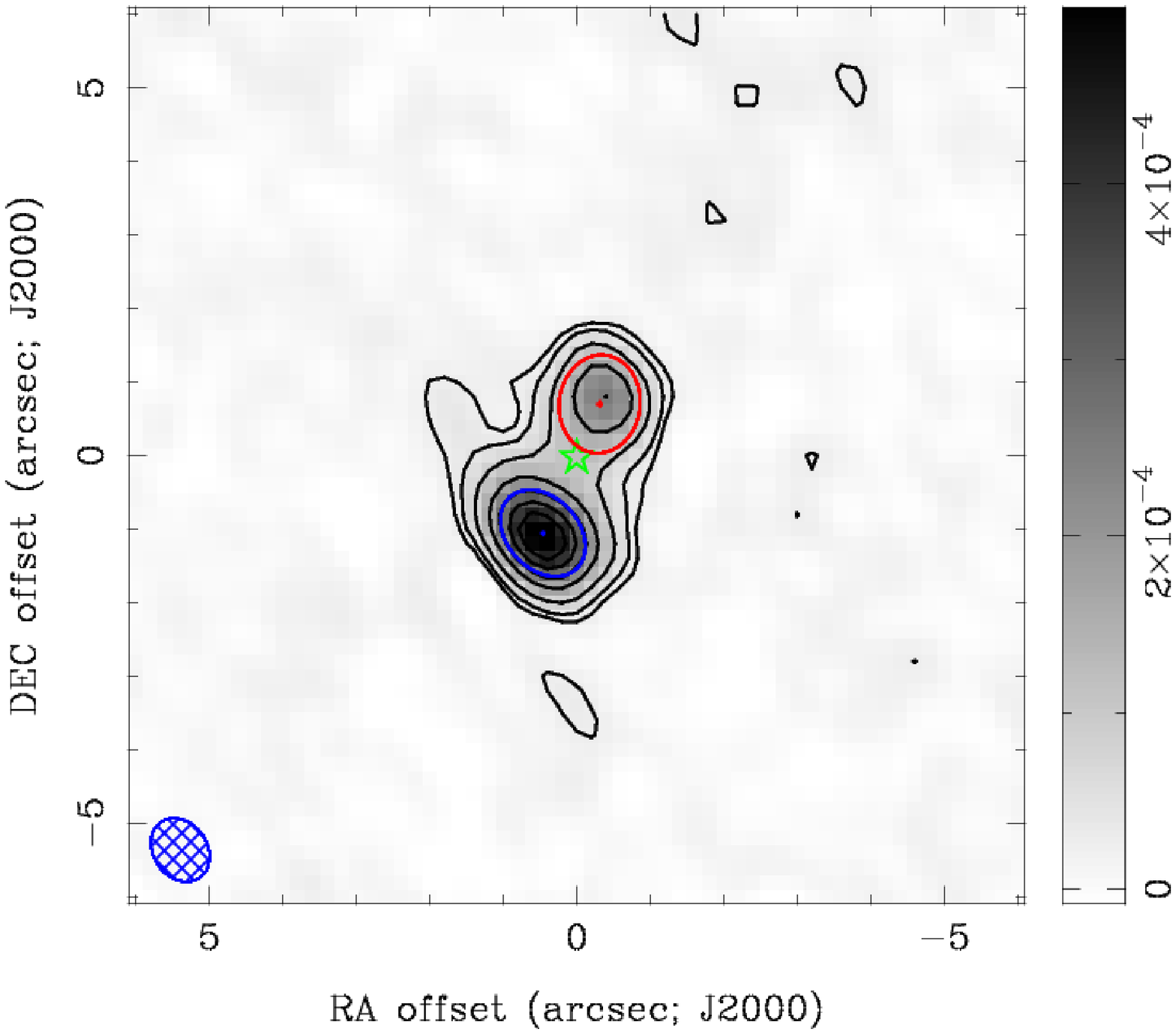}
    \includegraphics[width=0.31\textwidth]{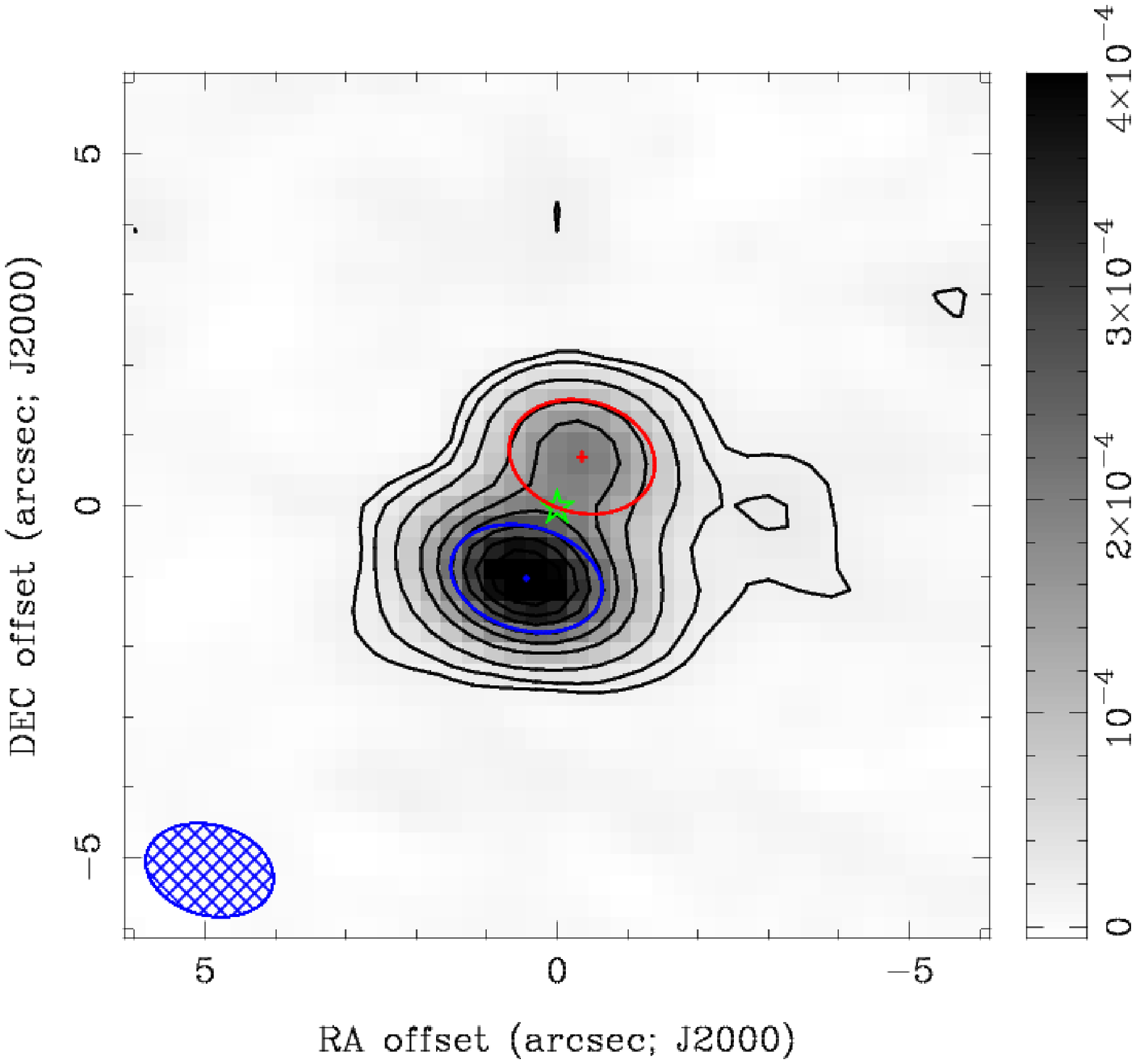}
    \includegraphics[width=0.31\textwidth]{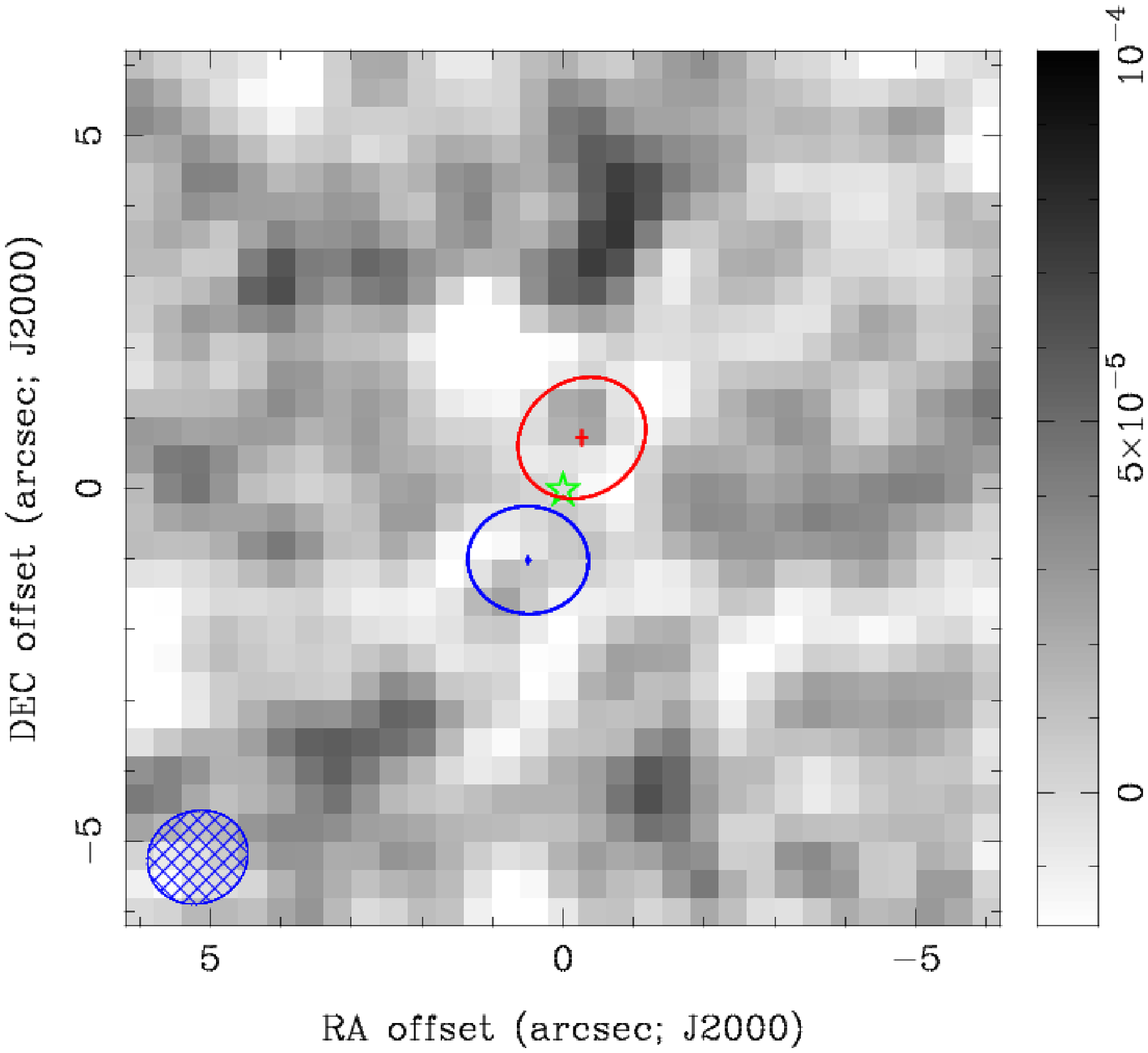}
    \includegraphics[width=0.31\textwidth]{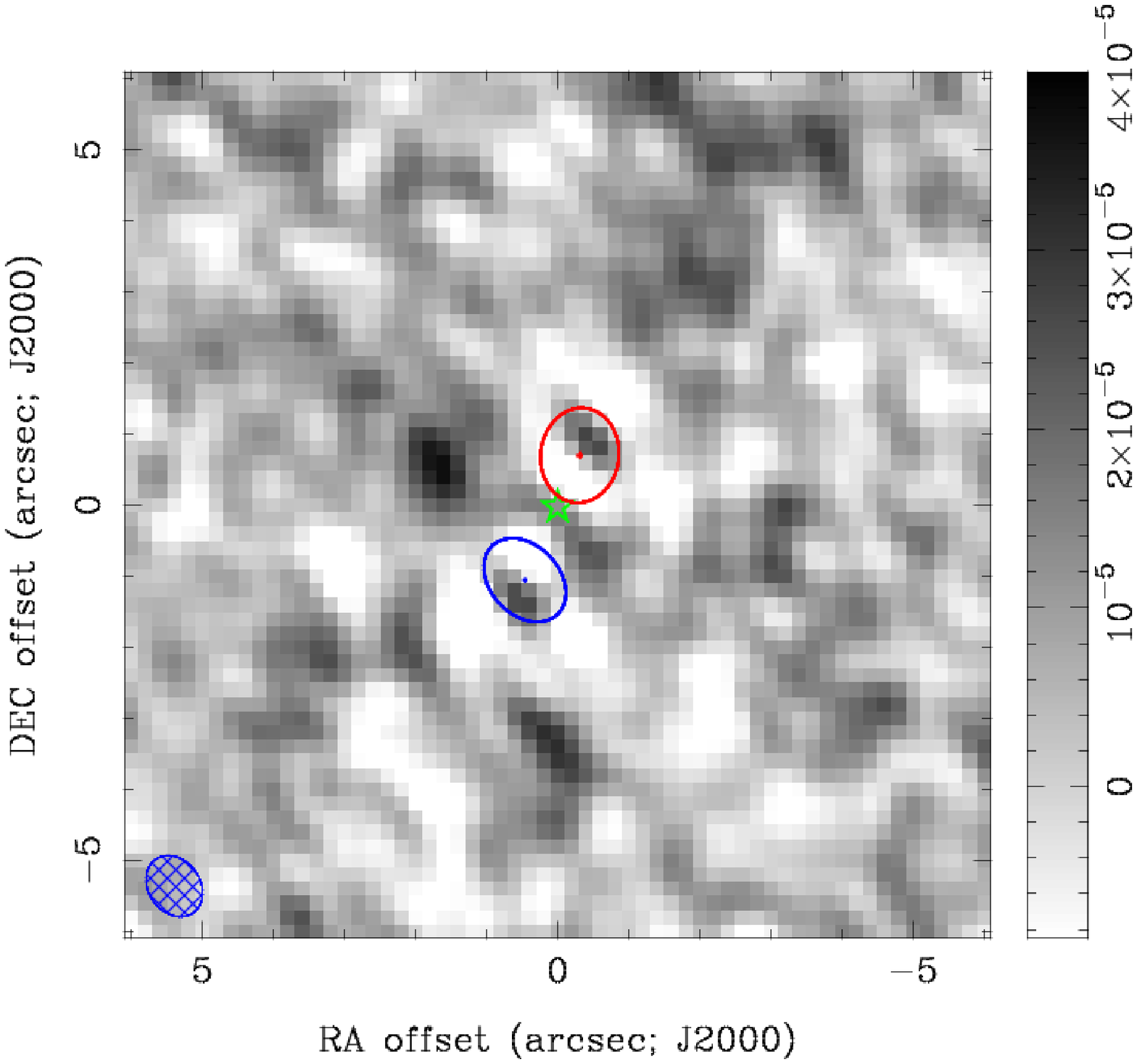}
    \includegraphics[width=0.31\textwidth]{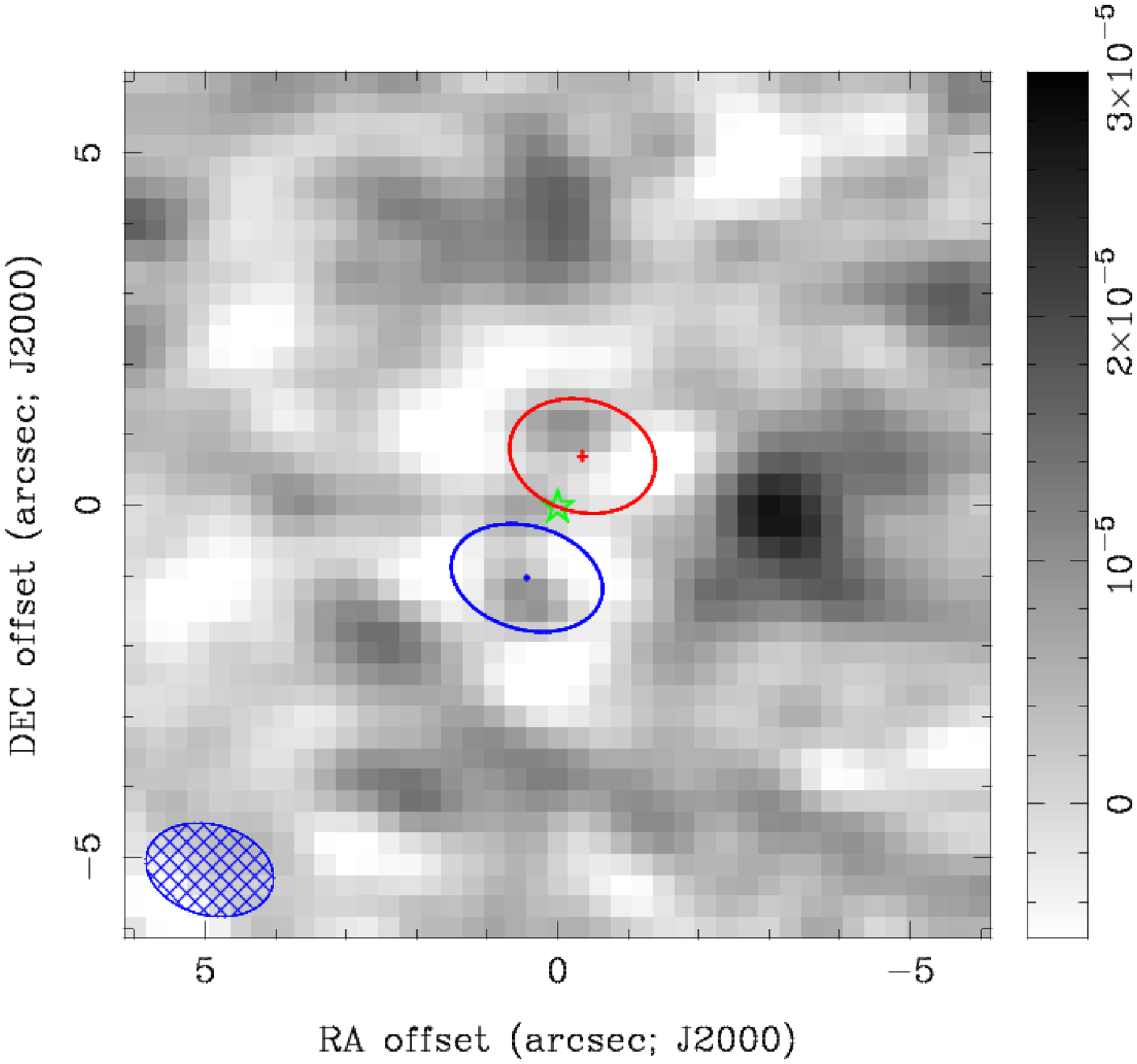}
\end{center}
\caption{L- (1-2~GHz, {\rm left}), S- (2-4~GHz, {\rm middle}) and C- (4-8~GHz, {\rm right}) band images of MaNGA $1-166919$.  The {\it top} figures are the images resulting from the deconvolution process described in \S\ref{sec:radio_img}), with the {\it bottom} is residual image after subtracting two Gaussians whose properties are given in Table \ref{tab:12arcsec_wideband_imfit}.   In the {\it top} figures, the black contours indicate intensity $3,5,10,\added{20,30,40,50},60 \times \mathrm{RMS}$ of the image (given in Table \ref{tab:radio_img_prop}), while in all figures the intensity scale is given in \replaced{Janskys}{Jy/beam}, the hatched ellipse indicates the size and orientation of the beam, the green star indicates the optical center of the galaxy, the red (blue) ellipse the size, orientation, and position of the NW (SE) lobe, and the red (blue) cross the center of the NW (SE) lobe.  In these images 1 arcsec corresponds to 1.4~kpc.}
\label{fig:wideband_maps}
\end{figure*}

\begin{figure}[tbh]
\centering
\includegraphics[width=0.4\textwidth,trim=0 0.3cm 0 1.0cm,clip]{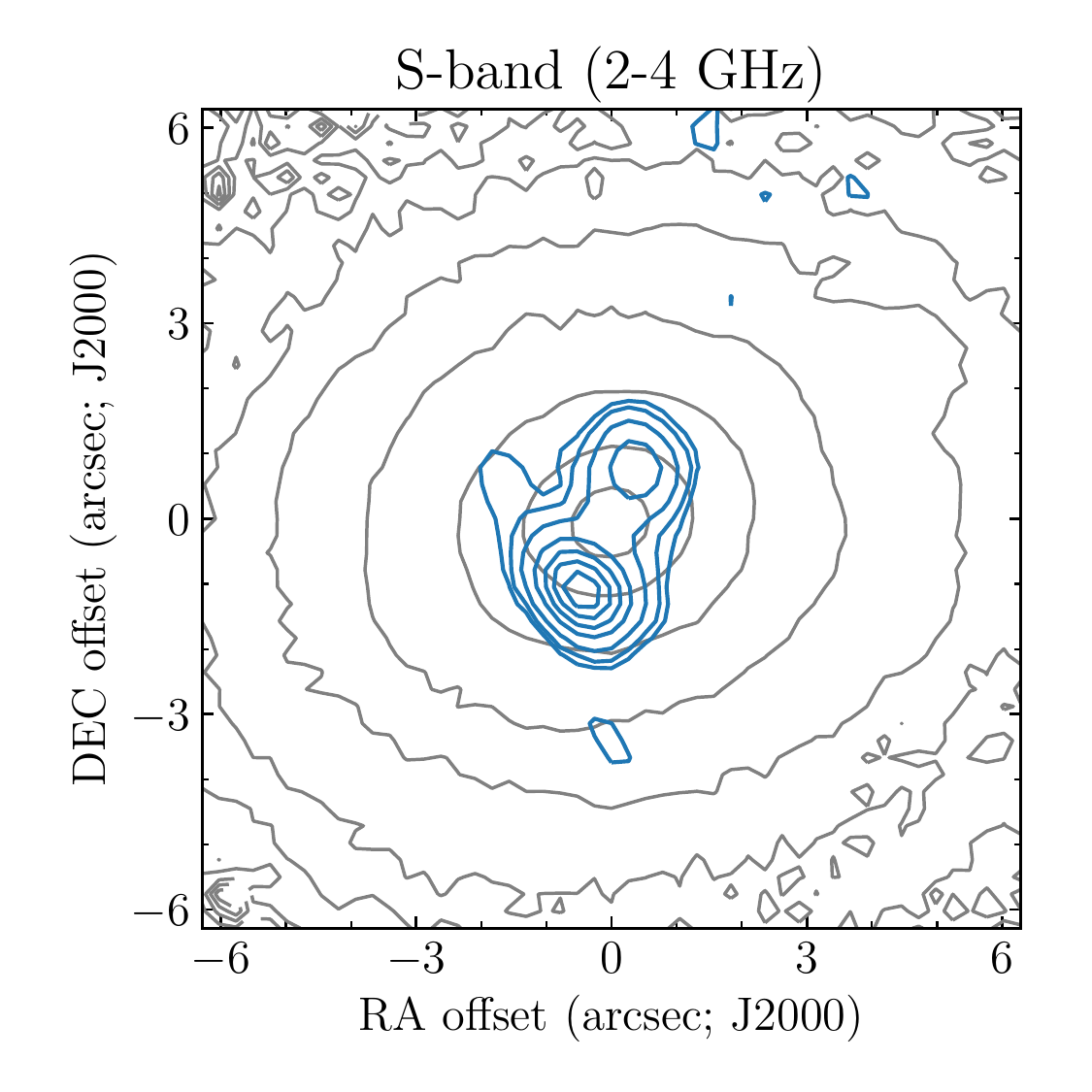}
\caption{Superposition of optical $z$-band MzLS image isophotes (gray color) and our highest spatial resolution radio image in S band (in blue). Optical image has a spatial resolution of $0\farcs84$, while 
S-band radio data -- $0\farcs9$. Radio isophotes are identical to that shown in Fig.~\ref{fig:wideband_maps}.
\label{fig:image_radio_isophotes}}
\end{figure}

As shown in Figure \ref{fig:wideband_maps}, in all three bands the radio emission consists of two lobes on opposite sides of the optical center of the galaxy, with the SE component consistently brighter than the NW.
Furthermore, the extent of the radio emission is considerably smaller than the optical size of this galaxy (see Figure~\ref{fig:image_radio_isophotes}). \added{The optical half light radius $R_{\rm eff} \approx 4\arcsec$ (Table \ref{tab:galprop}) is considerably smaller than the $9\arcsec - 18\arcsec$ largest angular scale of these observations (Table \ref{tab:radio_blob_obs}). As a result, this non-detection of larger scale radio emission is significant.} To measure the properties of lobes components, we fit the intensity distribution in the central $12\arcsec\times12\arcsec$ of each image with two Gaussians using the {\sc miriad} \citep{miriad} task {\tt imfit}.  The resultant properties of both components are listed in Table \ref{tab:12arcsec_wideband_imfit}, with the error in integrated flux density calculated using Equation 7 in the documentation of the \href{https://www.cv.nrao.edu/nvss/catalog.ps}{NVSS Source Catalog}.  The rms of the residual images (Table \ref{tab:12arcsec_wideband_imfit}) are comparable to that of the entire image (Table \ref{tab:radio_img_prop}), suggesting that no additional components are necessary.  This conclusions is supported by the lack of significant structures in the residual images (Figure \ref{fig:wideband_maps}) -- with the possible exception at $4-8$~GHz (C-band) where there is a $\sim5\sigma$ excess $\sim4\arcsec$ W of the center of the galaxy.  Furthermore, the centers and (deconvolved) extents of the two lobes are consistent across all three bands, suggesting these results are robust.

\begin{table*}[tb]
\caption{Parameters of two Gaussian fits to the radio emission of MaNGA 1-166919}
\label{tab:12arcsec_wideband_imfit}
\begin{center}
\begin{tabular}{cccc}
\hline
\hline
Band (Frequency) & L (1-2 GHz) & S (2-4 GHz) & C (4-8 GHz) \\
\hline
 & \multicolumn{3}{c}{\it SE Lobe} \\
Peak Flux Density $S_{\rm peak}$ $\left[ \frac{\rm mJy}{\rm beam} \right]$ & $1.22\pm0.05$ & $0.52\pm0.02$ & $0.45\pm0.01$ \\
Integrated Flux Density $S_{\rm Int}$ [mJy] & $1.68\pm0.11$ & $1.01\pm0.05$ & $0.62\pm0.03$ \\
$x-$offset\tablenotemark{a} & $0\farcs50\pm0\farcs03$ & $0\farcs46\pm0\farcs02$ & $0\farcs44\pm0\farcs03$ \\
$y-$offset\tablenotemark{a} & $-1\farcs02\pm0\farcs05$ & $-1\farcs06\pm0\farcs02$ & $-1\farcs033\pm0\farcs029$ \\
Major Axis $\theta_{\rm M}$ & $1\farcs71\pm0\farcs06$ & $1\farcs33\pm0\farcs04$ & $2\farcs19\pm0\farcs06$ \\
Minor Axis $\theta_{\rm m}$ & $1\farcs52\pm0\farcs09$ & $0\farcs97\pm0\farcs03$ & $1\farcs47\pm0\farcs06$  \\
Position Angle $\theta_{\rm PA}$ & $86^\circ\pm14^\circ$ & $43^\circ\pm4^\circ$ & $75^\circ\pm3^\circ$ \\
Deconvolved Size & $0\farcs99\times0\farcs69$ & $0\farcs97\times0\farcs64$ & $1\farcs18\times0\farcs71$ \\
Physical (deconvolved) size & $1.4~{\rm kpc} \times1.0~{\rm kpc}$ & $1.3~{\rm kpc} \times0.9~{\rm kpc}$ & $1.6~{\rm kpc} \times 1.0~{\rm kpc}$ \\
\hline
 & \multicolumn{3}{c}{\it NW Lobe} \\
Peak Flux Density $S_{\rm peak}$ $\left[\frac{\rm mJy}{\rm beam}\right]$ & $0.58\pm0.04$ & $0.24\pm0.02$ & $0.20\pm0.01$ \\
Integrated Flux Density $S_{\rm int}$ [mJy] & $0.95\pm0.04$ & $0.53\pm0.02$ & $0.29\pm0.01$ \\
$x-$offset\tablenotemark{a} & $-0\farcs27\pm0\farcs07$ & $-0\farcs31\pm0\farcs03$ & $-0\farcs35\pm0\farcs06$ \\
$y-$offset\tablenotemark{a} & $0\farcs72\pm0\farcs10$ & $0\farcs70\pm0\farcs04$ & $0\farcs69\pm0\farcs07$ \\
Major Axis $\theta_{\rm M}$ & $1\farcs89\pm0\farcs20$ & $1\farcs34\pm0\farcs09$ & $2\farcs10\pm0\farcs21$ \\
Minor Axis $\theta_{\rm m}$ & $1\farcs63\pm0\farcs13$ & $1\farcs10\pm0\farcs07$ & $1\farcs59\pm0\farcs09$ \\
Position Angle $\theta_{\rm PA}$ & $-55^\circ\pm27^\circ$ & $-6^\circ\pm13^\circ$ & $75^\circ\pm10^\circ$ \\
Deconvolved Size & $1\farcs22\times0\farcs96$ & $1\farcs06\times0\farcs71$ & $1\farcs00\times0\farcs94$ \\
Physical (deconvolved) size & $1.7~{\rm kpc} \times1.3~{\rm kpc}$ & $1.5~{\rm kpc} \times 1.0~{\rm kpc}$ & $1.4~{\rm kpc} \times1.3~{\rm kpc}$ \\
\hline
Residual Image RMS $\left[\frac{\rm \mu Jy}{\rm beam}  \right]$ & 23.4 & 10.9 & 7.0 \\
\hline
\hline
\end{tabular}
\tablenotetext{a}{Measured from the center of the field, $\alpha_{\rm J2000}=09:46:50.18, \delta_{\rm J2000} = +43:25:25.83$}
\end{center}
\end{table*}

\subsection{Radio Spectrum}
\label{sec:radio_spec}

In order to measure the physical properties of the radio emitting plasma, it is first necessary to identify the underlying emission mechanism.  This, in turn, requires determining the spectrum of the radio source, which we do using two methods:  measuring the flux density of both components in narrow-band images of this galaxy (\S\ref{sec:narrow_band}), and the spectral index $\alpha$ ($S_\nu \propto \nu^\alpha$) maps within each band produced by the MTMFS deconvolution described in \S\ref{sec:radio_img} (\S\ref{sec:spec_maps}). 

\subsubsection{Narrow Band Images}
\label{sec:narrow_band}

The integrated flux densities of the SE and NW lobes, as measured from the wideband images discussed above, differ significantly between the three observed bands.  To better measure how the flux density of these components changes with frequency, we first imaged contiguous subsets of the spectral windows (SPWs) within each band, and then -- as in \S\ref{sec:radio_img}, and fit the resultant image with two Gaussians to measure the integrated flux density of each lobe.  The SPWs were grouped such that there would be a $\gtrsim3\sigma$ change in the flux density of the fainter NW lobe assuming its continuum radio spectrum in this frequency radio is well-described by a single power law with spectral index $\alpha \sim -0.9$, the value resulting from fitting a power-law the flux densities derived from the wideband images (Table \ref{tab:12arcsec_wideband_imfit}).  These images were also produced using the {\sc casa} task {\tt tclean}, as in \S\ref{sec:radio_img}, again using Natural weighting and the same pixel and image size as before, using the ``multiscale'' deconvolved algorithm \citep{cornwell08} since the decreased fractional bandwidth of the dataset made an additional spectral term unnecessary.  We again used the {\sc miriad} task {\tt imfit} to fit the central $12\arcsec \times 12\arcsec$ region of each image with two Gaussians.  In these fits, the peak flux, size, and orientation of both ellipses were allowed to vary, \added{but the positions of the centers were fixed to the size obtained from wideband images given in Table \ref{tab:12arcsec_wideband_imfit}.  In general, the morphological properties of the two lobes derived from these fits were consistent with the values derived from the wideband images.}  The resultant integrated flux densities of both the SE and NW lobes are given in Table \ref{tab:narrow_fluxdens}.

\begin{table}[tbh]
\caption{Flux density of SE and NW lobes derived from narrow-band radio images}
\label{tab:narrow_fluxdens}
\begin{center}
\begin{tabular}{cccccc}
\hline
\hline
Band & SPW\tablenotemark{a} & $\nu$\tablenotemark{b} & $\Delta \nu$\tablenotemark{c} & $S_{\rm int}^{\rm SE}$\tablenotemark{d} & $S_{\rm int}^{\rm NW}$\tablenotemark{e} \\
$\cdots$ & $\cdots$ & [GHz] & [GHz] & [mJy] & [mJy] \\
\hline
L & $0-5$ & 1.200 & 0.384 & $1.76\pm0.02$ & $1.08\pm0.02$ \\
L & $6-9$ & 1.519 & 0.192 & $1.70\pm0.02$ & $0.90\pm0.02$ \\
L & $10-15$ & 1.839 & 0.352 & $1.60\pm0.02$ & $0.91\pm0.02$ \\
S & $0-3$ & 2.244 & 0.512 & $1.02\pm0.01$ & $0.60\pm0.01$ \\
S & $4-6$ & 2.691 & 0.384 & $0.96\pm0.01$ & $0.54\pm0.01$ \\
S & $7-10$ & 3.126 & 0.488 & $0.97\pm0.01$ & $0.51\pm0.01$ \\
S & $11-15$ & 3.691 & 0.600 & $0.92\pm0.01$ & $0.48\pm0.01$ \\
C & $0-3$ & 4.231 & 0.512 & $0.59\pm0.01$ & $0.33\pm0.01$ \\
C & $4-9$ & 4.871 & 0.768 & $0.61\pm0.01$ & $0.31\pm0.01$ \\
C & $10-16$ & 5.679 & 0.848 & $0.62\pm0.003$ & $0.30\pm0.003$ \\
C & $17-23$ & 6.551 & 0.896 & $0.56\pm0.01$ & $0.26\pm0.01$ \\
C & $24-31$ & 7.511 & 1.024 & $0.55\pm0.01$ & $0.23\pm0.01$ \\
\hline
\hline
\end{tabular}
\end{center}
\tablenotetext{a}{Range of Spectral Windows (SPWs) used in the associated Band.}
\tablenotetext{b}{Central Frequency of sub-band.}
\tablenotetext{c}{Range of frequency within sub-band.}
\tablenotetext{d}{Integrated Flux Density of the SE lobe.}
\tablenotetext{e}{Integrated Flux Density of the NW lobe.}
\end{table}

\begin{figure}[tbh]
    \centering
    \includegraphics[width=0.475\textwidth]{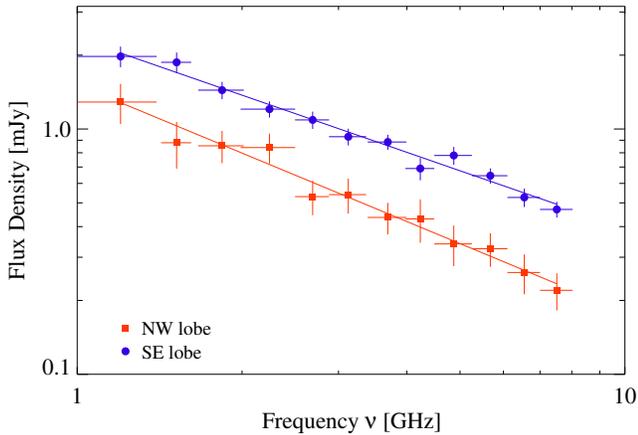}
    \caption{Radio spectrum of the NW and SE lobes as measured in the narrow-band radio images of this galaxy. The integrated flux densities of both components are given in Table \ref{tab:narrow_fluxdens} and the lines indicate the power-law fit whose parameters are given in Table \ref{tab:radspec_pow}.}
    \label{fig:spectrum}
\end{figure}

The resultant radio spectrum is shown in Figure \ref{fig:spectrum}, with the parameters derived from fitting a single power-law to the integrated flux densities of both the NW and SE radio lobes given in Table \ref{tab:radspec_pow}.  As shown in Figure \ref{fig:spectrum}, this model does a good job of reproducing the observed flux densities.  We also attempted to fit these flux densities with both a broken power-law (as expected if synchrotron cooling is important at higher frequencies) and a power-law with exponential cutoff at lower frequencies (as expected from free-free absorption along the line of sight), but these more complicated models did not produce significantly improved fits to the data.

\begin{table}[tbh]
    \caption{Parameters of Power-law Fits to Integrated Flux Density of NW and SE Radio Lobes }
    \label{tab:radspec_pow}
    \begin{center}
    \begin{tabular}{ccc}
    \hline
    \hline
    Parameter & SE lobe & NW lobe  \\
    \hline
    $S_{\rm 1.0}$\tablenotemark{a} & $2.35_{-0.26}^{+0.29}$~mJy & $1.52_{-0.32}^{+0.40}$~mJy \\
    $\alpha$ & $-0.78\pm0.09$ & $-0.93\pm0.18$ \\
    \hline
    \hline
    \end{tabular}     
    \end{center}
    \tablenotetext{a}{1~GHz integrated flux density}
\end{table}

\subsubsection{Spectral Index Maps}
\label{sec:spec_maps}
The technique used to get the spectral index maps assumes that the flux density $S$ at a particular frequency $\nu$ and sky position $(\alpha,\delta)$ can be accurately expressed as:
\begin{eqnarray}
\label{eqn:mtmfs}
S_\nu(\alpha,\delta) & = & S_{\nu_0}(\alpha,\delta) + \frac{\Delta S_{\nu_0}}{\Delta \nu} (\nu-\nu_0),
\end{eqnarray}
and then iteratively solves for the value of $S_{\nu_0}$ and $\frac{\Delta S_{\nu_0}}{\Delta \nu}$ at each location on the sky.  As implemented in the {\sc casa} command {\tt widebandpbcor}, the derived value of $\frac{\Delta S_{\nu_0}}{\Delta \nu}$ is used to calculate the spectral index $\alpha$ within the frequency range of the input data in each pixel of the resultant image.

\begin{figure*}[tbh]
    \begin{center}
    \includegraphics[height=0.2\textheight]{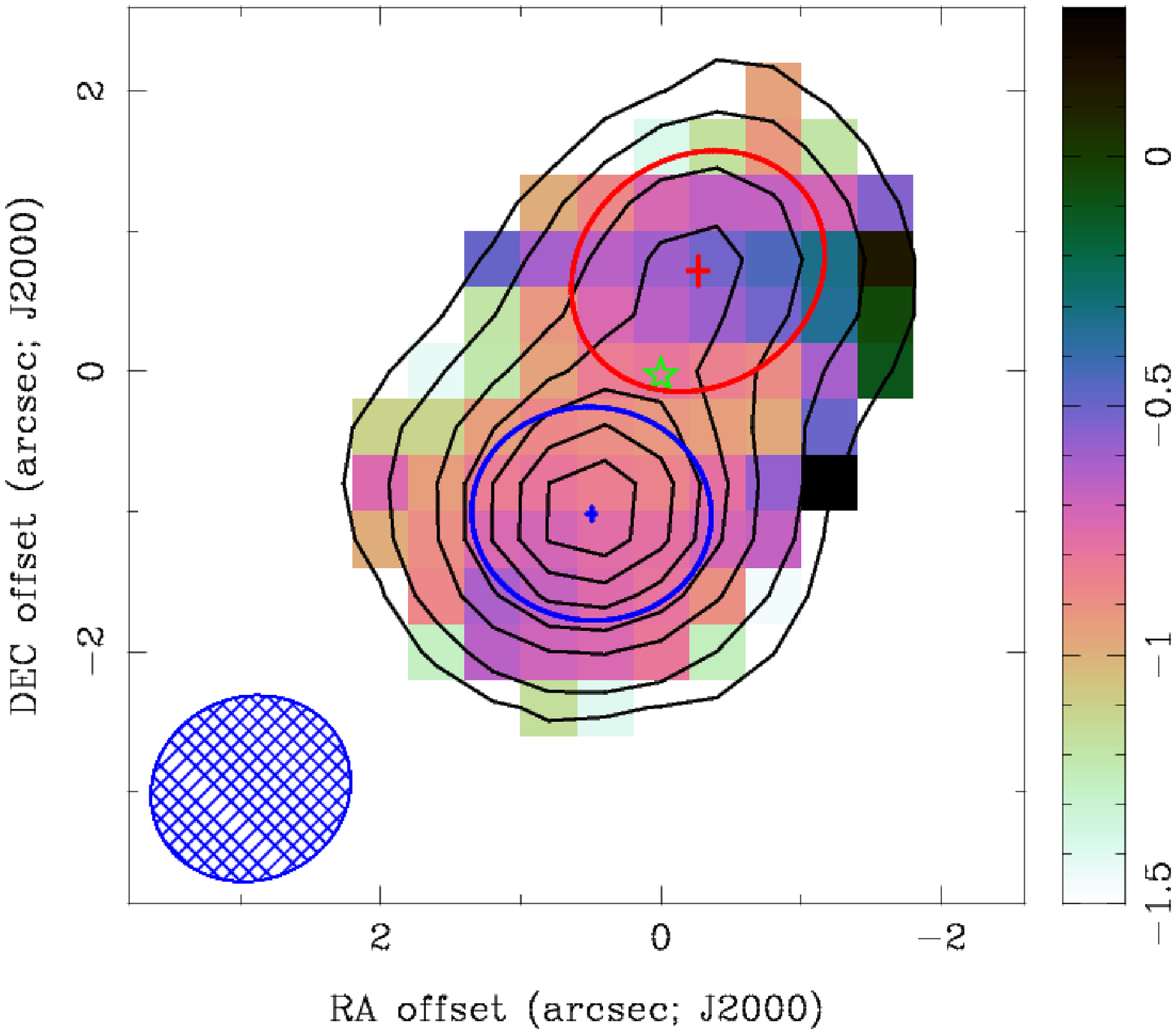}
    \includegraphics[height=0.2\textheight]{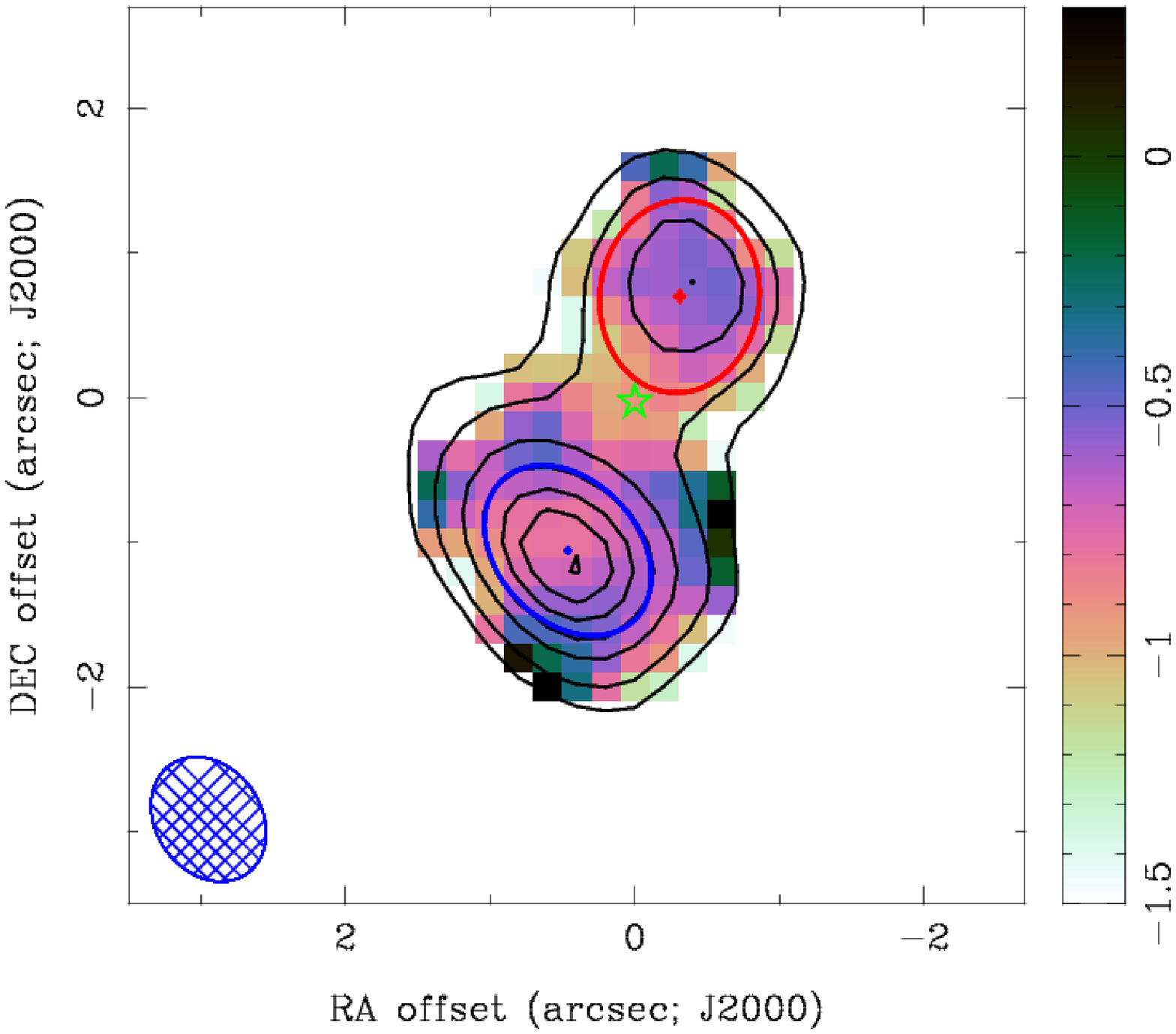}
    \includegraphics[height=0.2\textheight]{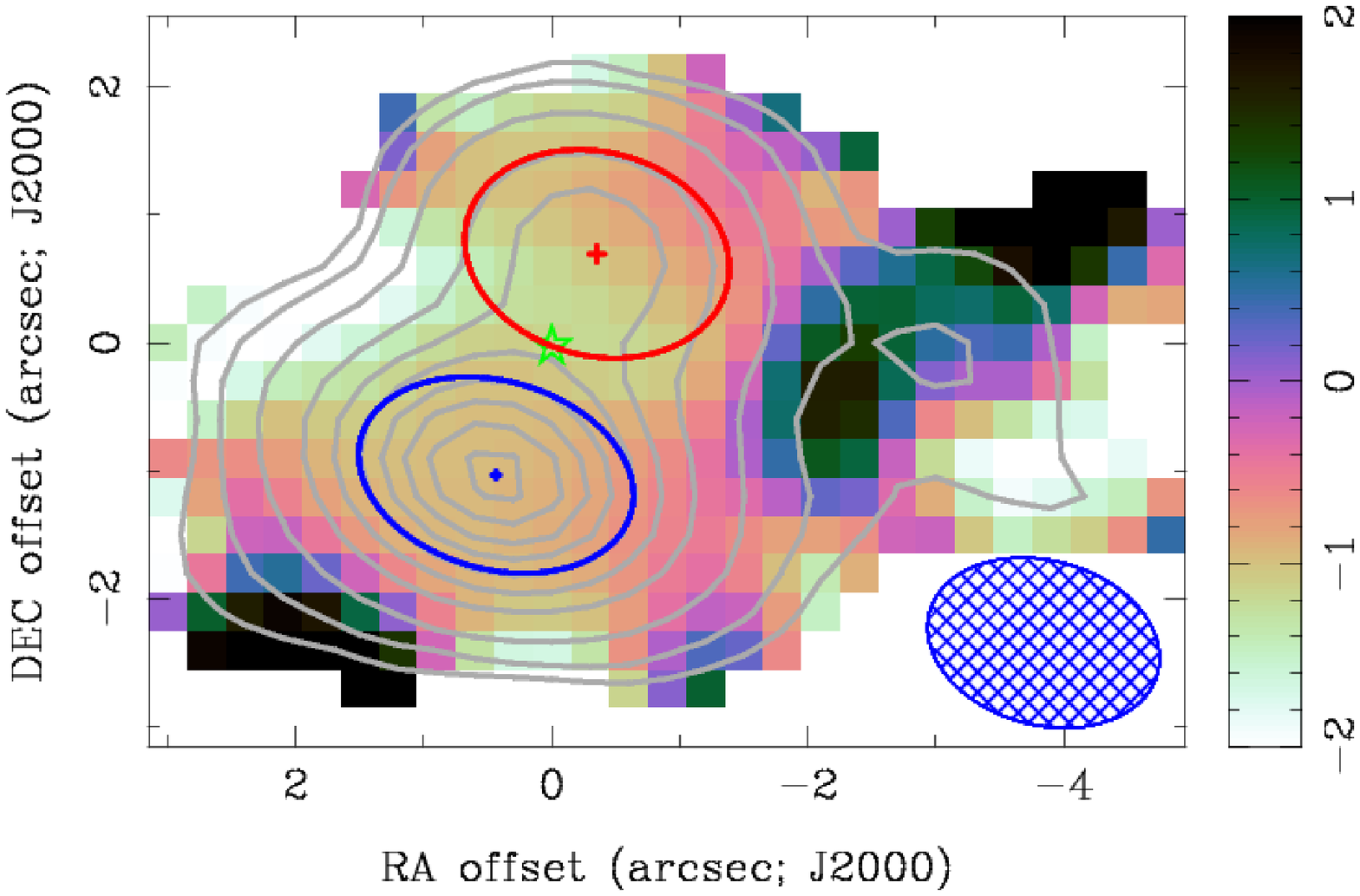}
    \end{center}
    \caption{Spectral index $\alpha$ within L- (1-2~GHz; {\it left}), S- (2-4~GHz; {\it middle}), and C- (4-8~GHz; {\it right}) band, derived using the procedure described in \S\ref{sec:spec_maps}.  In all three images the blue hatched ellipse shows the size and orientation of the beam, the green star indicates the optical center of the galaxy, the red (blue) ellipse shows the positions, size, and orientation of, and the red (blue) cross indicates the center of, the Gaussian component corresponding to the NW (SE) lobe as given in Table \ref{tab:12arcsec_wideband_imfit}.  The black contours in the {\it left} and {\it middle} images indicate flux densities $5,10,20,30,40,50,60\times \mathrm{RMS}$ of the image, while the grey contours in the {\it right} image indicate $3,5,10,20,30,40,50,60,70,80\times \mathrm{RMS}$ of the image, as given in Table \ref{tab:radio_img_prop}.}
    \label{fig:alpha_map}
\end{figure*}

As shown in Figure \ref{fig:alpha_map}, in all three bands the spectral index of pixels in the SE and NW are $\alpha \lesssim -0.5$, consistent with the value derived from the analysis described in \S\ref{sec:narrow_band} (Table \ref{tab:radspec_pow}).  However, in all three bands the values of  $\alpha$ in the SE lobe are, in general, steeper (more negative) than those in the NW lobe, with a difference in spectral index $\Delta \alpha \sim 0.1-0.2$ (Figure \ref{fig:alpha_map}) which may not be statistically significant.  
However, in C-band, this analysis indicates the presence of flat spectrum ($\alpha \gtrsim 0$) radio emission.  For example, such a spectral index is measured for the $\sim3-5\sigma$ peak located W of the two lobes. This suggests this component has a different physical origin than the two lobes which will be discussed in \S\ref{sec:hostgal}.

\section{Integral Field Unit Data Analysis}
\label{analysis_optical}

As mentioned in \S\ref{sec:intro}, previous studies of MaNGA 1-166919 suggest that it contains a kpc-scale outflow (e.g. \citealt{Wylezalek2017}).  In this section, we analyze data taken on this source during two Integral Field Unit (IFU) observations, one at the Apache Point Observatory as part of the Sloan Digital Sky Survey IV (SDSS-IV; \citealt{Blanton2017AJ....154...28B}) Mapping Nearby Galaxy at Apache Point Observatory (MaNGA; \citealt{Bundy2015ApJ...798....7B}) project, the other using Gemini-North telescope with the Multi-Object Spectrograph (GMOS; \citealt{gmos, gmos-ifu}).  As listed in Table~\ref{tab:IFUs_MaNGA_GMOS}, these two datasets are complimentary: the MaNGA data spans a wider range of $\lambda$ and covers a larger fraction of the galaxy, while the GMOS data has better angular and spectral resolution.    While results from both datasets were previous derived by \citet{Wylezalek2017}, we have used a different technique to analyze the MaNGA (\S\ref{sec:manga}) and GMOS (\S\ref{sec:gmos}) data as described below -- which in general agrees with the earlier work by \citet{Wylezalek2017}. 

\begin{table}[tbh]
\caption{Properties of Analyzed Integral Field Unit datasets}
\label{tab:IFUs_MaNGA_GMOS}
\begin{center}
\begin{tabular}{lcc}
\hline
\hline
Property & MaNGA & GMOS \\
\hline
Wavelength range [\AA] & $3600-10000$ & $4000-7000$ \\
Spectral resolution\tablenotemark{a} $R$ & $\approx1900$ & $\approx4000$ \\
Field-of-View [\arcsec]  & $17.5 \times 19$\tablenotemark{b} & $3.5 \times 5$ \\
Field-of-View [kpc] & $24.4 \times 26.5$ & $4.9 \times 7$ \\
Spatial resolution [\arcsec] & $\approx2.5$\ & $\approx0.9$ \\
\hline
\hline
\end{tabular}
\end{center}
\tablenotetext{a}{For the MaNGA data, this is defined as $\lambda/\delta \lambda$ at the observed position of \Ha\ line.  For this GMOS data, this value was derived by \citet{Wylezalek2017}.}
\tablenotetext{b}{Corresponds to the largest dimensions of the hexagonal MaNGA Field of View in the final spectral cube after accounting for dithering during the observation.}
\end{table}

\subsection{MaNGA data}
\label{sec:manga}

The MaNGA survey consists of IFU \citep{Drory2015AJ....149...77D} observations of 10000 galaxies in the nearby Universe chosen to collective sample a wide range of stellar mass and color \citep{Wake2017AJ....154...86W}.  Each galaxy was observed using bundles of 2\arcsec\ fibers covering (1.5-2.5)$\times$ the effective high-light radius of the target, with each galaxies observed with 3 dithered exposures to fill in the gaps between the fibers in a bundle \citep{Law2015AJ....150...19L, Yan2016AJ....152..197Y}.  These data were then calibrated using the procedure described by \citep{Yan2016AJ....151....8Y}, and reduced using the pipeline developed by \citet{Law2016AJ....152...83L}.  The flux-calibrated MaNGA spectral cube for MaNGA 1-166919 was made publicly available in the Fifteenth Data Release 
of the Sloan Digital Sky Survey (DR15; \citet{Aguado2019ApJS..240...23A}), as well as results derived from data analysis pipeline described by \citet{Westfall2019AJ....158..231W} -- which includes measurements of the emission line properties made using the procedure described by \citet{Belfiore2019AJ....158..160B}.  While these results can be accessed using the Marvin toolkit \citep{Cherinka2019AJ....158...74C}, we analyzed these dataset using the procedure described below aimed at better measuring the properties of the outflowing material.  In \S\ref{sec:manga_stars}, we describe how we measured the properties of the stellar population of MaNGA 1-166919, and in \S\ref{sec:emision_line_fitting} we describe the method using the measure properties of the ionized gas in this galaxy.

\subsubsection{Stellar fit}
\label{sec:manga_stars}

\begin{figure*}[bth]
\begin{center}
    \includegraphics[width=\textwidth,trim=0 0 0 0.65cm,clip]{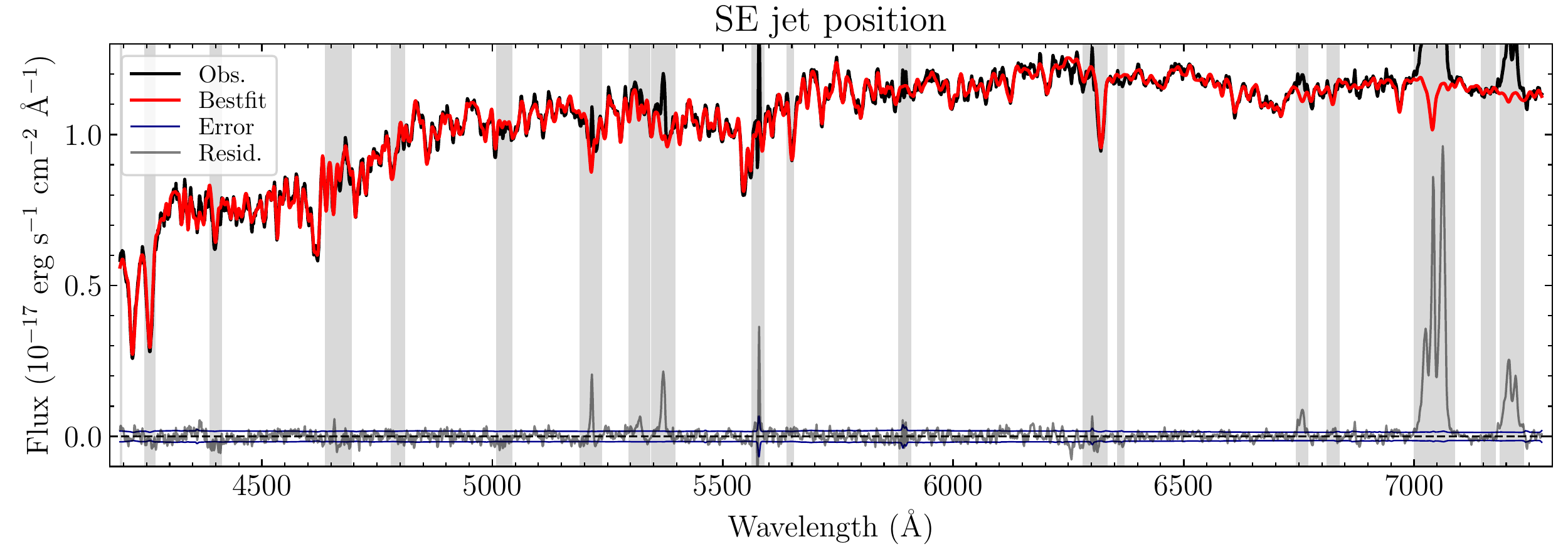}    
\end{center}
\caption{The observed ({\it black}) and best fit stellar spectrum ({\it red}) at the center of the SE radio lobe (Table \ref{tab:12arcsec_wideband_imfit}).  The grey boxes indicates regions excluded from our fits since data at these wavelengths are contaminated by galaxy emission lines, strong night sky lines, or NaD absorption feature poorly reproduced by stellar population models.}
\label{fig:spectrum_se}
\end{figure*}

We used the {\sc NBursts} full spectral fitting package \citep{nburst_a, nburst_b} to both derive the properties of the stellar population of this galaxy, as well as determine the stellar contribution to its spectrum.  This method uses a $\chi^2$ minimization algorithm to fit the spectrum in each (spatial) pixel with a model derived from broadening the spectrum predicted from a stellar population model with a Gauss-Hermite parameterized distribution \citep{vanderMarel1993ApJ...407..525V} of the line-of-sight velocity at this position.  To avoid systematically biasing the resultant parameters, we masked wavelengths corresponding to strong emission lines (e.g. Figure \ref{fig:spectrum_se}).  The stellar spectra were chosen from a grid of PEGASE.HR high-resolution simple stellar population (SSP) models \citep{LeBorgne+04} based on the ELODIE3.1 empirical stellar library \citep{Prugniel07} assuming a Salpeter initial mass function  \citep{salpeter55}, pre-convolved with the line spread function provided within the MaNGA datacube in order to account for instrumental broadening. 
While the derived properties of the ionized gas does depend on the choice of stellar models, the high signal-to-noise of our data suggests this will be a small effect \citep{chen18}.  An example of the results of this procedure is shown in Figure \ref{fig:spectrum_se}.

\begin{figure*}[hbt]
\centering
\includegraphics[trim=0.00cm 0 0.5cm 0,clip,scale=0.57]{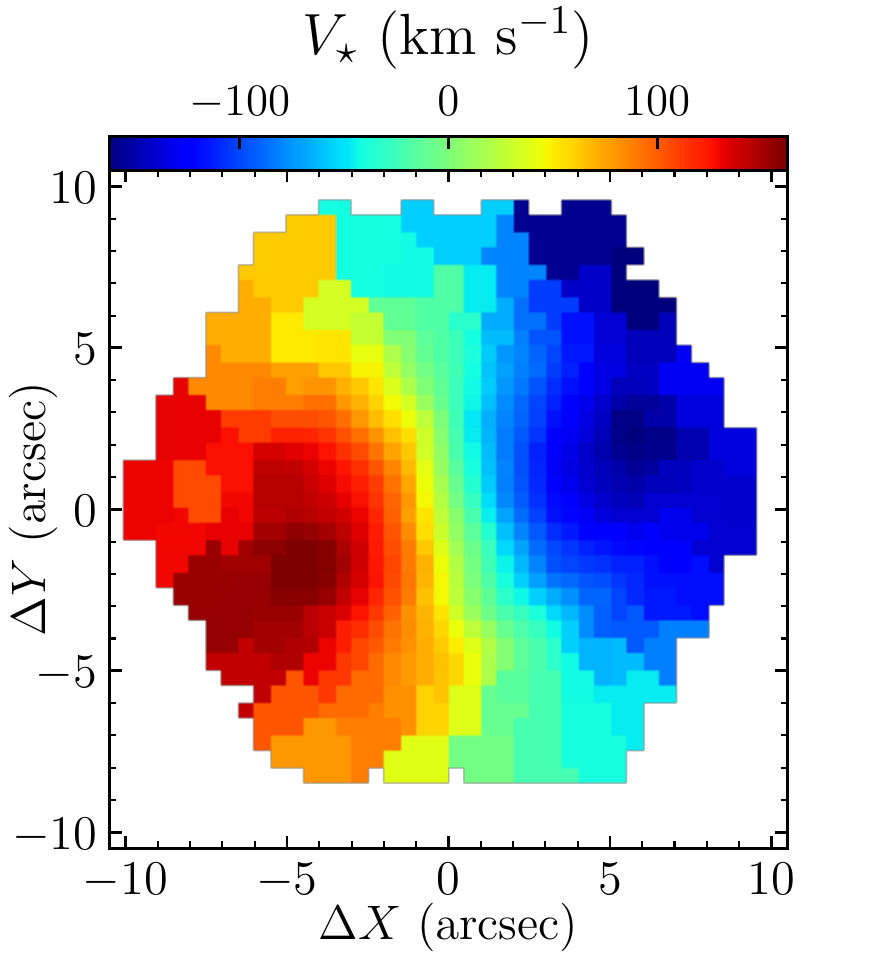}
\includegraphics[trim=0.96cm 0 0.5cm 0,clip,scale=0.57]{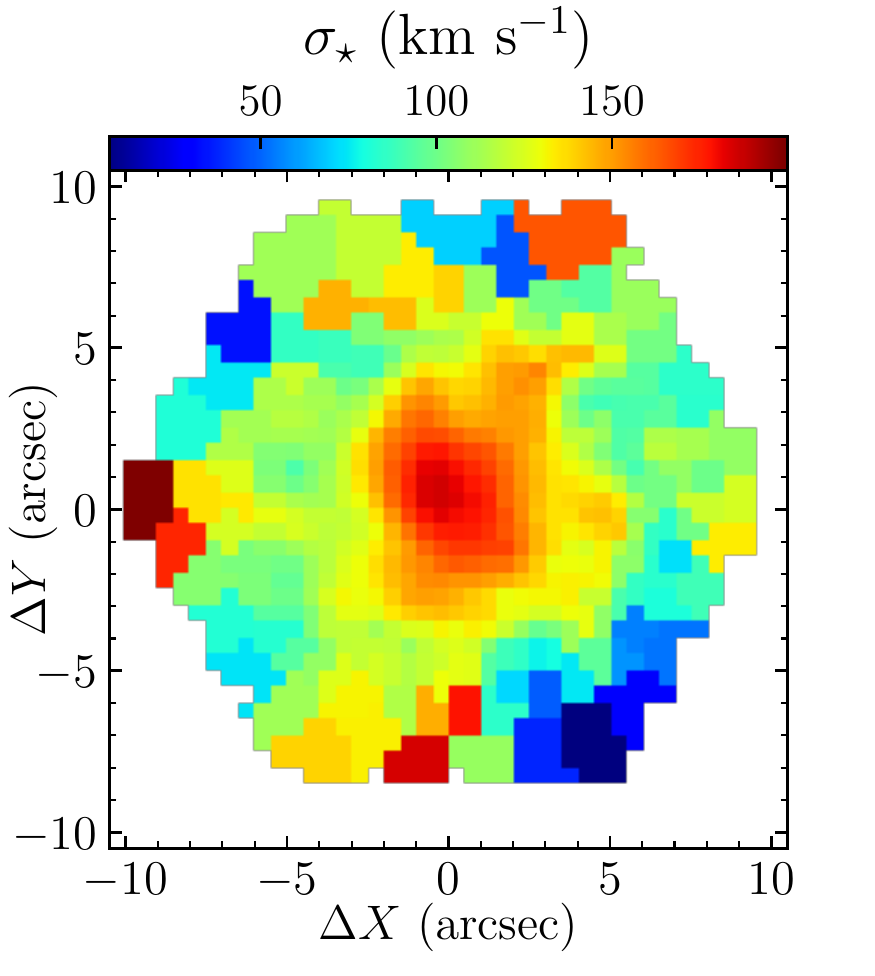}
\includegraphics[trim=0.96cm 0 0.5cm 0,clip,scale=0.57]{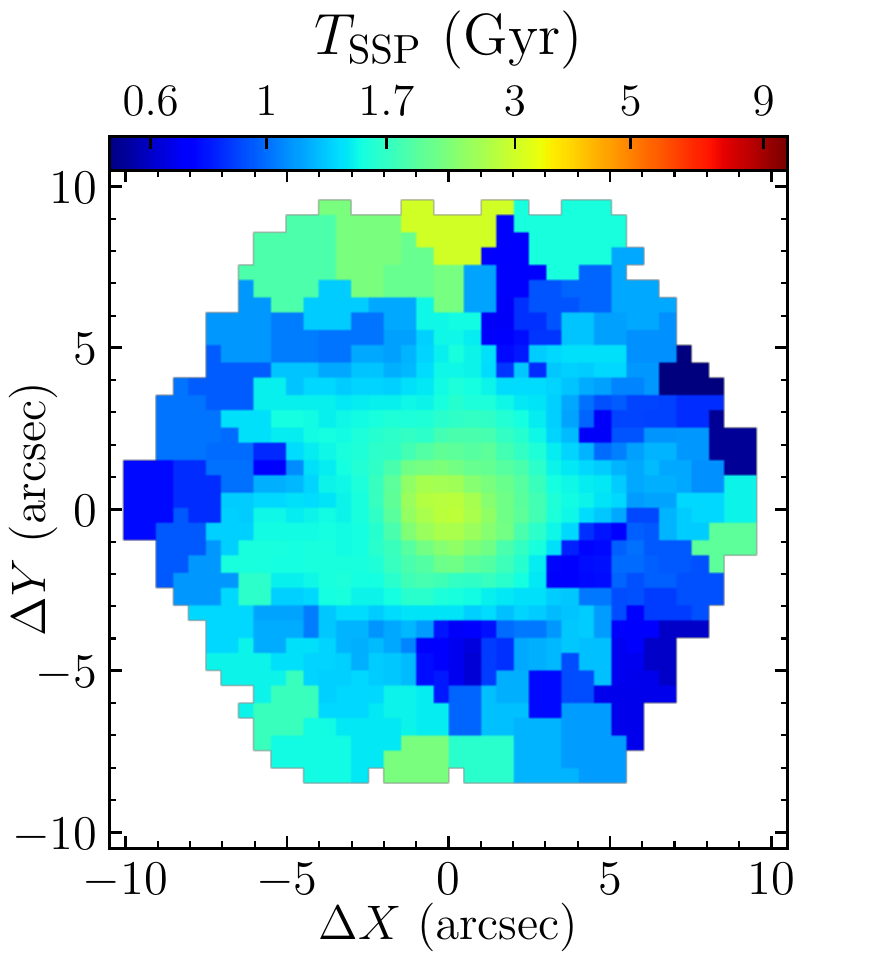}
\includegraphics[trim=0.96cm 0 0.5cm 0,clip,scale=0.57]{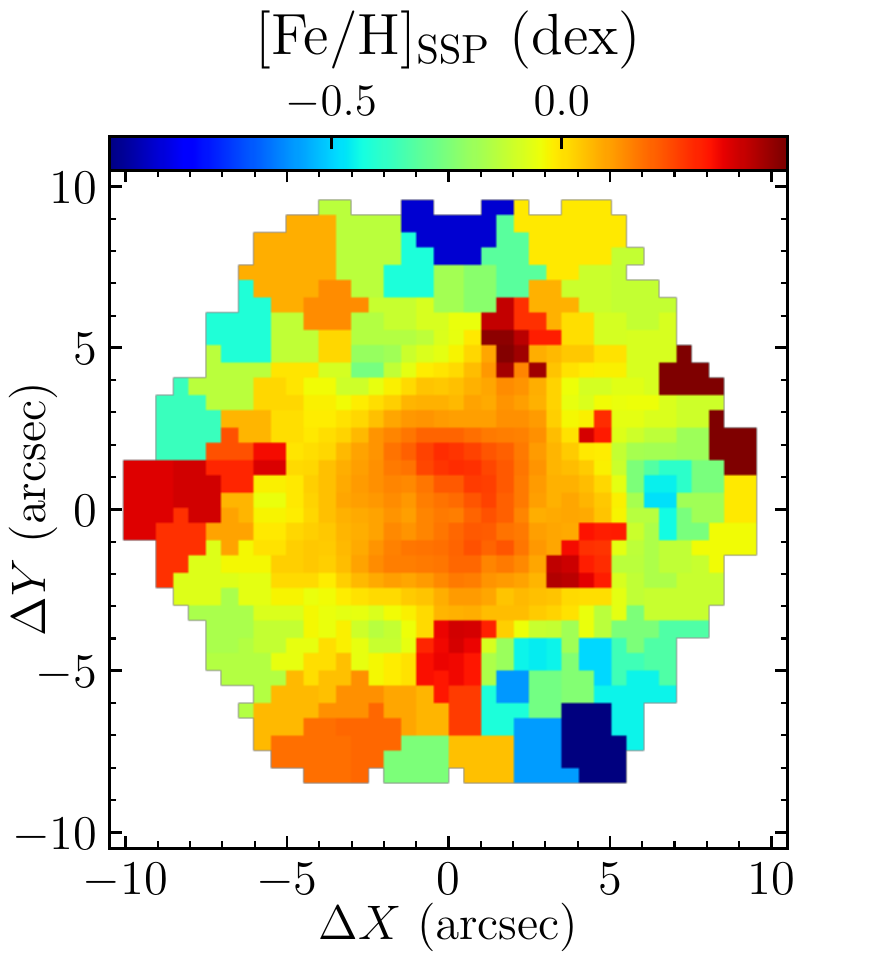}
\caption{Stellar line-of-sight velocity $V_\star$, velocity dispersion $\sigma_\star$, age ($T_{\rm SSP}$, and metallicity  $[{\rm Fe/H}]_{\rm SSP}$ recovered from the MaNGA spectral cube using \textsc{NBursts} full spectral fitting tool \citep{nburst_a} and grid of the simple stellar population (SSP) models \textsc{pegase.hr} \citep{LeBorgne+04}.
\label{fig:manga_stellarprop}}
\end{figure*}

\begin{figure*}
    \centering
    \includegraphics[width=\textwidth]{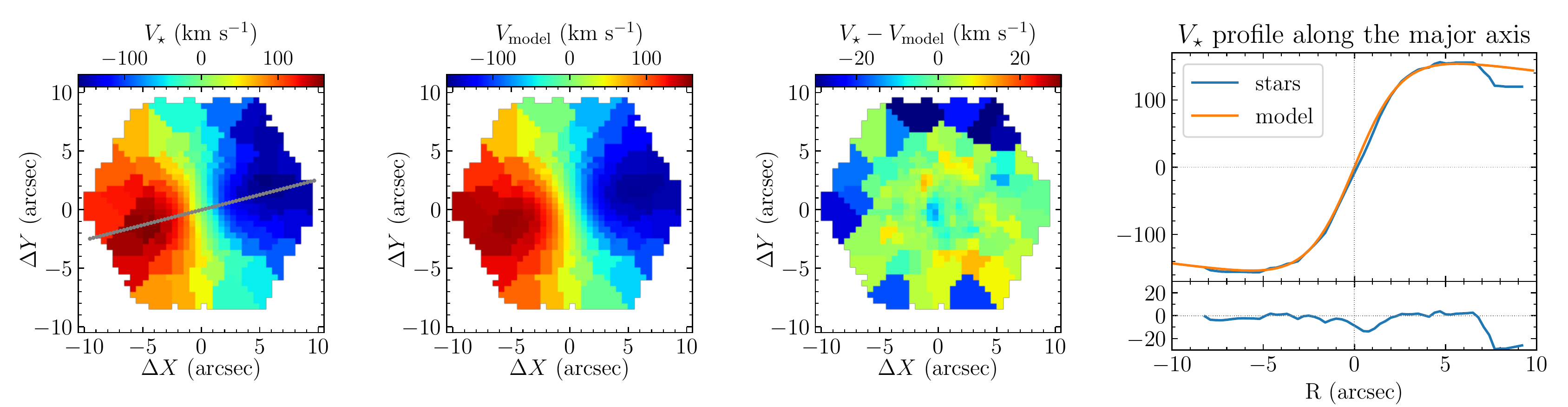}
    \caption{\textit{Left to Right}: stellar velocity field derived from MaNGA spectral cube, the bestfit model of the velocity field using our parameterization of the rotation curve (Equations~\ref{eqn:vlos} \& \ref{eqn:vphi}) maps of residuals.  The last panel shows the stellar line-of-sight velocity $V_\star$ (blue line) along the major kinematical axis (gray line in the first panel) along with predicted velocity profile from our best fit parameters (orange). The small bottom panel in the last figure shows the difference between the observed $V_\star$ profile and the model.}
    \label{fig:Vstar_tanh_compare}
\end{figure*}

For each spectrum, this model returns the
stellar line-of-sight velocity $V_\star$, velocity dispersion $\sigma_\star$, and the equivalent stellar age $T_{\rm SSP}$ and metallicity [Fe/H]$_{\rm SSP}$ of the best-fit SSP.  To ensure the derived parameters are reliable, we binned all spaxels with a signal-to-noise ratio (estimated in the stellar continuum spectrum in a narrow 10\AA\ spectral window centered on 5100~\AA\ in the galaxy's restframe) ${\rm SNR} > 1$ into spatial regions with ${\rm SNR} \geq 20$ using the adaptive Voronoi algorithm developed by \citet{cappellari_copin2003}.  As shown in Figure \ref{fig:manga_stellarprop}, the spatial distribution of stellar velocity $V_\star$ is suggestive of a regularly rotating stellar disk -- consistent with the spiral morphology inferred from its optical morphology (e.g., 88\% of Galaxy Zoo users classified this source as a spiral galaxy; \citealt{lintott08, lintott11}).  Furthermore, the increased stellar velocity dispersion $\sigma_\star$ and stellar age $T_{\rm SSP}$ observed towards the center of galaxy is indicative of a stellar bulge, with a peak velocity dispersion of $\sigma_\star \sim 170~{\rm km~s}^{-1}$ and light-weighted velocity distribution (here we averaged $\sqrt{v_\star^2 + \sigma_\star^2}$ values within an elliptical aperture of 4\arcsec\ size using an ellipticity $\epsilon=1-b/a=0.12$ ($b$ and $a$ - minor and major semi-axes) estimated from the optical image isophotes) within the central 4\arcsec\ of $\sigma_\star = 161.8\pm0.4~{\rm km~s}^{-1}$. 

As shown in Figure \ref{fig:manga_stellarprop}, the distribution of the stellar line-of-sight velocity of the stellar component $V_\star$ is indicative of a regular rotating stellar population.  We modelled this stellar velocity field  by assuming that, for a spaxel located at a  particular $(x,y)$ measured relative to the center of the galaxy, the emitting stars have a line-of-sight velocity:
\begin{eqnarray}
    \label{eqn:vlos}
    V_{\rm LOS}(x, y) & = & V_{\rm sys} + V_\phi(x, y) \frac{\cos \phi \sin i}{g},
\end{eqnarray}
where the azimuthal rotational velocity in the center of the disk (its ``galactic plane'') is:
\begin{eqnarray}
    \label{eqn:vphi}
    V_\phi(R) & = & V_0 \left( \tanh{\pi \frac{R}{R_0} } + c \frac{R}{R_0} \right),
\end{eqnarray}
with  $V_{\rm sys}$ is the systemic velocity of the galaxy, $g = \sqrt{\sec^2 i - \cos^2 \phi \tan^2 i}$ is a geometrical factor converting the projected sky distance $r \equiv \sqrt{x^2+y^2}$ between a spaxel at the center of the galaxy to the the distance along the galactic plane $R = g r$, and $i$ is the inclination angle of the disk, $R_0$ is a radius where velocity riches a constant maximum value $V_0$ in case of $c=0$, and $c$ describes the growth ($c>0$) or decline ($c<0$) of $V_\phi$ for $R > R_0$.  We then determined the values of $V_{\rm sys}$, $V_0$, $R_0$, $c$, $i$ and the orientation of the galactic disk on the plane of the sky using a $\chi^2$ minimization routine.  \added{This parameterization of a regularly rotating disk is similar to that presented by \citet{Chung2020_RotationCurve}.  As shown in Figure~\ref{fig:Vstar_tanh_compare}, this model is able to reproduce both the observed 1D and 2D stellar velocity distributions.}

\subsubsection{Emission-line fit}
\label{sec:emision_line_fitting}

\begin{figure*}[tbh]
    \begin{center}
    \includegraphics[trim=0 0cm 0 0.75cm,clip, width=\textwidth]{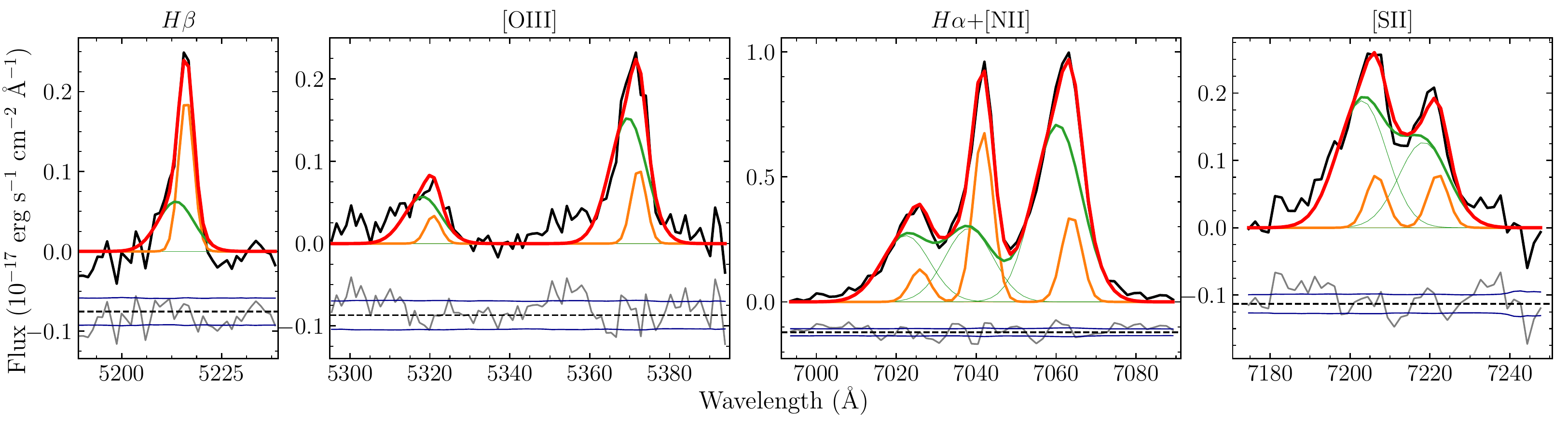}
    \end{center}
    \caption{{\it Left to Right}: H$\beta$, [O\iii] doublet lines, [N\ii]+H$\alpha$ and [S\ii] emission lines at the position of the SE lobe (Table \ref{tab:12arcsec_wideband_imfit}), as derived using the procedure described in \S\ref{sec:emision_line_fitting}.  The predicted spectrum of the main and outflow component are shown in orange and green, respectively, while the red line indicates the sum of the two.}
    \label{fig:lines_se}
\end{figure*}

\begin{figure}[hbt]
    \centering
    \includegraphics[width=0.475\textwidth,trim=0 0 0 1.2cm,clip]{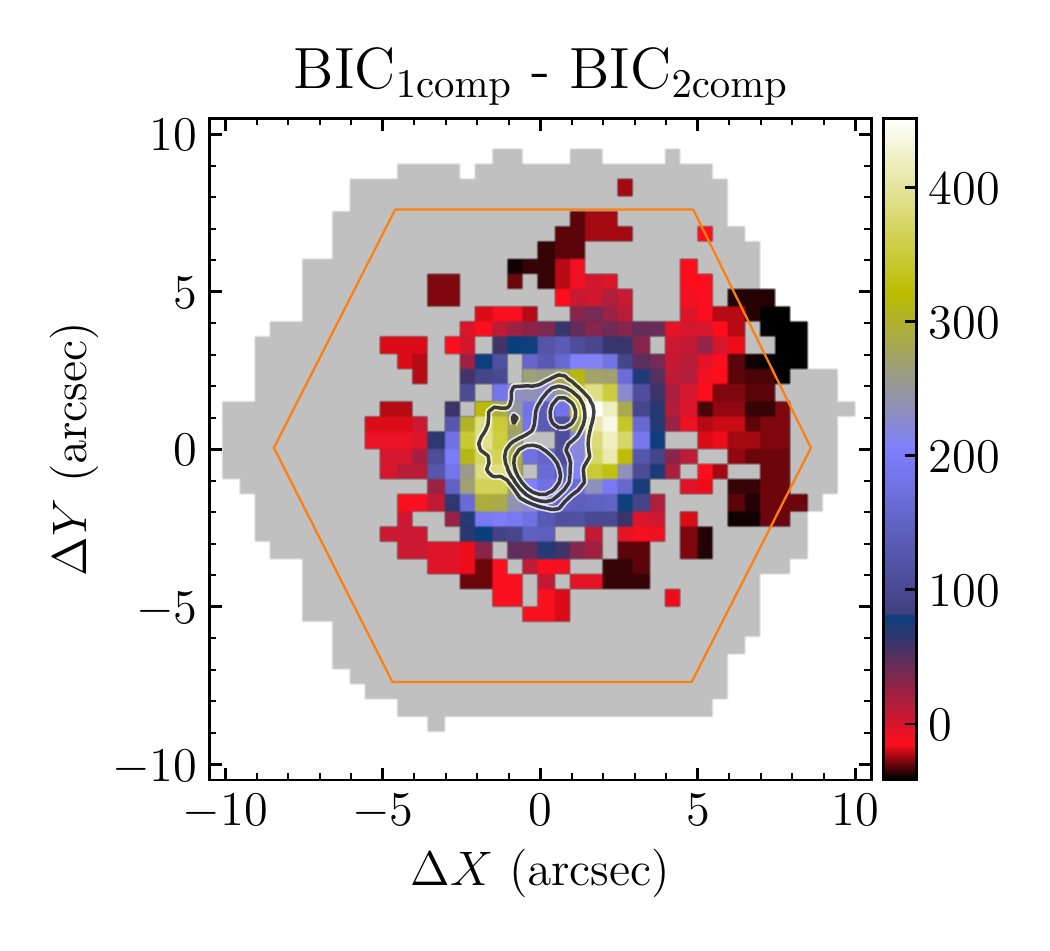}
    \vspace*{-0.5cm}
    \caption{Map of the difference between the Bayesian Informational Criteria (BIC; Equation \ref{eqn:bic}) derived from one- and two-Gaussian model fits ($\Delta{\rm BIC} = {\rm BIC}_1 - {\rm BIC}_2$) to the emission line spectra of this galaxy. $\Delta \mathrm{BIC}<0$ values suggests only the ``main'' component is needed to describe the emission line spectrum in the spaxel, while $0<\Delta \mathrm{BIC}<50$ suggests the addition of the second ``outflow'' component results in a marginal improvement (such spaxels are marked by transparent colors in other parameter maps.)  The contours indicate 3 GHz (S-band) emission  5, 10, 50$\times$ the rms of the image shown in Figure \ref{fig:wideband_maps}.}
    \label{fig:delta_bic_map}
\end{figure}

To measure the properties of emission lines produced by ionized gas in this galaxy, we first substracted the stellar continuum, as derived in \S\ref{sec:manga_stars}, from the observed spectrum in each region, weighting appropriately the contribution of the consituent spaxels.
An example of the resulting emission line spectrum is shown in Figure \ref{fig:lines_se}.  We then estimate the SNR of the resultant emission line spectra in each spaxel using the total flux in the ${\rm H}\alpha$+[N\ii] lines, and removed from further analyses all spaxels with ${\rm SNR} < 30$.  As shown in Figure \ref{fig:delta_bic_map}, this requirement primarily excluded spaxels in the outer regions of this galaxy -- beyond the observed extent of its radio emission (Figure \ref{fig:wideband_maps}).  We then used the Voronoi algorithm developed by \citet{cappellari_copin2003} to spatially bin the remaining spaxels into regions with ${\rm SNR}\geq50$.

As shown in Figure \ref{fig:lines_se}, the profiles of the emission lines in a particular spaxel were not always well-described by a single Gaussian.  As a result, we modelled the emission line spectrum in each spaxel assuming two Gaussian components.  Unfortunately, using a $\chi^2$ minimization routine to model the emission line spectrum in each spaxel with two independent Gaussians yielded unreliable results due -- in large part -- to the degeneracies inherent in this model.  As a result, we developed a procedure to fit the emission line spectra in all of the spaxels as the sum of two components:
\begin{itemize}
    \item a ``main'' component dominated by regularly rotating gas in the disk of this galaxy, and
    \item an ``outflow'' component.
\end{itemize}
In this decomposition, we accounted for the per-locus disperson, and assumed that the spatial distribution of the kinematic properties of gas in the ``main'' component are well described by the prescription for a regularly rotating disk given in Equations \ref{eqn:vlos} \& \ref{eqn:vphi}.

Initial parameters were derived assuming the main component dominated the emission in each spaxels, but final values resulted from simultaneously fitting the emission line spectra for the ``main'' and ``outflow'' component, as described below.

We then refit the emission line spectrum in every spatial bin, assuming the profile of each spectral line is described by two Gaussians.  The free parameters in this model are the:
\begin{itemize}
\item line-of-sight velocity $V_{\rm los}$ and intrinsic velocity dispersion (the observed velocity dispersion of a line $\sigma_{\rm obs}^2 = \sigma_{\rm gas}^2 + \sigma_{\rm inst}^2$ where  $\sigma_{\rm gas}$ is the intrinsic velocity dispersion and $\sigma_{\rm inst}$ is the instrumental resolution) $\sigma_{\rm gas}$  of the emitting gas,
\item $H\alpha$ flux,
\item Balmer decrement $H\alpha/H\beta$,
\item $\log$~[N\ii]6584/$H\alpha$,
\item $\log$~([S\ii]6717+[S\ii]6731)/$H\alpha$,
\item $\log$~[O\iii]5007/$H\beta$, and 
\item $\log n_e$, as determined from [S\ii]6717/[S\ii]6731 ratio using the methods described by \citealt{OsterbrockFreland2006, Proxauf2014},
\end{itemize}
of both the ``main'' and ``outflow'' components.  To determine the values of these quantities in each spatial region, we used the Levenberg-Marquardt  minimization method as implemented by the Python-based \textsc{lmfit} package \citep{lmfit} to determine the combination of values which minimized the $\chi^2$.  Furthermore, we required that -- for both components -- our fits returned values within the following ranges:
\begin{itemize}
    \item $2.75 \leq \frac{\mathrm{H}\alpha}{\mathrm{H}\beta} \leq 10 $
    \item $-1.2 \leq \log\left( \frac{[\mathrm{O}\textsc{iii}]5007}{\mathrm{H}\beta} \right) \leq 1.3$
    \item $-1.0 \leq \log \left( \frac{[\mathrm{N}\textsc{ii}]6584}{\mathrm{H}\alpha} \right) \leq 0.4$
    \item $-0.9 \leq \log \left( \frac{[\mathrm{S}\textsc{ii}]6717+[\mathrm{S}\textsc{ii}]6731}{\mathrm{H}\alpha} \right) \leq 0.25$ 
    \item $0.475 \leq \left( \frac{[\mathrm{S}\textsc{ii}]6717}{[\mathrm{S}\textsc{ii}]6731} \right) \leq 1.425$,
\end{itemize}
as expected from the physical processes governing these emission lines and observations of large samples of other galaxies (e.g, \citealt{baldwin81, OsterbrockFreland2006, Proxauf2014}). We further required that, for each spatial region, the fitted value $V_{\rm LOS}$ of the ``main'' was within $50~{\rm km}~{\rm s}^{-1}$ of the value for the ``main'' component derived from the initial analysis described above.  Using this procedure, we simultaneously fit for the properties of the ``main'' and ``outflow'' contribution to the emission line spectrum in each spaxel.    An example of the results from this fitting procedure is shown in Figure \ref{fig:lines_se}.

To assess the statistical signficance of the ``outflow'' component in a given spaxel, we calculated the Bayesian Information Criterion (BIC) statistic \citep{Schwarz1978_BIC, Liddle2007_BIC}:
\begin{eqnarray}
\label{eqn:bic}
\mathrm{BIC} & = & N_\mathrm{data} \ln \frac{\chi^2}{N_\mathrm{data}} + N_\mathrm{vars} \ln N_\mathrm{data},
\end{eqnarray} 
where $N_{\rm data}$ is the number of data points, $N_{\rm vars}$ is the number of free parameters in the model, and $\chi^2$ is the result from fitting the data with said model, resulting from fitting the emission line spectrum of a given region with a single Gaussian (BIC$_1$) and two Gaussian (BIC$_2$).  As shown in Figure \ref{fig:delta_bic_map}, BIC$_1$ is substantially higher than BIC$_2$ in the innermost spaxels, strongly implying that the ``outflow'' component is significant in these regions.  These spaxels are also coincident with the radio emission detected from this galaxy, suggesting a physical connection between the ionized gas ``outflow'' and radio-emitting plasma.  In the spaxels beyond the radio emission, BIC$_1$ is either slightly larger or smaller than BIC$_2$ -- implying that either the ``outflow'' component is not present or a marginal fraction of the ionized gas at these locations.  As a result, in 2D maps of the parameters of the ``outflow'' component, we mask spaxels with $\mathrm{BIC}_1<{\rm BIC}_2$, while those with ${\rm BIC}_1 - {\rm BIC}_2 < 50$ are shown in transparent color.  Furthermore, in the parameter maps of main component, we present values from the one-component fit for spaxels with ${\rm BIC}_1 \leq {\rm BIC}_2$, and values from the two-component fit for those spaxels where ${\rm BIC}_1 > {\rm BIC}_2$. 

\begin{figure}[tb]
    \centering
    \includegraphics[width=0.45\textwidth]{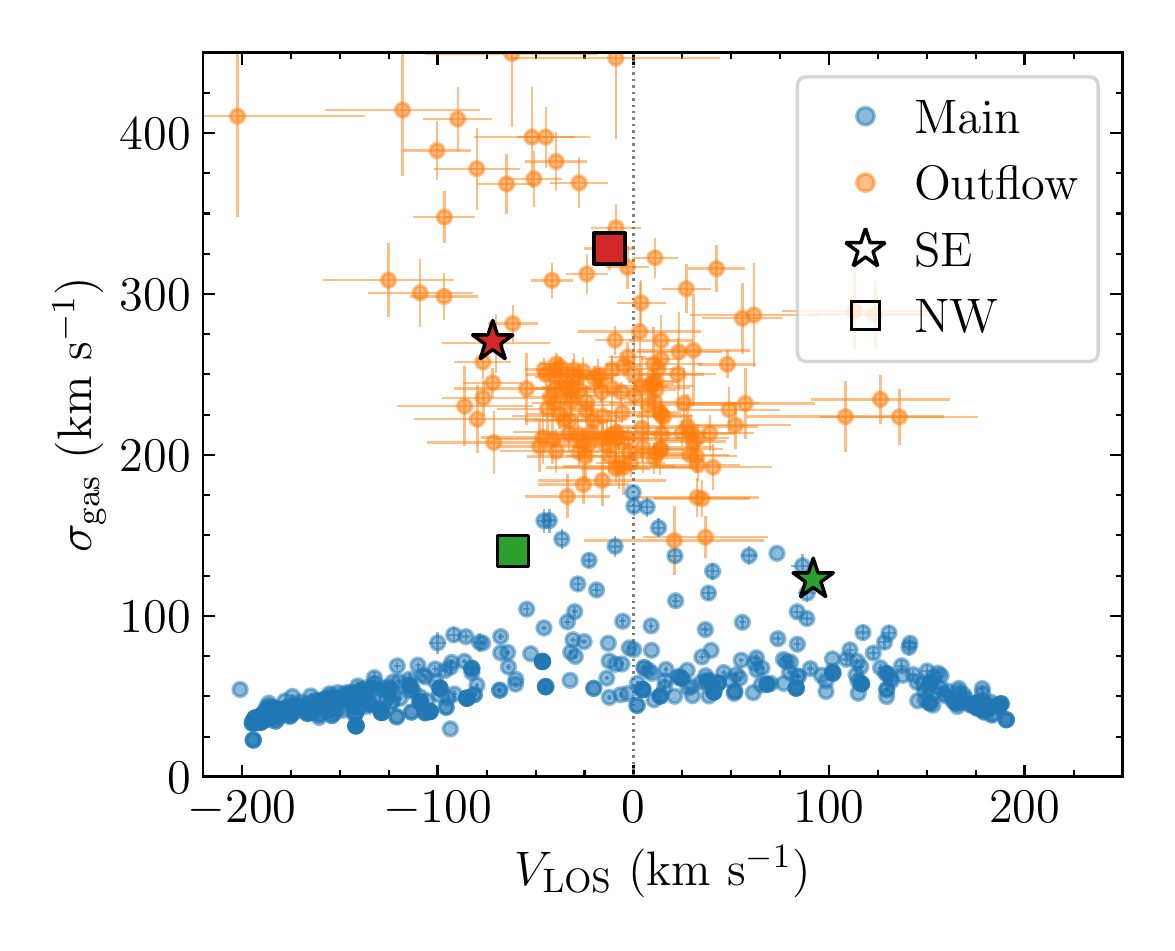}
    \caption{Velocity - velocity dispersion diagram of the main (blue) and outflowing (orange) components. Star and square symbols demonstrate kinematical properties of the both components in the positions of SE and NW radio jets, respectively.}
    \label{fig:vvd}
\end{figure}

\begin{figure*}[tbh]
\begin{center}
    \includegraphics[trim=0.55cm 0.5cm 0.5cm 0.5cm,clip,width=0.475\textwidth]{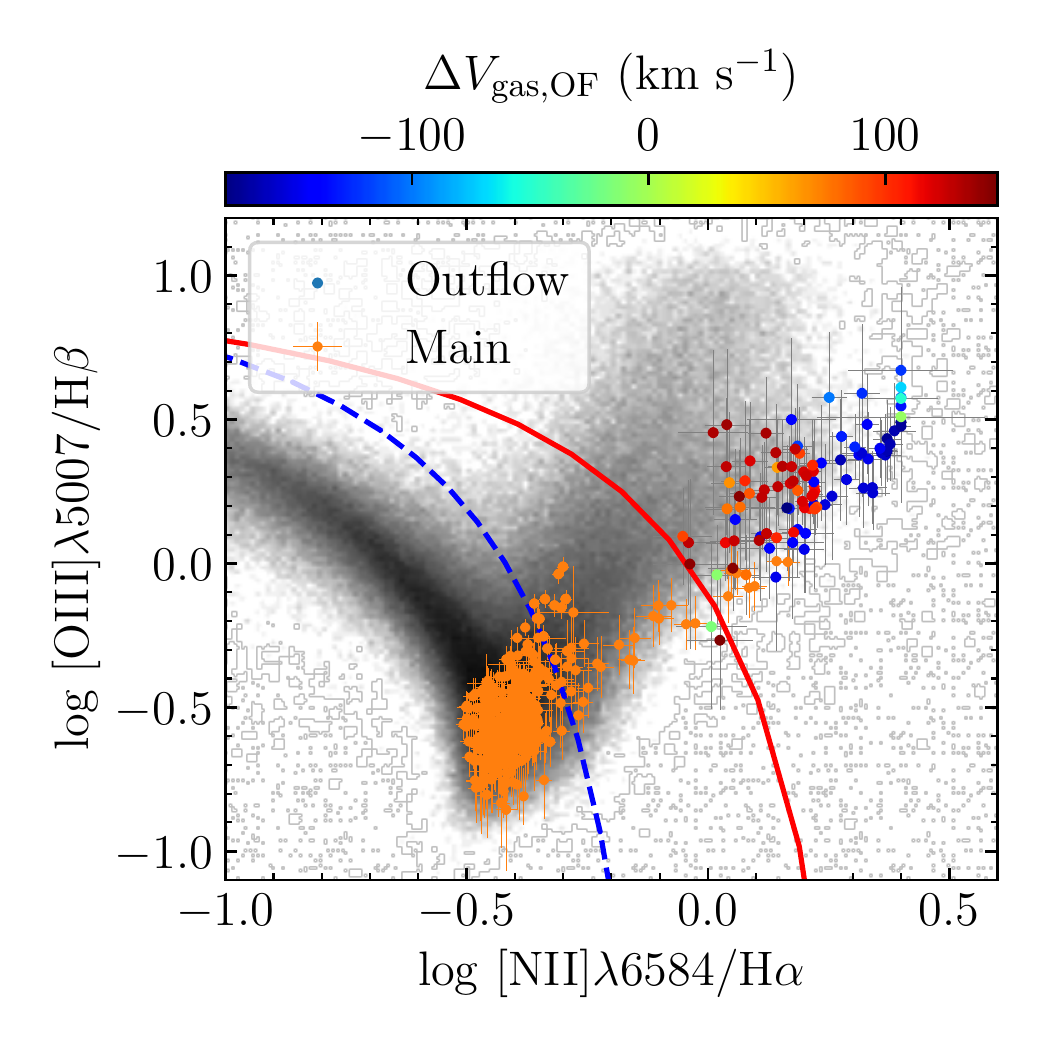}
    \includegraphics[trim=2.2cm 0.5cm 0.5cm 0.5cm,clip,height=0.475\textwidth]{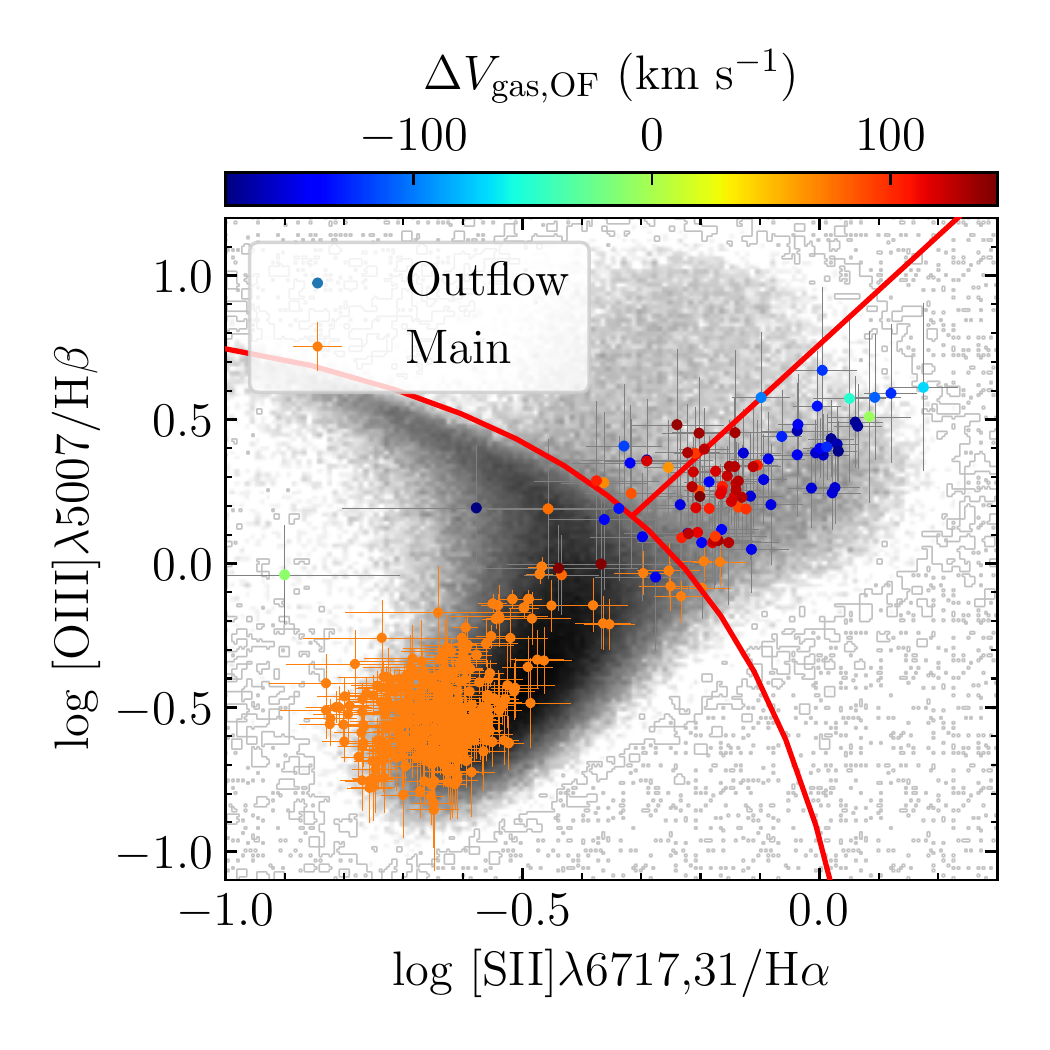}
\end{center}
\caption{The Baldwin, Philips, and Telervich (BPT) diagrams of the ionized gas in the ``main'' and ``outflow'' components of each fitted spaxel, as derived using the procedure described in \S\ref{sec:emision_line_fitting}. Points corresponding to the ``main'' component are shown in orange, while the points corresponding to the ``outflow'' component are color coded by the relative line-of-sight velocities $\Delta V_{\rm gas, OF} = V_{\rm los,OF} - V_{\rm los,main}$ at the particular spaxel.  In the left panel, the source of the ionizing photons are believed to be emission from young stars below the blue dotted line, AGN, shocks, and/or post-AGB stars above the red solid line, and a mix of the two between them \citet{Kauffmann2003MNRAS.346.1055K, Kewley2001ApJ...556..121K}.  In the right panel, points below the curved red line are believe to be primarily photoionized by emission from young stars, with the region above the curved red line are separated into Low Ionization Emission Regions (LIERs, below) and photoionization by an AGN (e.g., \citealt{kewley06}).}
\label{fig:manga_bpt}
\end{figure*}

The difference between the ``main'' and ``outflow'' components manifests themselves not only in the statistical significance of the fits, but also in the derived properties of the ionized gas.  As shown in Figure \ref{fig:vvd}, the velocity dispersion $\sigma_{\rm gas}$ of the ``outflow'' component is in general higher than that of the ``main'' component -- even for spaxels with similar line-of-sight velocities.  The ``V'' shape of the ``outflow'' component on the $V_{\rm LOS} - \sigma_{\rm gas}$ diagram shown in Figure \ref{fig:vvd} is suggestive of a biconal geometry (e.g., \citealt{bae16}).  Furthermore, as shown in Figure \ref{fig:manga_bpt}, the line ratios measured for the ``main'' and ``outflow'' occupy very different regions on the Baldwin, Philips, and Telervich (BPT) diagrams -- indicating they are ionized by different mechanisms \citep{baldwin81}.  The physical implications of both results will be discussed further in \S\ref{sec:outflow_kin}.

\subsection{GMOS data}
\label{sec:gmos}

In addition to using MaNGA data to measure the properties of the ionized gas in this galaxy (\S\ref{sec:manga}), we also analyzed the spectrum obtained in a recent GMOS IFU observations -- whose results were previously presented by \citet{Wylezalek2017}.  The GMOS data were taken on Gemini-North in one-slit mode, and covered the central $3\farcs5 \times 5\arcsec$ region of this galaxy (Figure \ref{fig:desi_img}).  The angular resolution of this dataset is limited by the atmospheric seeing during this observation, estimated to be $\approx0\farcs9$, and these observations measured the spectrum between $\lambda \approx 4000-7000$~\AA\ with a spectral resolution $R\approx3000$, corresponding to an instrumental dispersion $\sigma_\mathrm{inst}\approx45$~\kms \citep{Wylezalek2017} (Table \ref{tab:IFUs_MaNGA_GMOS}).  This data was reduced following the procedure described by \citet{Wylezalek2017}.  The primary differences between our analysis of this data and that presented by \citet{Wylezalek2017} is the modeling of the stellar contribution to observed spectrum and a different spatial binning of the inferred emission line spectra -- as described below.

Just as the case for the MaNGA data \S\ref{sec:manga}, we first determined the stellar contribution to the observed spectrum at given sky location.  We again used the {\sc NBursts} package to fit the observed spectrum with that predicted by the SSP models described in \S\ref{sec:manga_stars}.  Due to the relatively low SNR in this region, we fixed the equivalent stellar age $T_{\rm SSP}$ and metallicty [Fe/H]$_{\rm SSP}$ in a particular GMOS spaxel to the values derived in the MaNGA spaxel at the same sky position.  As a result, this fitting returned the stellar line-of-sight velocity $V_\star$ and velocity dispersion $\sigma_\star$ for each GMOS sky pixel.  We then subtracted the predicted stellar contribution in each sky pixel of the GMOS data cut to determine the emission line spectrum at each position.  \added{To determine the absolute flux calibration of the GMOS data, we compared the total \Ha+[N\ii] flux within a 3\arcsec\ radius inferred from a single Gaussian fit to the value in the \href{http://rcsed.sai.msu.ru/}{Reference Catalog of galaxy Spectral Energy Distributions} (RCSED; \citet{Chilingarian2017rcsed}) -- which measured this quantity from an earlier SDSS spectrum using a similar methodology for determining the properties of the emission lines.  We found that the flux inferred from the GMOS data was $1.4\times$ lower than that in the RCSED, and used this factor to adjust the measured fluxes and equivalent widths (EWs) of the aforementioned spectral lines.}

\begin{figure}[tbh]
    \centering
    \includegraphics[width=0.475\textwidth,trim=0.05cm 0 0.0cm 0,clip]{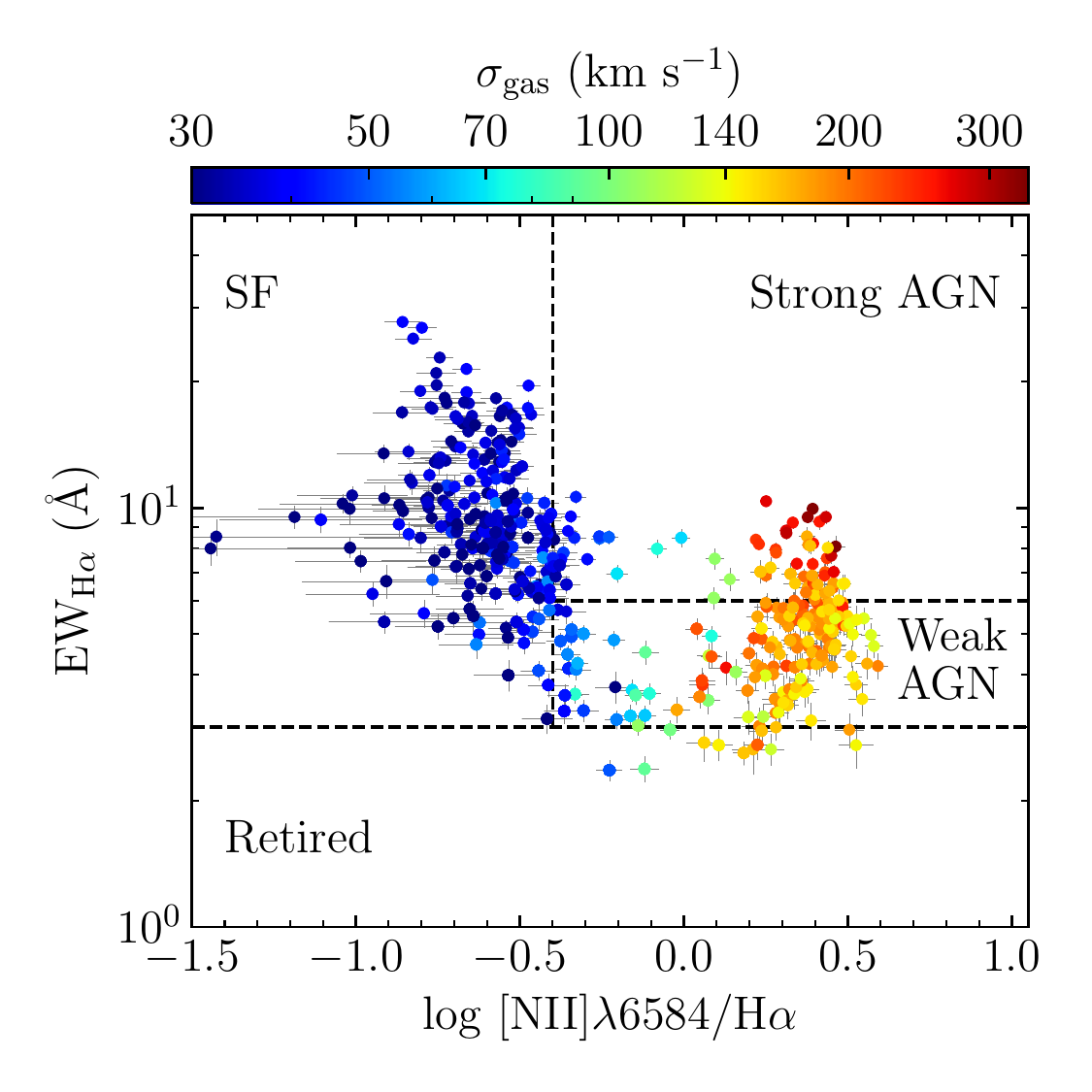}
\vspace*{-0.5cm}
\caption{WHAN diagram - equivalent width of H$\alpha$ versus [N\ii]/H$\alpha$ \citep{CidFernandes2011whan} for individual spaxels of GMOS IFU data, with the symbol colors indicating the gas velocity dispersion derived from our single Gaussian fits to the emission line spectra, as described in \S\ref{sec:gmos}.}
\label{fig:emission_gmos_WHAN}
\end{figure}

\begin{figure*}[tbh]
\begin{center}
    \includegraphics[width=0.475\textwidth]{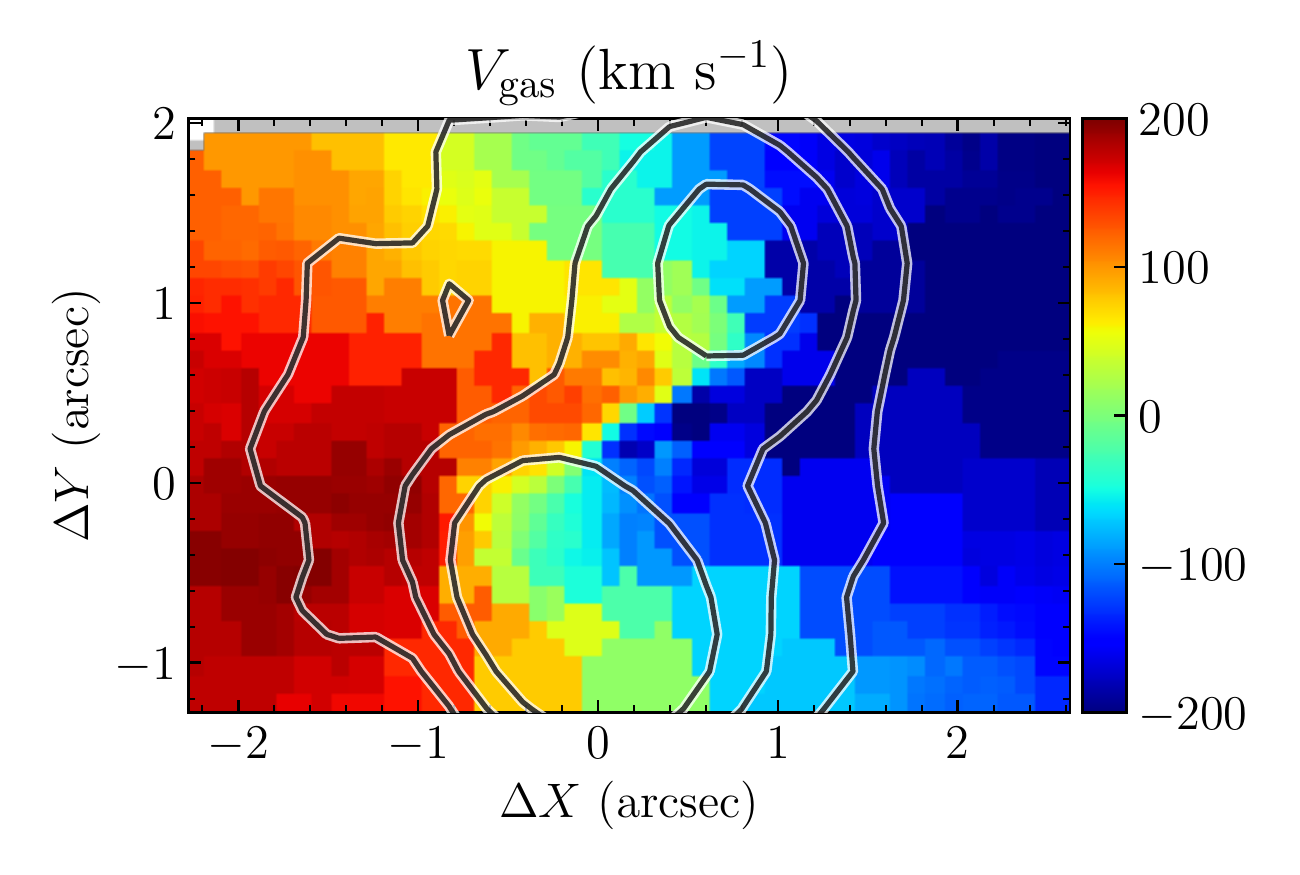}
    \includegraphics[width=0.475\textwidth]{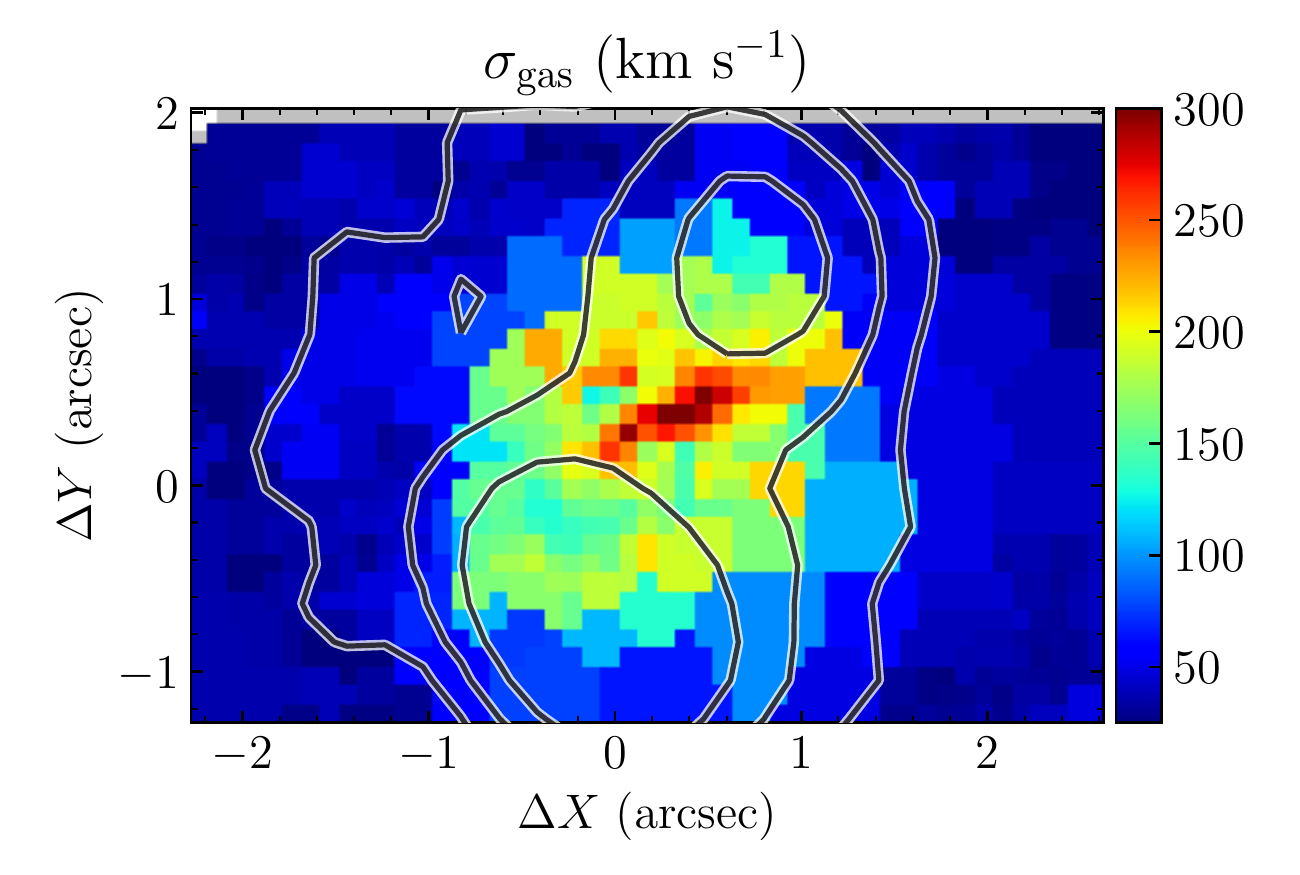}  
\end{center}
\vspace*{-0.75cm}
\caption{The line-of-sight velocity $V_{\rm LOS}$ and velocity dispersion $\sigma_{\rm gas}$ of the ionized gas in MaNGA 1-166919 as measured in our analysis of the GMOS data described in \S\ref{sec:gmos}.  In both images, the contours indicate 3 GHz (S-band) emission  5, 10, 50$\times$ the rms of the image shown in Figure \ref{fig:wideband_maps}.}
\label{fig:gmos_gaskin}
\end{figure*}

Before using these spectra to measure the properties of the ionized gas, it was necessary to first adaptively bin the emission line spectra into regions of sufficient signal-to-noise -- as done for the MaNGA datacube (\S\ref{sec:emision_line_fitting}).  We again used the adaptive Voronoi algorithm developed by \citet{cappellari_copin2003} to combine the spectra of adjacent spatial pixels using the maximum SNR per
channel measured within the \Ha+[N\ii] line complex, such that the combined spectra had an overall SNR $\gtrsim 3$.  We then used the Levenberg-Marquardt  minimization method implemented  by  the  Python-based {\tt lmfit} package \citep{lmfit} to simultaneously fit single Gaussians to the H$\alpha$, [N\ii] and [S\ii] emission lines (We excluded the H$\beta$ and [O\iii] lines from this analysis due to the low SNR at the edge of the GMOS band pass).  Again, we required that all three lines have the same line-of-sight velocity $V_{\rm los}$ and velocity dispersion $\sigma_{\rm gas}$ at a given sky position, and the resultant spatial distributions of these parameters are shown in Figure \ref{fig:gmos_gaskin}.

\added{As discussed in \S\ref{sec:emision_line_fitting}, the emission line spectrum measured by MaNGA suggests the ionized gas in this galaxy of two components: a ``main'' component comprised of material rotating within the galactic disk, and an ``outflow'' component, which we modeled using two Gaussians.  Unfortunately, the SNR of the GMOS spectra is too low to fit a two Gaussian model to the spectrum in each spaxel as done for the MaNGA data (\S\ref{sec:emision_line_fitting}).  However, as shown in Figure \ref{fig:vvd}, our analysis of the MaNGA data indicates the ``main'' ionized gas has a velocity dispersion $\sigma_{\rm gas} \la 100~{\rm km~s^{-1}}$, while $\sigma_{\rm gas}$ for outflow component is significantly larger.  Furthermore, as shown in Figure \ref{fig:emission_gmos_WHAN}, the properties of the emission lines in the GMOS data differ significantly between spaxels with $\sigma_{\rm gas} < 100~{\rm km~s}^{-1}$ and $\sigma_{\rm gas} > 100~{\rm km~s}^{-1}$.  Combined, this indicates that ionized gas in GMOS spaxels with low velocity dispersion ($\sigma_{\rm gas}<100$~\kms) are dominated by the ``main'' component, while the ionized in GMOS spaxels with high velocity dispersion ($\sigma_{\rm gas}<100$~\kms) are dominated by the ``outlfow'' component.   As defined using this criterion , the morphology of the GMOS ``outflow'' component is similar to the radio morphology of this galaxy (right panel of Figure \ref{fig:gmos_gaskin}), as was also the case for the MaNGA ``outflow'' component (Figure \ref{fig:delta_bic_map}; \S\ref{sec:emision_line_fitting}).

\begin{figure*}[tbh]
\begin{center}
    \includegraphics[height=0.25\textheight]{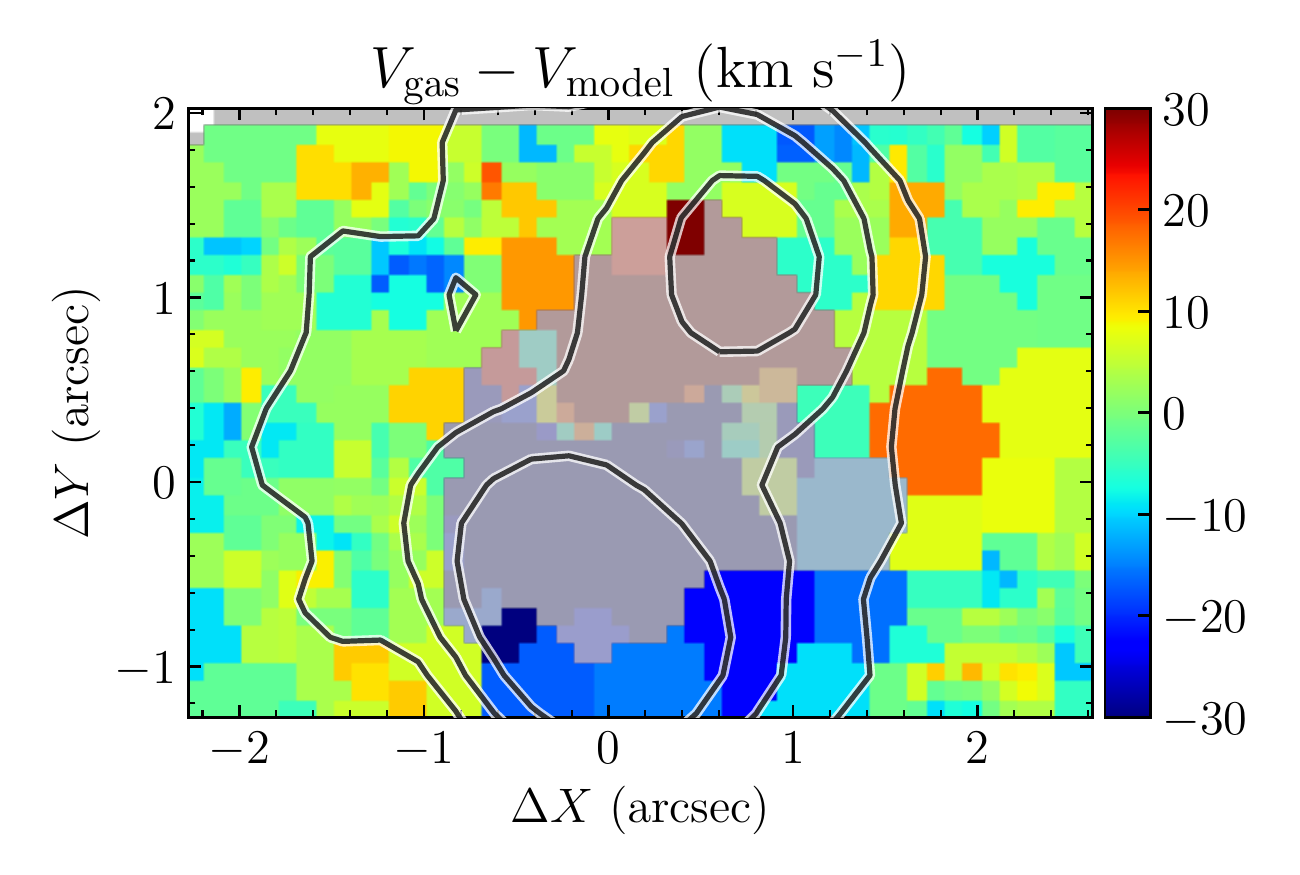}
    \includegraphics[height=0.23\textheight]{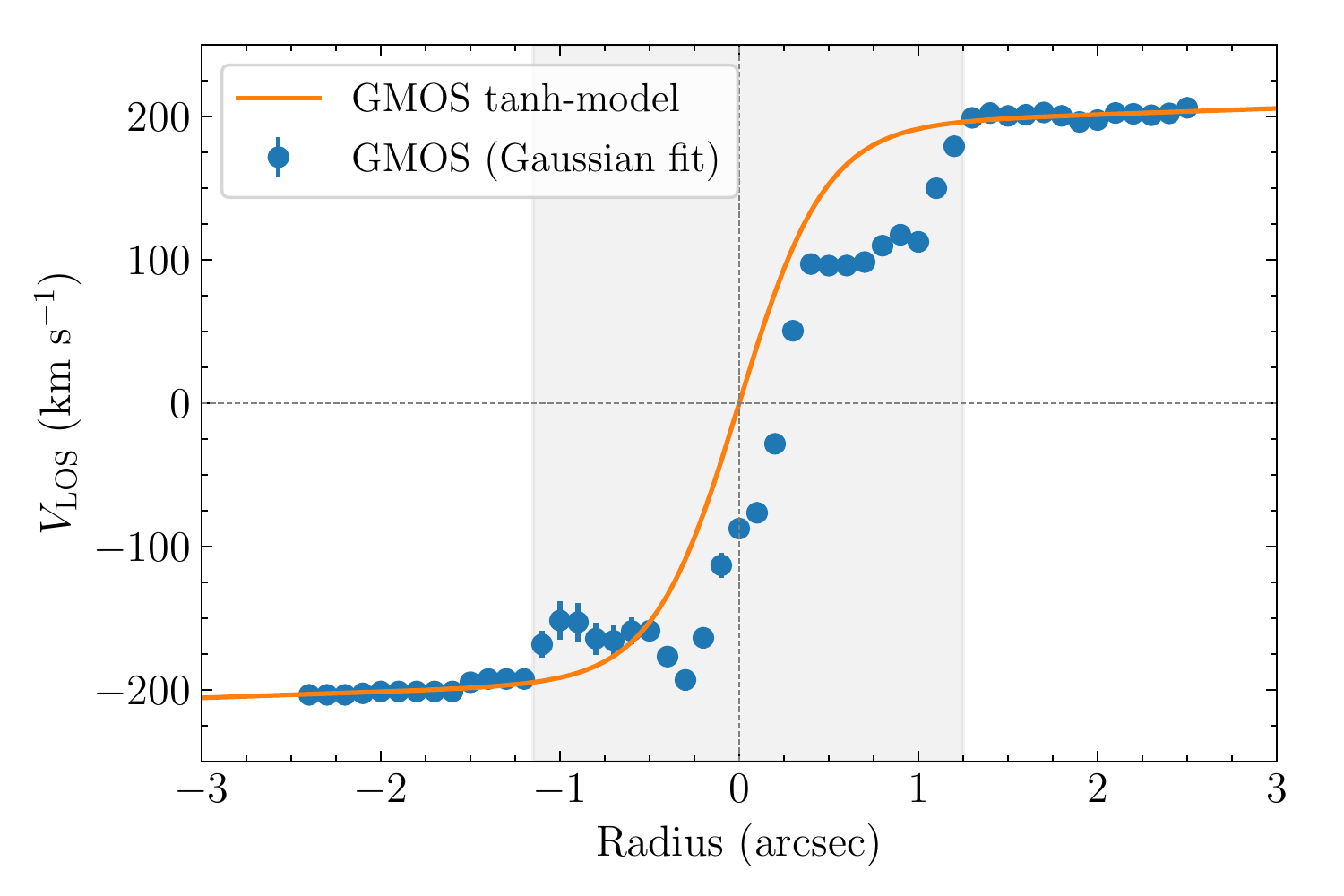}
\end{center}
\vspace*{-0.75cm}
\caption{ \added{{\it Left}: Residual between the measured line-of-sight velocity $V_{\rm gas}$ and the value predicted by the $\tanh$-model described in \S\ref{sec:gmos}.  GMOS spaxels with high velocity dispersion ($\sigma_{\rm gas} > 100$~\kms) are dimmed since emission in these regions were not used in the fits, as explained in the text.  {\it Right}: 1D velocity profile of $V_{\rm gas}$ measured in the GMOS data (left panel of Figure~\ref{fig:gmos_gaskin}) along the major axis derived from $\tanh-$model fitting described in \S\ref{sec:gmos} (blue point).  The orange line represents the 1D velocity profile predicted by the $\tanh$-model, and the grey area corresponds to the high velocity dispersion regions excluded from this modeling.}}
\label{fig:tanh_model_residuals_and_1d_profile}
\end{figure*}

\begin{figure*}[tbh]
\begin{center}
    \includegraphics[height=0.25\textheight,trim=1.0cm 0 0.0cm 0,clip]{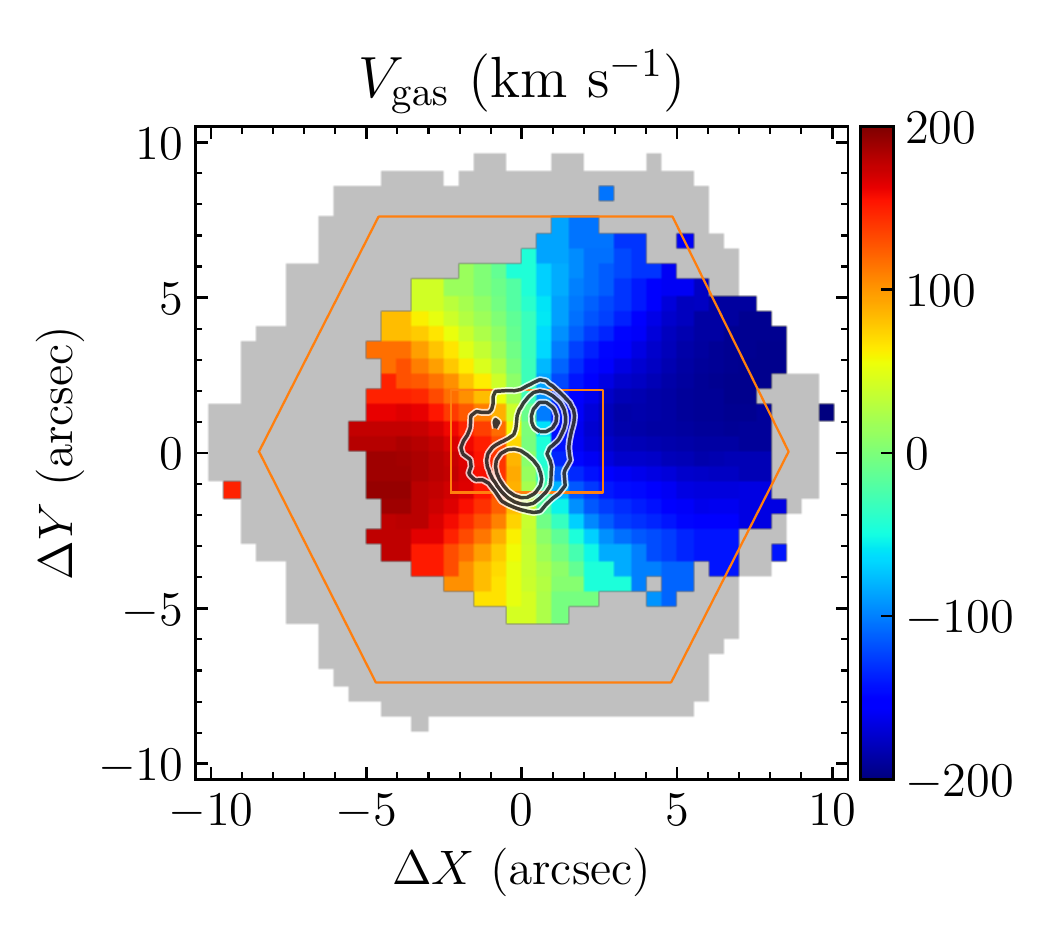}
    \includegraphics[height=0.25\textheight]{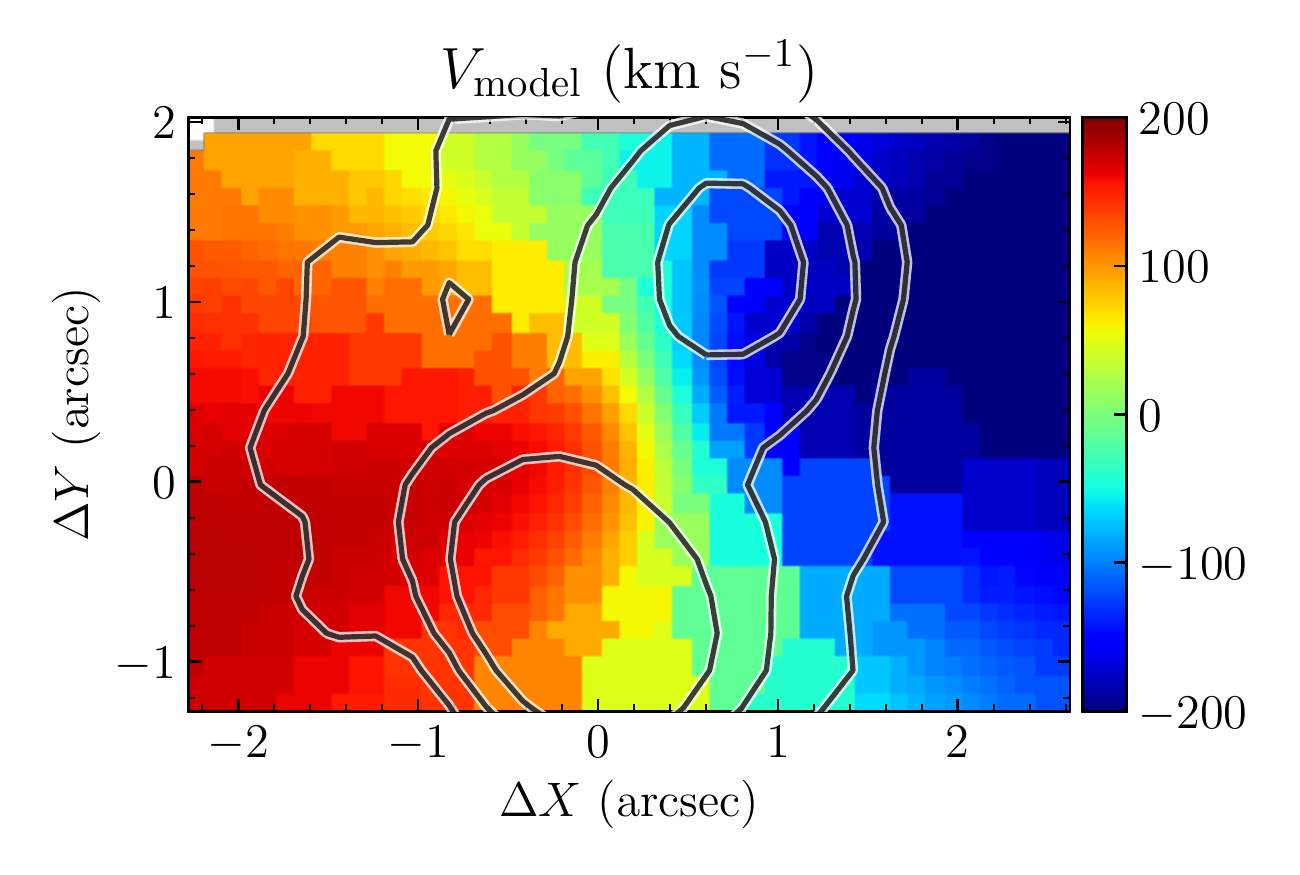}
\end{center}
\vspace*{-0.75cm}
\caption{Line-of-sight velocity of the regularly rotating ``main'' component of the ionized gas in MaNGA 1-166919 as derived from our analysis of the MaNGA ({\it left}; \S\ref{sec:emision_line_fitting}) and GMOS ({\it right}; \S\ref{sec:gmos}) observations of this galaxy.  In both figures the contours indicate 3 GHz (S-band) emission $5, 10, 50\times \mathrm{RMS}$ of the image shown in Figure \ref{fig:wideband_maps}.}
\label{fig:main_rot}
\end{figure*}

To determine the kinematics of the ``main'' component in the GMOS data, we fit the $V_{\rm LOS}$ of all spaxels with $\sigma_{\rm gas} < 100$~\kms\ using the mathematical model for a regularly rotating disk defined in Equations \ref{eqn:vlos} \& \ref{eqn:vphi}.  As shown in Figure \ref{fig:tanh_model_residuals_and_1d_profile}, this model accurately reproduce both 1D and 2D profiles of $V_{\rm LOS}$ in these spaxels.  Furthermore, as shown in Figure \ref{fig:main_rot}, this fit to the GMOS data suggests a regularly rotating disk whose orientation and line-of-sight velocities are similar to what was derived from the MaNGA data.}

\section{Physical Interpretation}
\label{sec:physics}

In this section, we use the properties measured in the JVLA radio (\S\ref{analysis_radio}) and IFU (\S\ref{analysis_optical}) observations analyzed above to measure the properties of material associated with the outflow (\S\ref{sec:outflow}) and active galactic nucleus (AGN; \S\ref{sec:agn}) in this galaxy, as well as the impact the outflow has on its host galaxy (\S\ref{sec:hostgal}).

\subsection{Outflow}
\label{sec:outflow}

As previously mentioned the observed radio emission from this galaxy is spatially coincident with the regions where two Gaussians are needed to accurately model the emission line spectra (Figure \ref{fig:delta_bic_map}) derived from the MaNGA data cube (\S\ref{sec:emision_line_fitting}) as well as regions where the ionized gas has a high velocity dispersion (Figure \ref{fig:gmos_gaskin}) as derived from the GMOS data cube (\S\ref{sec:gmos}).  As a result, in the discussion below we assume that the radio and optical emission are produced by two different, but related, components of the outflowing material.  By doing so we are able to study the relativistic content (\S\ref{sec:outflow_cr}), kinematics (\S\ref{sec:outflow_kin}), and thermal content (\S\ref{sec:outflow_thermal}) in this outflow.

\subsubsection{Relativistic Component}
\label{sec:outflow_cr}

As described in \S\ref{sec:radio_spec}, the spectrum of the both the SE and NW radio lobes is well described by a power-law  with spectral index $\alpha \approx -0.7$, consistent with optically thin synchrotron emission resulting from relativistic electrons (and positrons) interacting with a magnetic field (e.g., \citealt{pacholczyk70, condon16}).  In this case, the observed radio luminosity $L_{\rm rad}$ depends on the size $R$, the relativistic electron $u_{\rm e}$, and the magnetic $u_{\rm B}$ energy density of the emitting region.  However, since the synchrotron power $P_{\rm syn}$ radiated by an electron of energy $E$ in a magnetic field of strength $B$ is (e.g., \citealt{pacholczyk70,rybicki86}):
\begin{eqnarray}
\label{eqn:pow_synch}
P_{\rm synch}(E) & = & \frac{4e^4}{9m_e^4c^7}B^2E^2,
\end{eqnarray}
where $e$ and $m_e$ are, respectively, the charge and mass of the electron and $c$ is the speed of light, for a given radio luminosity and size there is not a unique solution for $u_{\rm e}$ and $u_{\rm b}$.  However, there is a minimum in the  combined (relativistic electron + magnetic field) energy required to power such a source when $u_e \equiv \frac{4}{3}u_{\rm B}$ (e.g., \citealt{pacholczyk70,rybicki86,condon16}.  The magnetic field strength $B_{\rm min}$ is (e.g., \citealt{pacholczyk70,condon16}):
\begin{eqnarray}
\label{eqn:b_min}
B_{\rm min} & = & [4.5(1+\eta)c_{\rm 12}L_{\rm rad}]^\frac{2}{7}R^{-\frac{6}{7}}~{\rm Gauss}
\end{eqnarray}
where $\eta$ is the ion to electron energy ratio, $c_{\rm 12}$ is a ``constant'' whose value depends on  $\nu_{\rm min}$, $\nu_{\rm max}$, and $\alpha$ (for $\nu_{\rm min} = 10^6~{\rm Hz}$, $\nu_{\rm max} = 10^{10}~{\rm Hz}$, and $\alpha \approx -0.7$ as derived for both the SE and NW radio lobes (Table \ref{tab:radspec_pow}), $c_{12} \approx 10^8$ in cgs units \citep{condon16}), and $R$ is the radius of the assumed spherical emitting region, and relativistic particle (electrons + ions) energy (e.g., \citealt{pacholczyk70,condon16}):
\begin{eqnarray}
\label{eqn:e_min}
E_{\rm min} & = & c_{\rm 13}[(1+\eta)L_{\rm rad}]^{\frac{4}{7}}R^{\frac{9}{7}}~{\rm ergs},
\end{eqnarray}
where $c_{13}$ is another constant whose value depends on  $\nu_{\rm min}$, $\nu_{\rm max}$, and $\alpha$ (for $\nu_{\rm min} = 10^6~{\rm Hz}$, $\nu_{\rm max} = 10^{10}~{\rm Hz}$, and $\alpha \approx -0.7$ as derived for both the SE and NW radio lobes (Table \ref{tab:radspec_pow}), $c_{13} \approx 3\times10^4$ in cgs units \citep{condon16}).  Since the 3D geometry of the radio emitting regions are not known, we assume that $R$ for a particular lobe is between physical radius inferred by the smallest deconvolved semi-minor axis and the largest deconvolved semi-major axis derived from our modeling of the wideband radio images of this source (\S\ref{sec:radio_img}), as reported in Table \ref{tab:12arcsec_wideband_imfit}.  The ``minimum energy'' magnetic field strengths and relativistic particle energies of both the SE and NW lobes inferred from our measurements of their radio morphology (\S\ref{sec:radio_img}) and spectrum (\S\ref{sec:radio_spec}) are reported in Table \ref{tab:rel_prop}.

\begin{table}[tbh]
    \caption{Physical Properties of SE and NW radio lobes}
    \label{tab:rel_prop}
    \begin{center}
    \begin{tabular}{ccc}
    \hline
    \hline
    {\sc Property} & SE lobe\tablenotemark{b} & NW lobe\tablenotemark{b} \\
    \hline
    $L_{\rm rad}~\left[\frac{\rm ergs}{\rm s} \right]$ \tablenotemark{a} & $(2.0\pm0.3)\times10^{39}$ & $1.6^{+0.9}_{-0.5} \times 10^{39}$ \\
    $R$ [kpc] & 0.45$-$0.82 & 0.49$-$0.85 \\ 
    $B_{\rm min}$ [$\mu$G]\tablenotemark{c} & $\sim20-40$ & $\sim20-40$ \\
    $E_{\rm min}$ [ergs]\tablenotemark{c} & $\sim(1-3)\times10^{54}$ & $\sim(1-3)\times10^{54}$ \\
    \hline
    \hline
    \end{tabular}
    \end{center}
    \label{tab:lobe_prop}
    \tablenotetext{a}{Calculated for $\nu_{\rm min} \equiv 10^6~{\rm Hz}$, $\nu_{\rm max} \equiv 10^{10}~{\rm Hz}$, and value of $d_{\rm L}$ given in Table \ref{tab:galprop}.}
    \tablenotetext{b}{Calculated using spectral properties given in Table \ref{tab:radspec_pow} and morphological properties given in Table \ref{tab:12arcsec_wideband_imfit}.}
    \tablenotetext{c}{Calculated assuming $\eta=0$, i.e., the emitting plasma is composed solely of electrons (no ions).}
\end{table}

With this information, we can estimate the energy of the radio-emitting electrons.  The synchrotron emission from an electron with energy $E$ in a magnetic field of strength $B$ peaks at a frequency (e.g., \citealt{pacholczyk70,rybicki86}):
\begin{eqnarray}
\label{eqn:nupeak}
\nu_{\rm peak} & = & 0.29 \times \frac{3}{2} \left(\frac{E}{m_e c^2}\right)^2 \frac{e B}{m_e c}.
\end{eqnarray}
As a result, for a particular $\nu_{\rm peak}$ and $B$, the energy of the emitting electron is:
\begin{eqnarray}
\label{eqn:e_peak}
E & \sim & 6 \left(\frac{\nu_{\rm peak}}{\rm 10^9~Hz} \right)^{\frac{1}{2}} \left( \frac{B}{\rm 1~\mu G}\right)^{-\frac{1}{2}} ~{\rm GeV}. 
\end{eqnarray}
For the observed frequency range of $\nu=1-8~{\rm GHz}$ and range of $B_{\rm min}$ given in Table \ref{tab:lobe_prop}, for both lobes the radio emission is dominated by  $E \sim 1-4~{\rm GeV}$ electrons.  The synchrotron cooling time for such particles $t_{\rm cool}$, is $t_{\rm cool} \equiv \frac{E}{P_{\rm synch}} \sim 100 - 200~{\rm Gyr}$ in both lobes for the estimated particle energies and magnetic field strengths.  This suggests that radiative cooling plays a minor role in the evolution of the radio emission from the relativistic particles in this outflow.

Furthermore, the synchrotron spectrum of the observed radio emission can be used to determine the spectrum and origin of the GeV-emitting electrons.  Optically-thin synchrotron radiation with a power-law spectrum $S_\nu \propto \nu^\alpha$ is the result of emission from particles with a power-law energy spectrum:
\begin{eqnarray}
\frac{dN}{dE}(E) & \propto & E^{-p},
\end{eqnarray}
where $\frac{dN}{dE}(E)$ is the number of particles per unit energy at particle energy $E$ and $p$ is the particle index ($p = 1-2\alpha$) (e.g., \citealt{rybicki86, condon16}).  For the spectral index $\alpha\sim -0.85$ measured for both the SE and NW lobes (\S\ref{sec:radio_spec}, Table \ref{tab:radspec_pow}), this suggests $p \sim 2.7$ -- the value expected from first order Fermi or Diffusive Shock Acceleration (DSA; e.g. \citealt{fermi49, fermi54}).  DSA requires that particles cross a shock multiple times (e.g., \citealt{bell78a, bell78b, blandford78}), gaining energy in each shock crossing.  The particle spectrum, and the spectral index $\alpha$ of its resultant synchrotron emission, generated from this process is dependent on the Mach number ${\mathcal M}$ of the shock, where (e.g., \citealt{berezhko99,guo14,digennaro18}):
\begin{eqnarray}
\label{eqn:mach}
{\mathcal M} & = & \sqrt{\frac{2\alpha-3}{2\alpha+1}}.
\end{eqnarray}
The spectral index $\alpha \sim -0.8 - (-0.9)$ observed from both the SE and NW lobes (\S\ref{sec:radio_spec}, Table \ref{tab:radspec_pow}) suggests the emitting particles in both components are accelerated in shocks with ${\mathcal M} \sim 2-3.5$. 

However, such shocks should also efficiently accelerate ions to high energies (e.g., \citealt{guo14}).  Recent simulations (e.g., \citealt{park15}) and observations of particles accelerated in shock with similar Mach numbers ${\mathcal M}$ (e.g., SNR 5.7$-$0.1; \citealt{joubert16}) suggest the ion to electron energy ratio $\eta > 100$.  If that also occurs in this galaxy, then minimum total relativistic particle energies in the SE and NW lobes is $\sim10\times$ higher than the values given in Table \ref{tab:lobe_prop}, or on the order of $\sim10^{55}~{\rm ergs}$.  Regardless of the true value of $\alpha$, the larger energy in the relativistic component of this outflow suggests it can have a significant impact on the host galaxy (e.g., \cite{mao18, hopkins20a}).

\subsubsection{Kinematics}
\label{sec:outflow_kin}

\begin{figure*}[tb]
    \begin{center}
    \includegraphics[height=0.275\textheight]{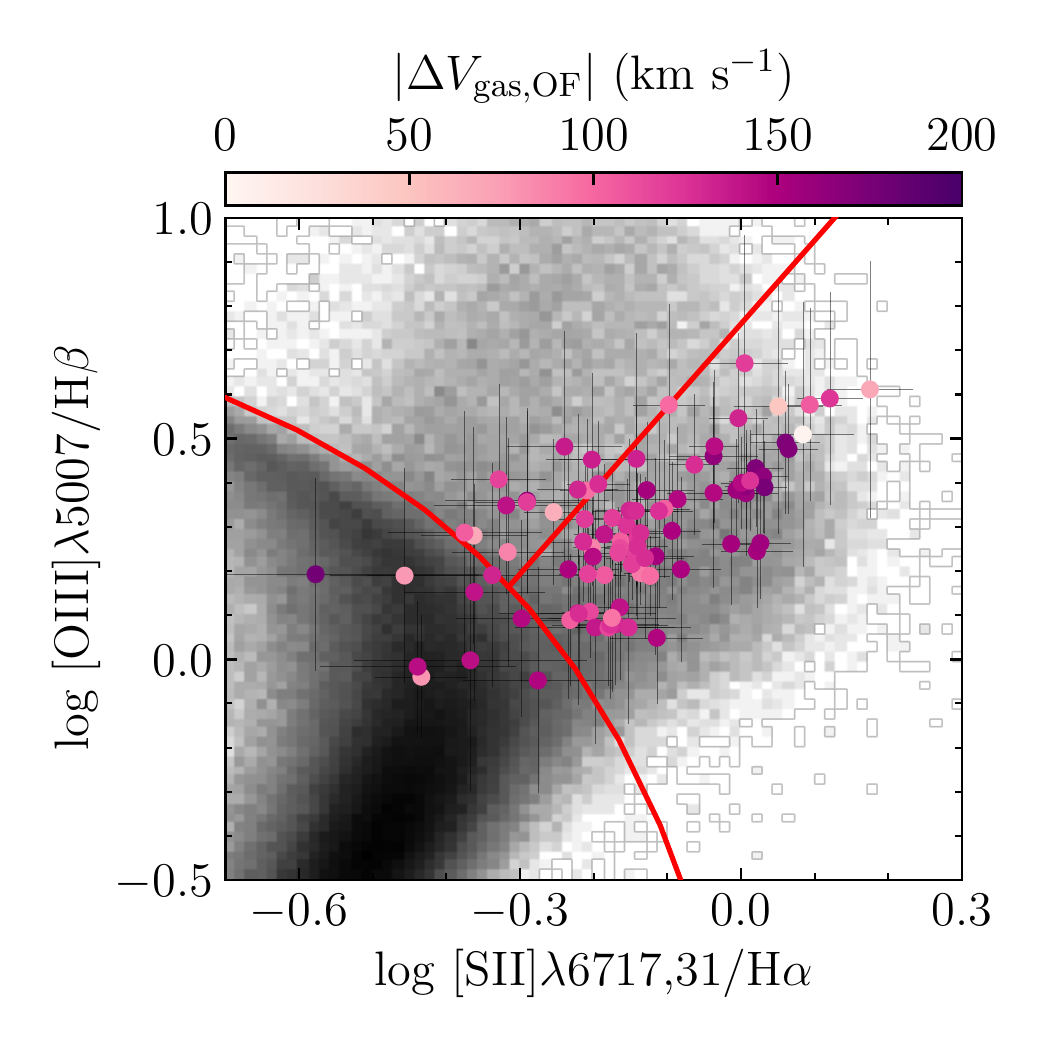}        
    \includegraphics[trim=0.5cm 0.55cm 0.5cm 0.4cm,clip,height=0.275\textheight]{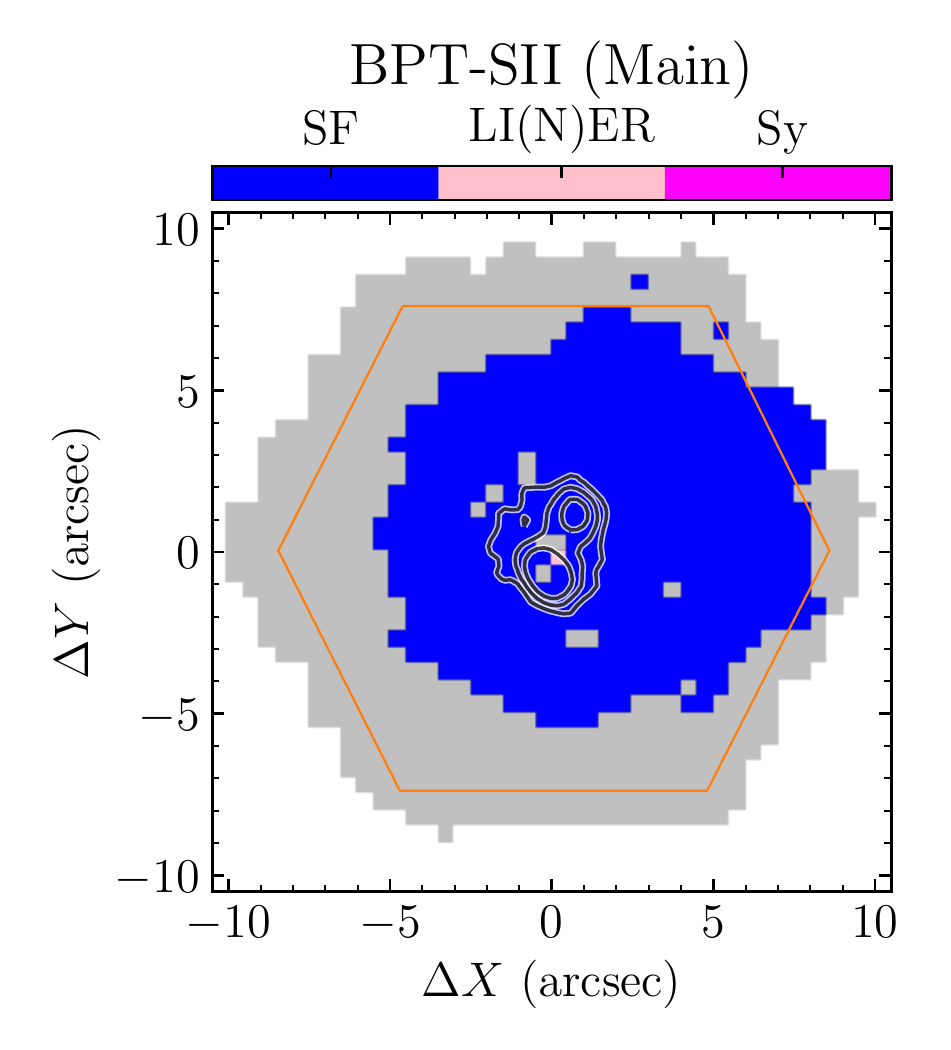}
     \includegraphics[trim=2.0cm 0.55cm 0.5cm 0.4cm,clip,height=0.275\textheight]{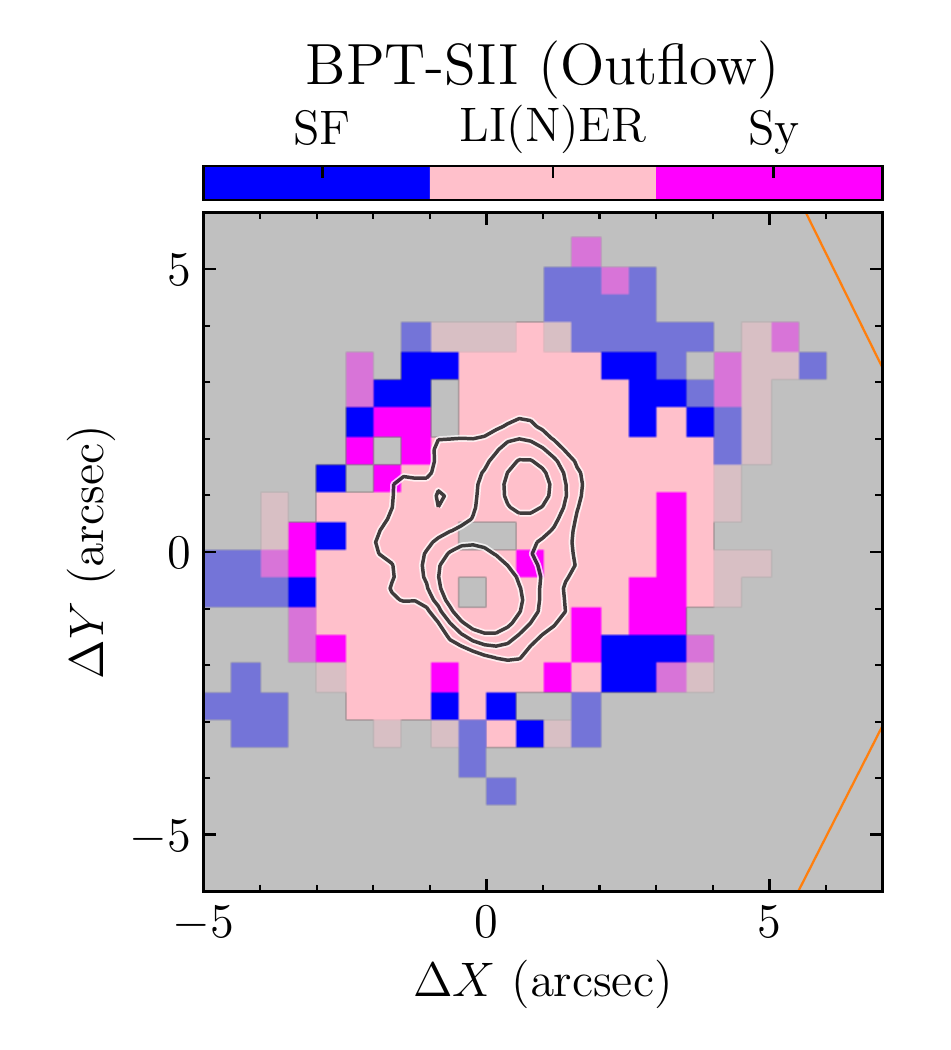}
    \end{center}
    \vspace*{-0.5cm}
    \caption{{\it Left}: [S{\sc ii}] BPT diagram of the ``outflow'' ionized gas in MaNGA 1-166919, zoomed in on the ``LI(N)ER'' region as defined by \citet{kewley06}.  The data points are the same as those shown in Figure \ref{fig:manga_bpt}, but are instead color coded by $|\Delta V_{\rm gas,OF}|$ -- the magnitude of the difference in line-of-sight velocity between the ``main'' and ``outflow'' components measured at a particular spaxel.  Spatial distribution of star formation (SF), LIER, and Seyfert-like photo-ionization of the ``main'' ({\it center}) and ``outflow'' ({\it right}) components, as determined from the location of the line-ratios measured in a particular spaxel on the [S{\sc ii}] BPT diagram shown in Figure \ref{fig:manga_bpt}.   In the {\it center} and {\it right} panels, the contours indicate 3 GHz (S-band) emission  5, 10, 50$\times$ the rms of the image shown in Figure \ref{fig:wideband_maps}.}
    \label{fig:bpt_sii}
\end{figure*}

As described in \S\ref{sec:outflow_cr}, the radio emission observed from MaNGA 1-166919 is believed to be produced by electrons accelerated by a shock propagating through this galaxy.  However, theoretical work suggests that $\lesssim10\%$ of the shocked material is accelerated to relativistic energies (e.g., \citealt{caprioli14, caprioli15}), with the bulk of the material heated to a temperature $T_{\rm shock}$ (e.g., \citealt{faucher12, caprioli14}):
\begin{eqnarray}
\label{eqn:tshock}
T_{\rm shock} & \sim & \frac{1}{2}\frac{m v_{\rm shock}^2}{k},
\end{eqnarray}
where $m$ is the mass of the particle, $k$ is Boltzmann's Constant, and $v_{\rm shock}$ is the velocity of the shock relative to the surroundings.  If $v_{\rm shock}$ is high enough, a copious amount of UV and soft X-ray photons will be generated at the shock front (e.g., \citealt{raymond76, allen08}).  The spectra from material photoionized by this radiation are expected to have emission line ratios (e.g., \citealt{dopita95b, allen08}) which lie within the Low Ionization Excitation Region (LIER) of the [S\ii] BPT diagram \citep{kewley06} (in the literature, this emission is often referred to arise from a ``Low Ionization Nuclear Excitation Region'' (LINER) since they were first and primarily identified in the centers of galaxies (e.g., \citealt{Heckman1980, Heckman1981}). However, subsequent work has found that such emission can be detected throughout a galaxy (e.g., \citealt{belfiore16}), and therefore use the more general term).  Indeed, as shown in Figures \ref{fig:manga_bpt} \& \ref{fig:bpt_sii}, the emission line ratios of the ``outflow'' material largely fall within the LIER region of such a diagram.  Furthermore, as shown in Figure \ref{fig:bpt_sii}, LIER-like emission is only detected in the center of this galaxy, and predominantly found in the ``outflow'' component of the ionized gas.  Therefore, it seems likely the ionized gas ``outflow'', as inferred from our analysis of the MaNGA (\S\ref{sec:emision_line_fitting}) and GMOS (\S\ref{sec:gmos}) data, is dominated by material photo-ionized by the shock which also accelerates the relativistic electrons responsible for the observed radio emission.  However, in many galaxies, such line ratios are instead thought to result from photo-ioninzation by post-AGB stars (e.g., \citealt{belfiore16, singh13, yan12}).  Such stars are expected to be prevalent in older ($\gtrsim1~{\rm Gyr}$) stellar populations -- as inferred for the central regions of this galaxy ($T_{\rm SSP} \sim 3$~Gyr; Figure \ref{fig:manga_stellarprop}) from our derivation of its stellar population as described in \S\ref{sec:manga_stars}.  

\begin{figure*}
    \begin{center}
    \includegraphics[height=0.3\textheight]{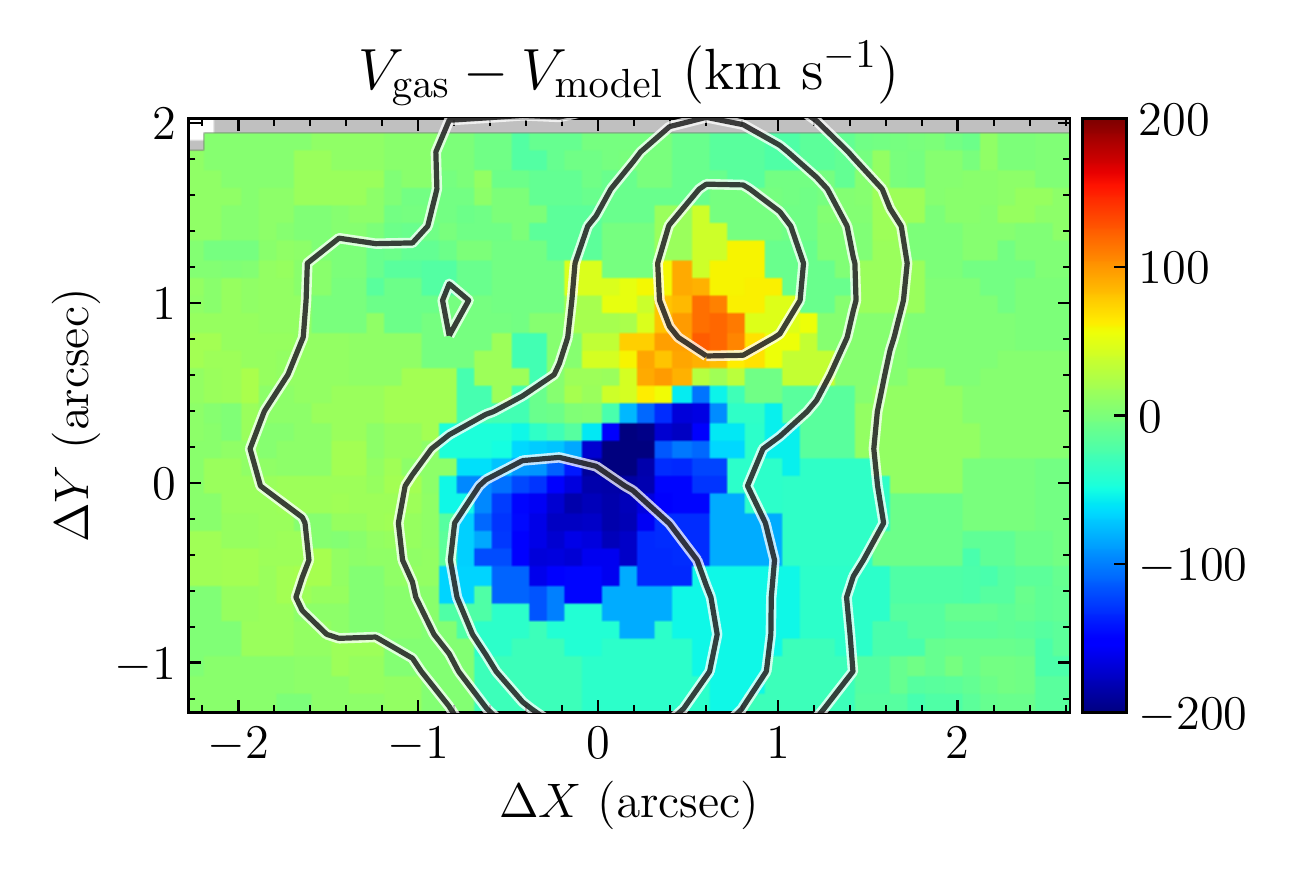}
    \includegraphics[height=0.3\textheight,trim=1.0cm 0 0.0cm 0,clip]{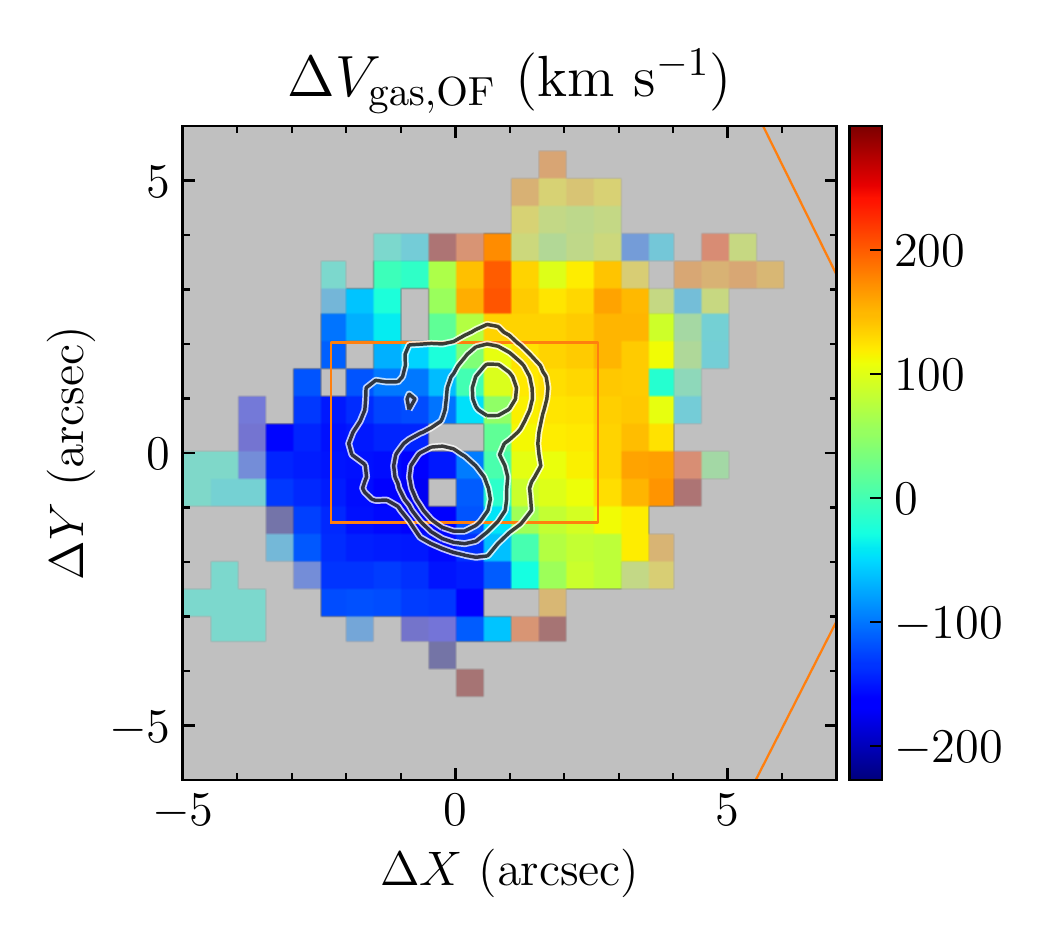}        
    \end{center}
    \vspace*{-0.75cm}
    \caption{The difference in line-of-sight velocity between the ``outflow'' and ``main'' (regularly rotating) ionized gas in MaNGA 1-166919 as derived from the GMOS ({\it left}) and MaNGA ({\it right}) datacubes.  In both figures, the contours indicate 3 GHz (S-band) emission  5, 10, 50$\times$ the rms of the image shown in Figure \ref{fig:wideband_maps}.}
    \label{fig:outflow_delv}
\end{figure*}

It is possible to distinguish between these models by measuring the kinematics of the putative ``outflow'' component of the ionized gas in this galaxy \added{, which we identify through deviations from a regularly rotating disk (as derived in \S \ref{sec:emision_line_fitting}.}  For the GMOS data, we estimated this by subtracting the line-of-sight velocity measured in a particular spaxel (left panel of Figure \ref{fig:gmos_gaskin}) with that predicted by our model for the regular rotating gas in this galaxy (right panel of Figure \ref{fig:main_rot}) while, for the MaNGA data, we calculated the difference in line-of-sight velocity between the ``main'' and ``outflow'' components ($\Delta V_{\rm gas, OF}$) as derived from the modelling described in \S\ref{sec:emision_line_fitting}.  As shown in Figure \ref{fig:outflow_delv}, the relative line-of-sight velocities show a clear spatial separation of ``red'' and ``blue'' components -- strongly suggesting a biconical ``outflow''.  This geometry is consistent with the $V_{\rm LOS} - \sigma_{\rm gas}$ of the ``outflow'' component (Figure \ref{fig:vvd}).  Furthermore, the correspondence between the kinematics of the ``outflow'' ionized gas and the SE and NW radio ``lobes'', which is particularly evident in the higher angular resolution GMOS data (left panel, Figure \ref{fig:outflow_delv}), strongly suggests a physical connection between the two.  Since the radio emission is produced by shock-accelerated particles, we therefore conclude these shocks are indeed responsible for producing the LIER-like emission observed from the ionized gas.  This conclusions is further supported by the observed dependence between $|\Delta V_{\rm gas,OF}|$ and the line-ratios of the ``outflow'' gas.  As shown in Figure \ref{fig:bpt_sii}, spaxels with higher values of $|\Delta V_{\rm gas,OF}|$ typically falling above and to the right of spaxels with lower $|\Delta V_{\rm gas,OF}|$ on the [S\ii] BPT diagram.  This trend is similar to that predicted by models for the emission of material photoionized by shock heated gas, which find that their location on the [S\ii] BPT diagrams moves up and to right as the shock velocity $v_{\rm shock}$ increases (e.g., \citealt{allen08}).

If correct, then the relative line-of-sight velocity between the ``main'' and ``outflow'' component provides a lower limit on $v_{\rm shock}$, since this quantity is not sensitive to differences in velocity between these components in the plane of the sky.  As a result, we estimate $v_{\rm shock}\gtrsim 100~{\rm km~s}^{-1}$ for the NW lobe and $v_{\rm shock} \gtrsim 200~{\rm km~s}^{-1}$ for the SE lobe -- sufficient to photoionize substantial amounts of material both ``downstream'' (post-shock) and ``upstream'' (pre-shock) of the shock (e.g., \citealt{Dopita1995, dopita96, raymond76, wilson99}).  Furthermore, the observed geometry, and differences in extent and shock velocity between the two components of the outflow, are consistent with that expected from outflows resulting from high-velocity material ejected from an AGN interacting with a clumpy interstellar medium (e.g., \citealt{mukherjee18, nelson19}).  In fact, simulations suggest that the bi-conical geometry of this outflow is the natural consequence of a central outflow being confined by the disk of a galaxy (e.g., \citealt{wagner12}).   In summary, both the relativistic and ionized component of this outflow appears to be the result of shocks driven into the surrounding ISM by a central engine.

\subsubsection{Ionized Gas Component}
\label{sec:outflow_thermal}

In \S\ref{sec:outflow_cr}, we presented our measurements for the energy contained in the relativistic component of this outflow.  However, this outflow also consists of several non-relativistic components, including ionized gas, atomic gas, and molecular material.  In this section, we use the emission line spectra derived from the MaNGA (\S\ref{sec:emision_line_fitting}) and GMOS (\S\ref{sec:gmos}) to measure the properties of its ionized component.  While studies of similar outflows in other galaxies suggest that atomic and molecular material may constitute the bulk of the entrained mass (e.g., \citealt{oosterloo17}), currently the observational data needed to measure the properties of these components in this galaxy are not available.

The mass of the ionized gas in this outflow $M_{\rm out}$, can be estimated as (e.g., \citealt{Soto2012, Baron2017}):
\begin{eqnarray}
    \label{eqn:m_out}
    M_{\rm out} & = & \mu m_{\rm H} V n_e f,
\end{eqnarray}
where $m_{\rm H}$ is mass of the Hydrogen atom, $\mu$ is the average atomic number of the emitting material (assumed to be Solar, such that $\mu \equiv 1.4$), $V$ is volume of the emitting region, $f$ is the filling factor, and $n_{\rm e}$ is the number density of electrons.  The H$\alpha$ luminosity of this region is equal to (e.g. \citealt{Baron2017}):
\begin{eqnarray}
    \label{eqn:h_alpha}
    L_{\rm H\alpha} & = & \gamma_{\rm H\alpha} n_\mathrm{e}^2 f V,
\end{eqnarray}
where $\gamma_{\rm H\alpha}$ is the H$\alpha$ emissivity of the ionized plasma.  In the case of highly ionized material with an electron temperature $T_e\approx10^4$~K and optically thick to Lyman line emission (``Case B''; e.g. \citealt{baker38,burgess58}), $\gamma = 3.56 \times 10^{-25}$~ erg cm$^3$ s$^{-1}$ \citep{OsterbrockFreland2006}.  As a result, we can calculate $M_{\rm out}$ by evaluating:
\begin{eqnarray}
    M_{\rm out} = \frac{\mu m_{\rm H} L_{\rm H\alpha}}{\gamma_{\rm H\alpha} n_{\rm e}}.
\end{eqnarray}
Therefore, to calculate this quantity, we first need to determine $n_e$ in the outflow as well as correct the observed H$\alpha$ emission for extinction along the line of sight.

\begin{figure*}[tb]
    \centering
    \includegraphics[trim=0.4cm 0 0.0cm 0,clip,height=0.24\textheight]{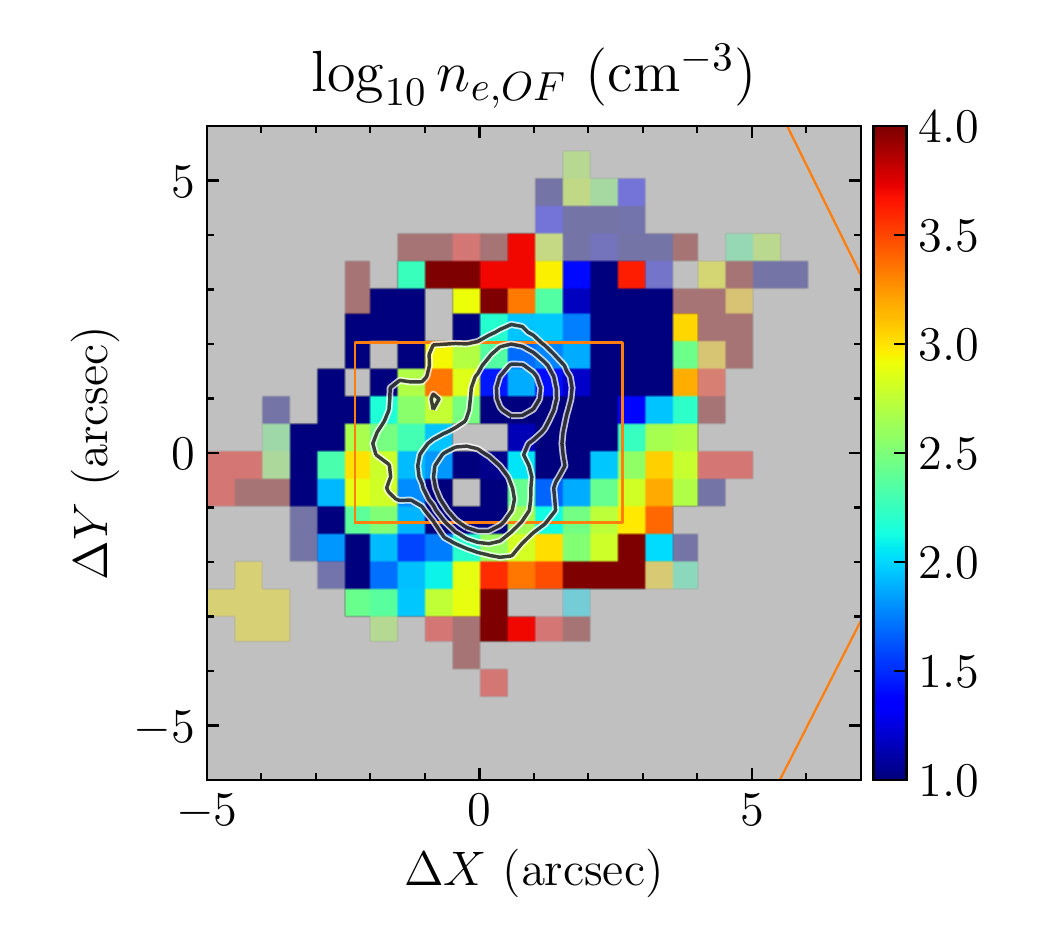}
    \includegraphics[trim=1.6cm 0 0.0cm 0,clip,height=0.24\textheight]{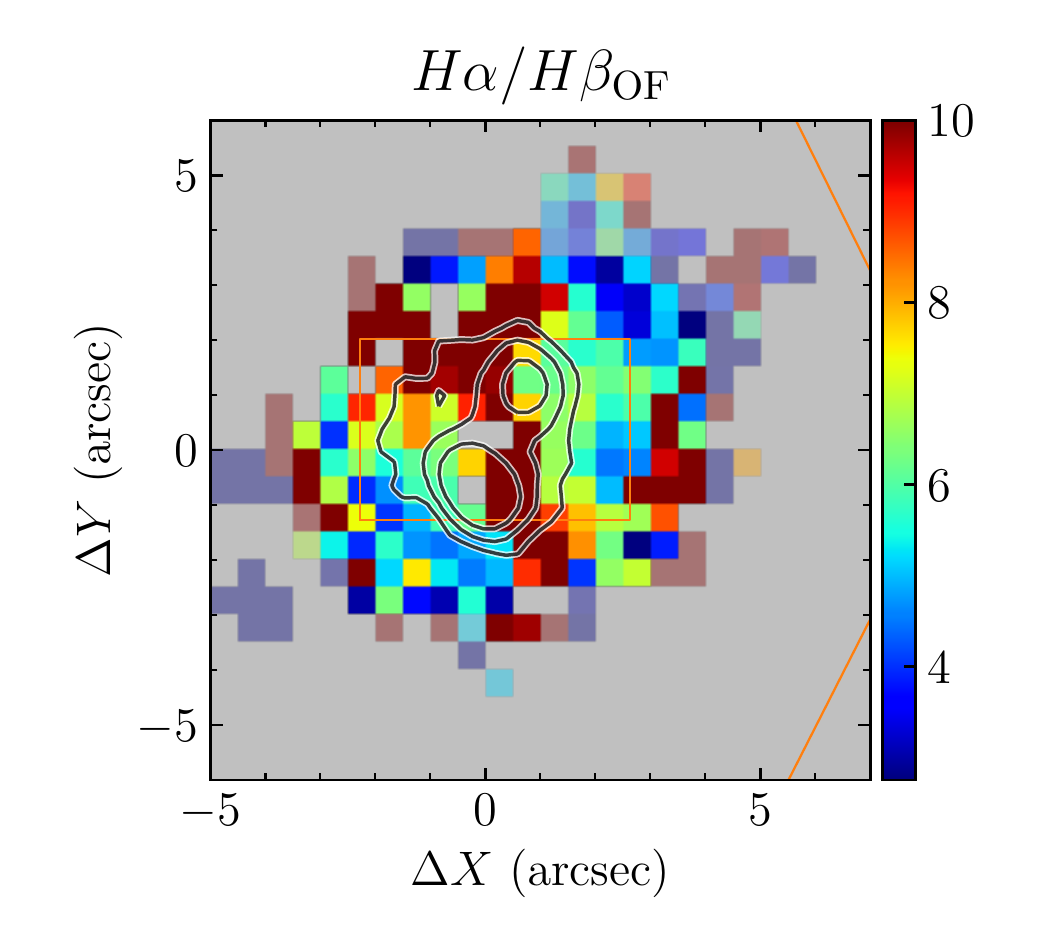}
    \includegraphics[trim=1.6cm 0cm 0 0cm,clip, height=0.24\textheight]{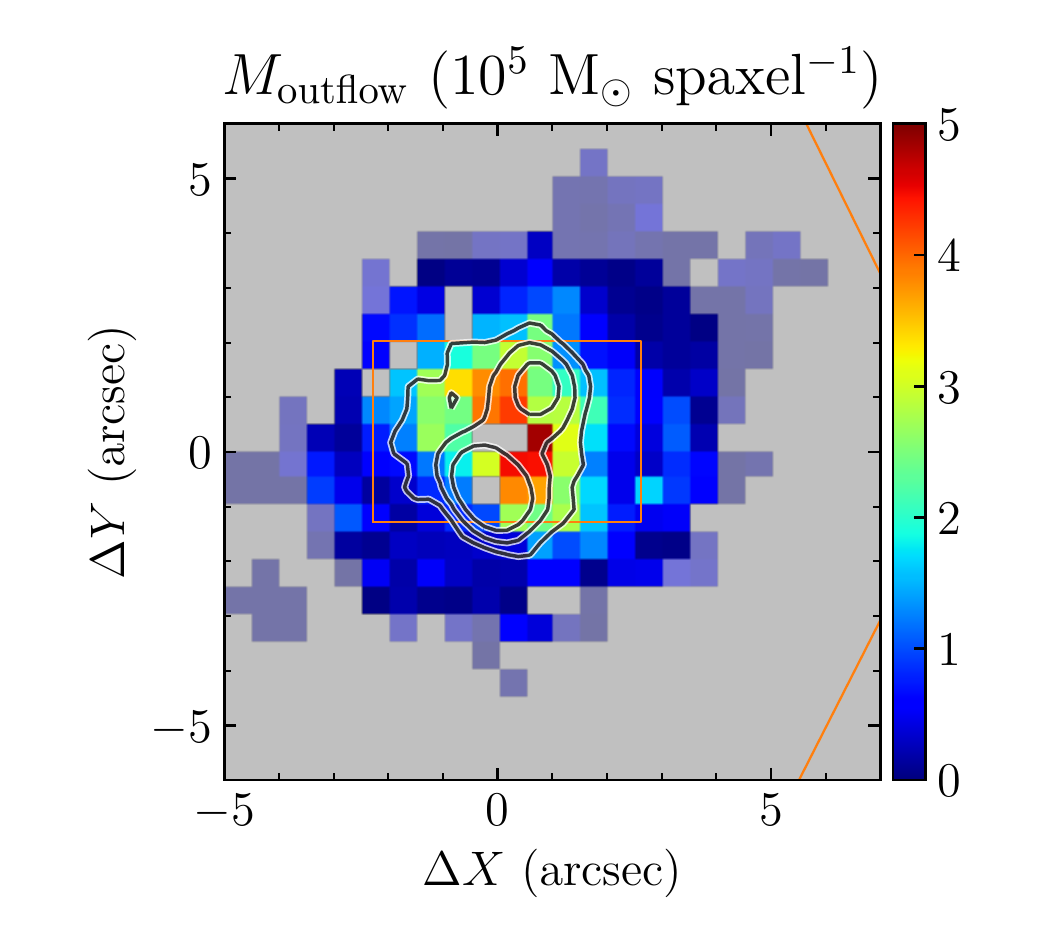}
    \vspace*{-0.5cm}
    \caption{Spatial distribution of the electron density $n_e$ ({\it left}), Balmer Decrement (H$\alpha$/H$\beta$ flux ratio, {\it center}), and ionized gas mass $M_{\rm out}$ of the ``outflow'' component to the ionized gas as measured using the MaNGA data (\S\ref{sec:emision_line_fitting}). In all panels the contours indicate 3 GHz (S-band) emission  5, 10, 50$\times$ the rms of the image shown in Figure \ref{fig:wideband_maps}.}
    \label{fig:m_out}
\end{figure*}

\begin{figure*}[tbh]
    \begin{center}
    \includegraphics[trim=1.6cm 0 0.0cm 0,clip,height=0.275\textheight]{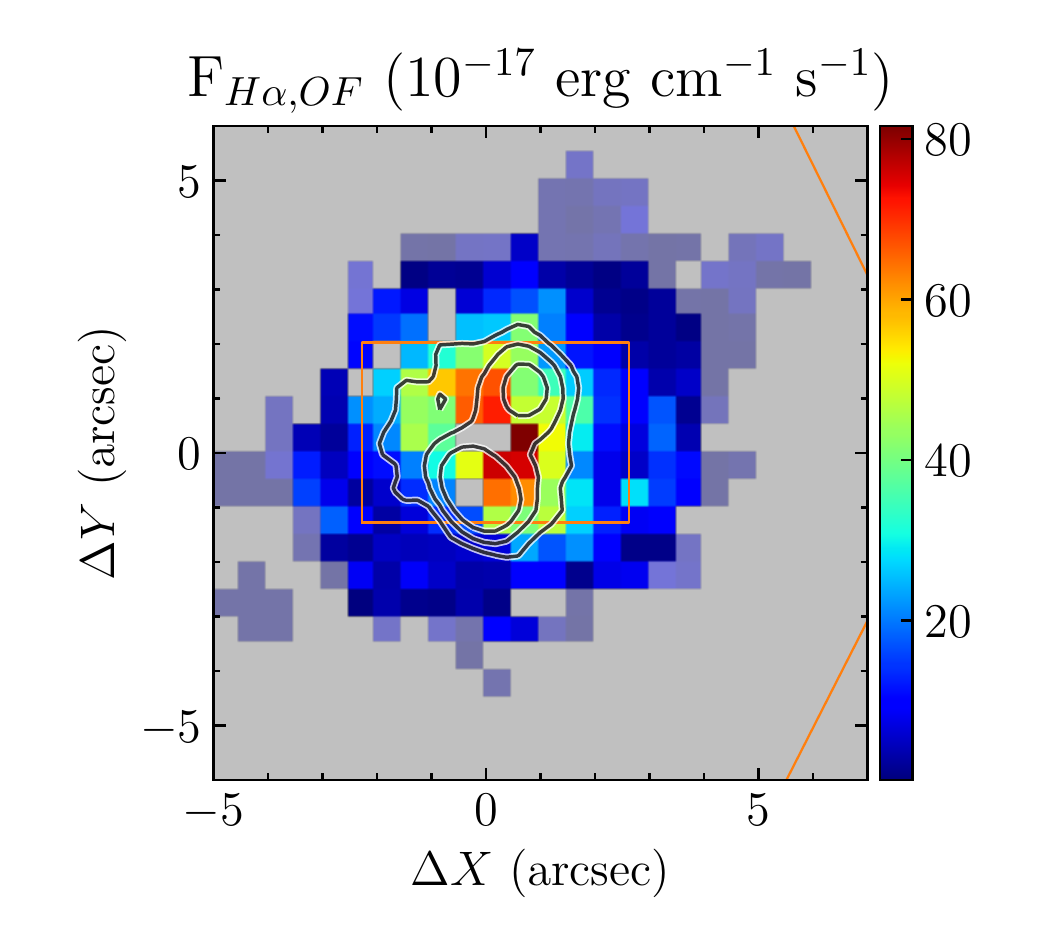}
    \includegraphics[height=0.275\textheight]{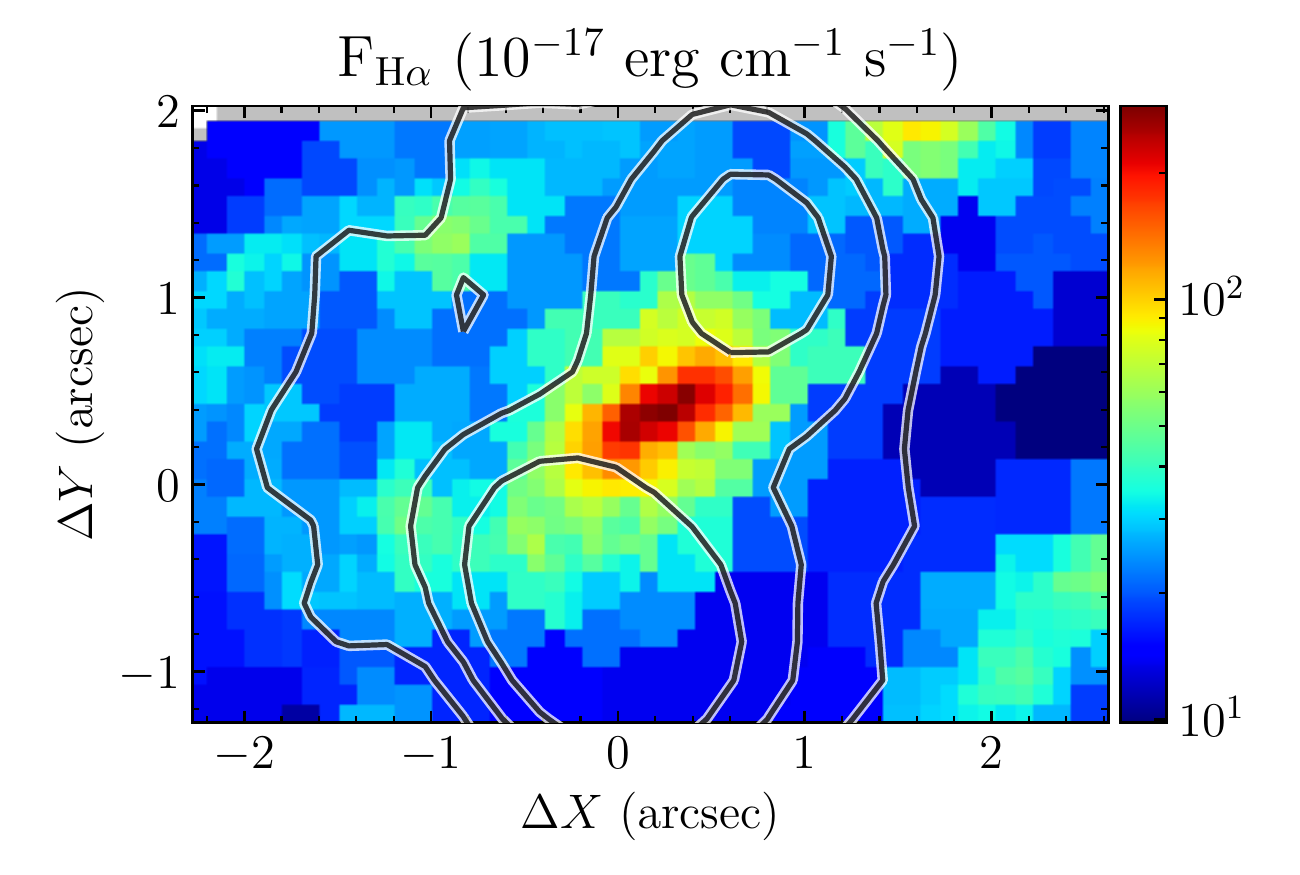}
    \end{center}
    \vspace*{-0.5cm}
    \caption{Extinction corrected H$\alpha$ flux of the ``outflow'' component to the ionized gas detected by MaNGA ({\it left}) and GMOS ({\it right}).  The contours indicate 3 GHz (S-band) emission  5, 10, 50$\times$ the rms of the image shown in Figure \ref{fig:wideband_maps}.}
    \label{fig:halpha_outflow}
\end{figure*}

We estimate $n_{\rm e}$ using the observed ratio of the [S\ii]6717/[S\ii]6731 emission lines \citep{OsterbrockFreland2006, Proxauf2014} separately for the ``outflow'' component of the MaNGA emission line spectrum and the GMOS data.  As shown in Figure \ref{fig:m_out} for the MaNGA ``outflow'', there are considerable variations in this parameter between adjacent spaxels due to the weakness of these lines and/or complexity of the line decomposition in many spaxels. We therefore calculated an average $n_{\rm e}$ for the outflow in each dataset, weighting the value in each spaxel by the total H$\alpha$+[N\ii] flux of its outflow component.  This yields a weighted average of $\log n_e \approx 1.9$ ($\approx80$~cm$^{-3}$) in the MaNGA data and $\log n_e \approx 2.0$ ($\approx100$~cm$^{-3}$) in the GMOS data, similar to the value inferred for outflows in other galaxies (e.g., \citealt{Harrison2014, Karouzos2016}) as well as in the previous analysis of this galaxy by \citet{Wylezalek2017} (Table \ref{tab:outflow_ion}).  However, this method for estimating the electron density preferentially returns values $10~{\rm cm}^{-3} \lesssim n_e \lesssim 10^4~{\rm cm}^{-3}$ (e.g., \citealt{Proxauf2014}).  Therefore, the central regions of the outflow where we estimate $n_{\rm e} \sim 10~{\rm cm}^{-3}$ may have a true density below this value.  Furthermore, the low ionization of the emitting gas suggests there is a significant neutral component to this material, whose density is not measured using this technique.  As a result, the total (neutral and ionized) gas density is likely to be significantly higher than the estimated value of $n_e$ (e.g., \citealt{dempsey18}).  Lastly, recent results suggests that the electron density estimated using [S\ii] are systematically lower than those using other emission lines (e.g., \citealt{davies20}) unfortunately not detected with sufficient spectral resolution or low signal-to-noise in our data. 

To estimate the extinction along the line-of sight towards the outflowing material in both datasets, we use the Balmer Decrement (the H$\alpha$/$H\beta$ flux ratio) of the outflow component measured in the MaNGA data.  This is because, as described in \S\ref{sec:gmos}, H$\beta$ is not detected in the GMOS data.  As shown in Figure \ref{fig:m_out}, this quantity varies significantly, and for the MaNGA outflow corrected the H$\alpha$ flux of the outflow in each spaxel with the corresponding Balmer Decrement measured for this component.  The differing angular resolutions of the MaNGA and GMOS data precludes us from making a similar spaxel by spaxel correction.  As a result, we estimate the average extinction towards the outflow in the GMOS data from the MaNGA data in two ways: the mean value, $\langle H\alpha/H\beta \rangle = 6.8$, and the \Ha+[N\ii] flux weighted averaged value $\langle H\alpha/H\beta \rangle_\mathrm{H\alpha+NII} = 8.6$.   In all cases, we use Balmer Decrement to correct the H$\alpha$ flux using the extinction law derived by \citet{Cardelli1989}, with the results shown in Figure \ref{fig:halpha_outflow}.

With these measurements of $\log n_e$ and extinction corrected H$\alpha$ flux in hand, it is now possible to measure the total mass of outflowing ionized gas in this galaxy.  For the MaNGA data, we do so by adding together the mass estimated in each spaxel (Figure \ref{fig:m_out}), deriving a total mass of $M_{\rm out} \sim 2.4\times10^7~{\rm M}_\odot$.  For the GMOS data, we use the total H$\alpha$ flux measured in the ``outflow'' region -- which, as described in \S\ref{sec:gmos}, corresponds to those spaxels with $\sigma_{\rm gas} > 100~{\rm km~s}^{-1}$, or the central $r_\mathrm{OF} \approx 1\arcsec \approx 1.4$~kpc (Figure \ref{fig:gmos_gaskin}) of this galaxy.  In this case, we estimate $M_{\rm out} \sim (2.2-3.9)\times10^7~{\rm M}_{\odot}$ -- in good agreement with the value derived from the MaNGA data alone.

\begin{table*}[tb]
\caption{Properties of the Ionized Gas Outflow and AGN}
\label{tab:outflow_ion}
\vspace*{-0.25cm}
\begin{center}
\begin{tabular}{cccc}
    \hline
    \hline
    {\sc Property} & \multicolumn{2}{c}{This Work} & \citet{Wylezalek2017} \\
    & MaNGA & GMOS\tablenotemark{a} &\\
    \hline
    $n_e$ & $\approx80~{\rm cm}^{-3}$ & $\approx100~{\rm cm}^{-3}$ & $\equiv100~{\rm cm}^{-3}$ \\  
    $M_{\rm out}$ & $2.4\times10^7~{\rm M}_\odot$ & $(2.2-3.9)\times10^7~{\rm M}_\odot$ & \\
    $K_{\rm ion}$ & $2.4\times10^{55}~{\rm ergs}$ & $(4.6-8)\times10^{54}~{\rm ergs}$ & \\
    $t_{\rm age}$ & $\approx 6$~Myr & 6.2~Myr &$2-3$~Myr \\
    $\dot{M}_{\rm out}$ & $\approx 4~\frac{{\rm M}_\odot}{\rm year}$ & $(3.5-6.3)~\frac{{\rm M}_\odot}{\rm year}$ & $66~\frac{{\rm M}_\odot}{\rm year}$ \\
    $\dot{E}_{\rm kin}$ & $\approx1.3\times10^{41}~\frac{\rm ergs}{\rm s}$ & $(2.4-4.1)\times10^{40}~\frac{\rm ergs}{\rm s}$ & $1.4\times10^{42}~\frac{\rm ergs}{\rm s}$ \\
    $L_{\rm bol}$ & $(0.3-9.1)\times10^{43}~\frac{\rm ergs}{\rm s}$ & & $(6-7.1)\times10^{43}~\frac{\rm ergs}{\rm s}$ \\
    \hline
    \hline
\end{tabular}
\end{center}
\tablenotetext{a}{Ranges reflect the difference resulting from the two ways of estimating the extinction along the line of sight, as described in \S\ref{sec:outflow_thermal}.}
\end{table*}

\begin{figure*}[tbh]
    \begin{center}
    \includegraphics[height=0.175\textheight]{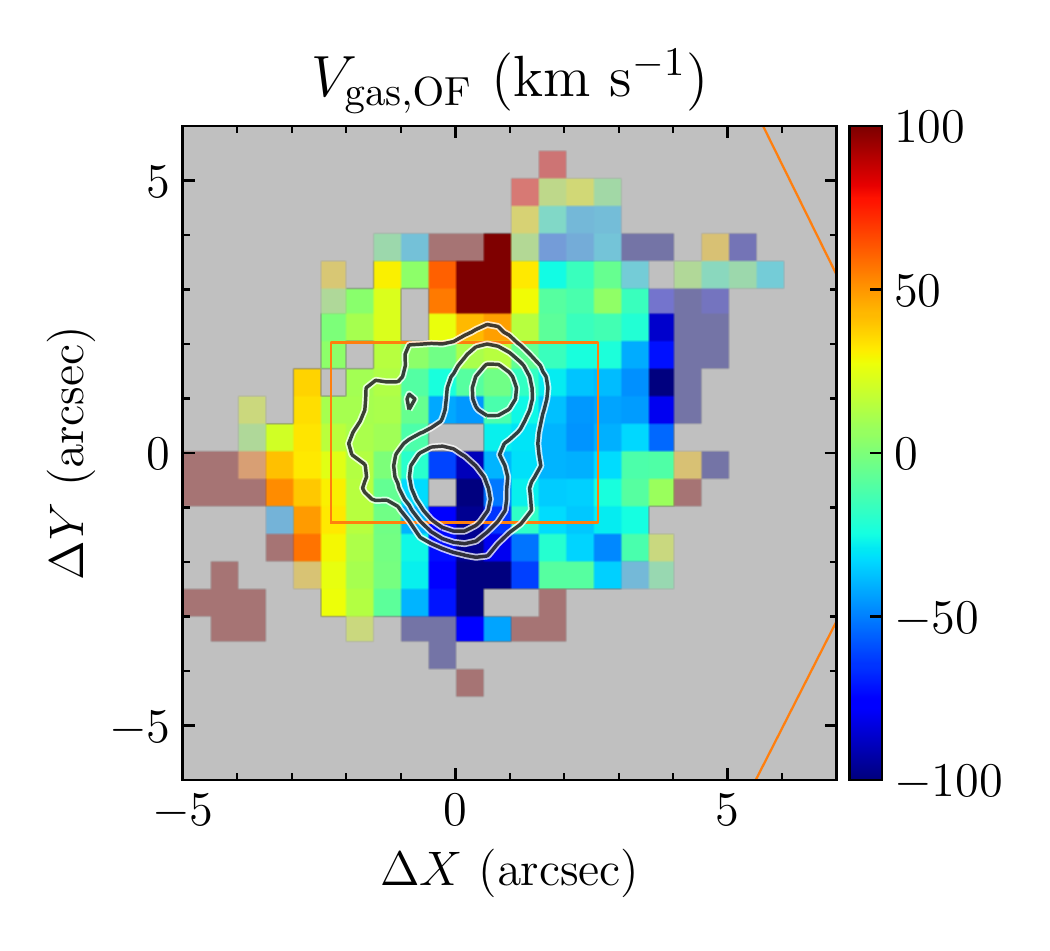}    
    \includegraphics[height=0.175\textheight,trim=1.3cm 0 0.0cm 0,clip]{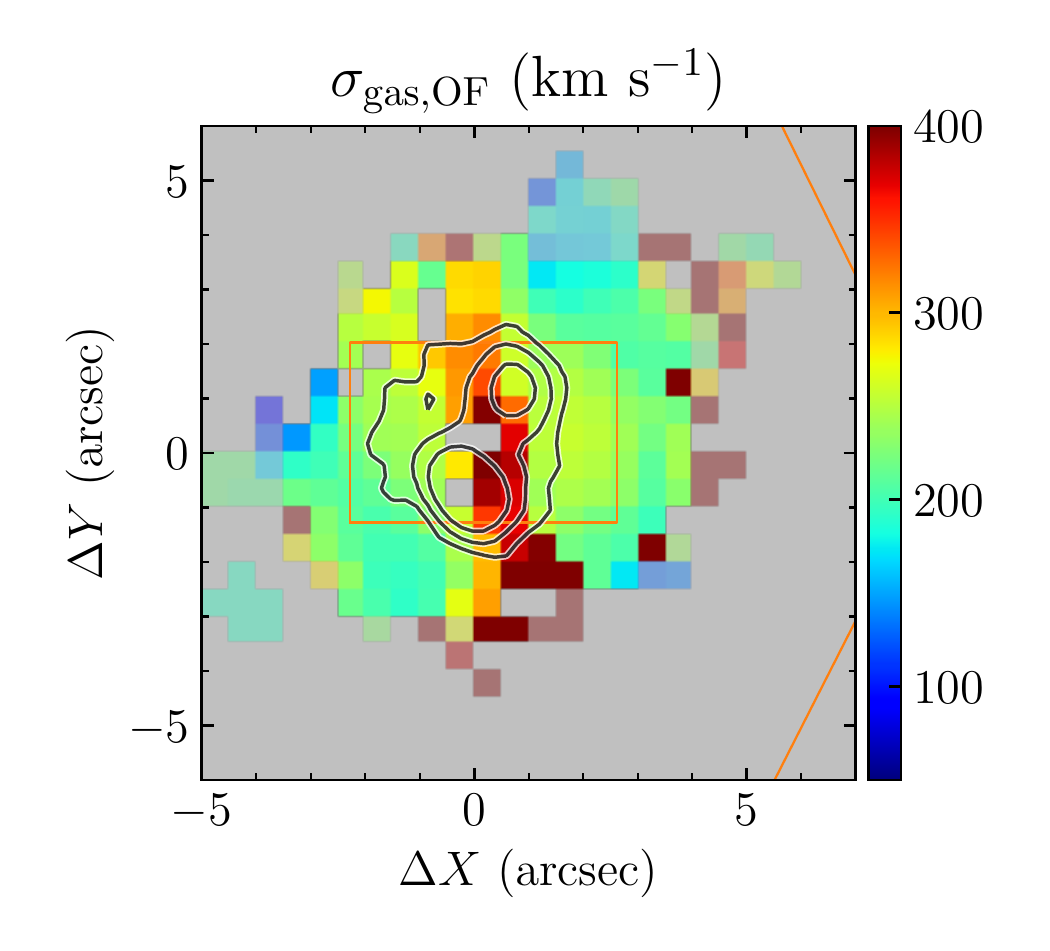}
    \includegraphics[height=0.175\textheight,trim=1.3cm 0 0.0cm 0,clip]{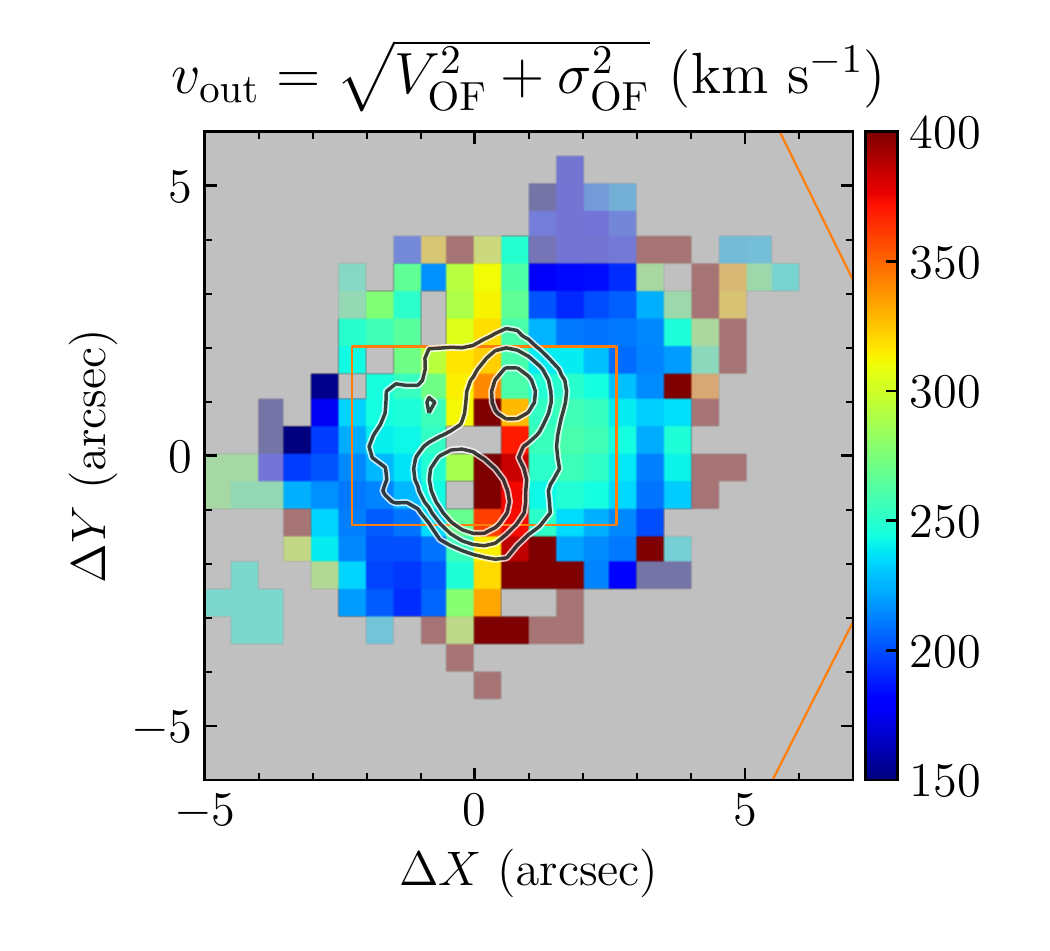}
    \includegraphics[height=0.175\textheight,trim=1.3cm 0 0.0cm 0,clip]{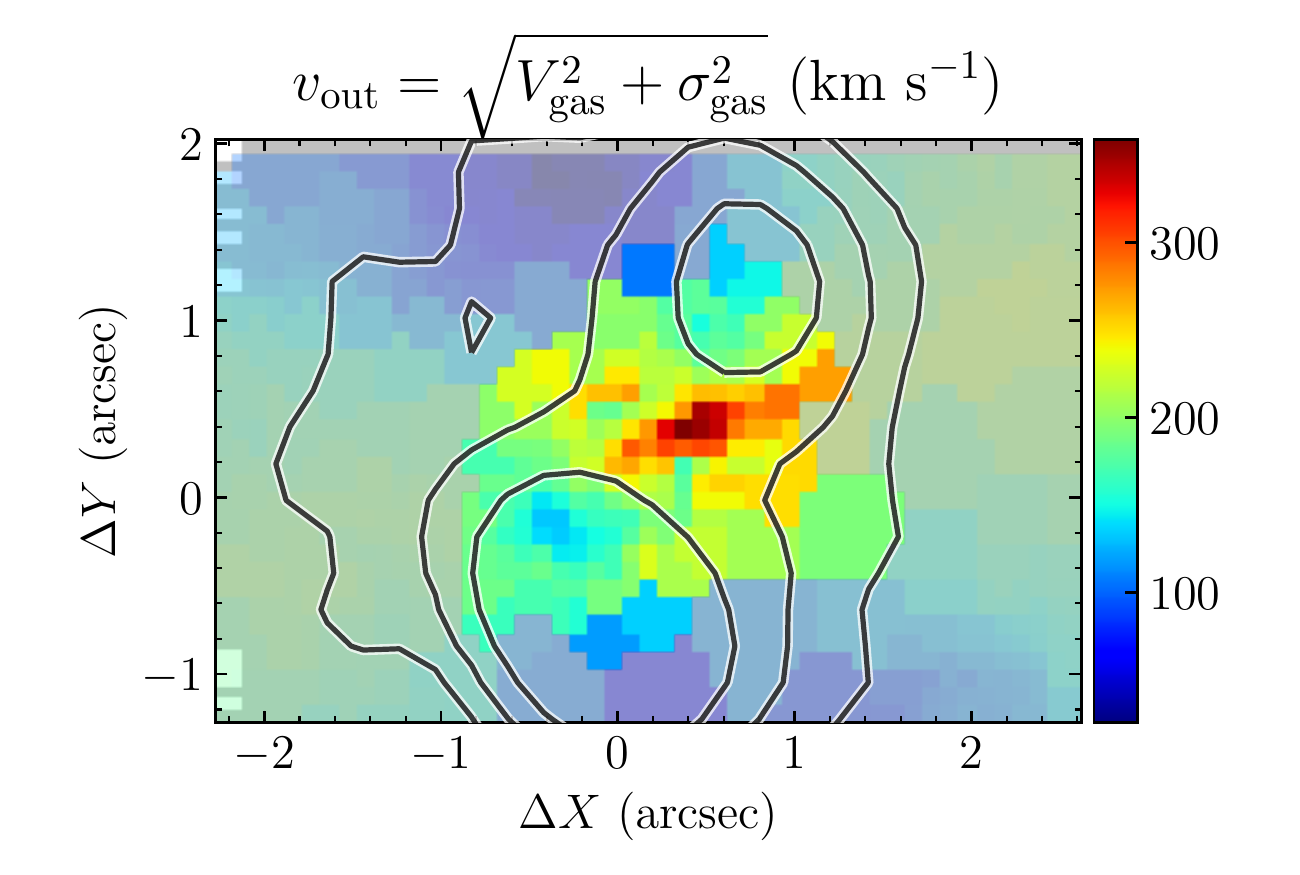}
    \end{center}
    \vspace*{-0.5cm}
    \caption{Line-of-sight velocity $V_{\rm gas, OF}$ ({\it left}) and velocity dispersion $\sigma_{\rm gas,OF}$ ({\it center left}) of the ``outflow'' component to the ionized gas as measured from the MaNGA emission line spectrum (\S\ref{sec:emision_line_fitting}). Outflow velocity $v_{\rm out}$ as derived from the ``outflow'' contribution to the emission line spectrum as measured by MaNGA ({\it center right}) and GMOS ({\it right}). In all panels the contours indicate 3 GHz (S-band) emission  5, 10, 50$\times$ the rms of the image shown in Figure \ref{fig:wideband_maps}.}
    \label{fig:v_out}
\end{figure*}

Furthermore, we can estimate the total velocity of the outflow $v_{\rm out}$ as (e.g., \citealt{Karouzos2016}):
\begin{eqnarray}
\label{eqn:v_out}
v_{\rm out} & = & \sqrt{ v_{\rm los}^2 + \sigma_{\rm gas}^2},
\end{eqnarray}
where the value of these quantities as measured from the GMOS data is shown in Figure \ref{fig:gmos_gaskin} and as derived for the ``outflow'' component in the MaNGA emission line spectrum is shown in Figure \ref{fig:v_out}.  Both data sets give similar values of $v_{\rm out}$, with the ionized gas moving near $\sim400$~\kms\ near the center of the galaxy and slowing to values of $\sim200-300$~\kms\ near the edge of the radio emission.  

\begin{figure}[tbh]
    \begin{center}
    \includegraphics[trim=1.6cm 0cm 0 0cm,clip, width=0.375\textwidth]{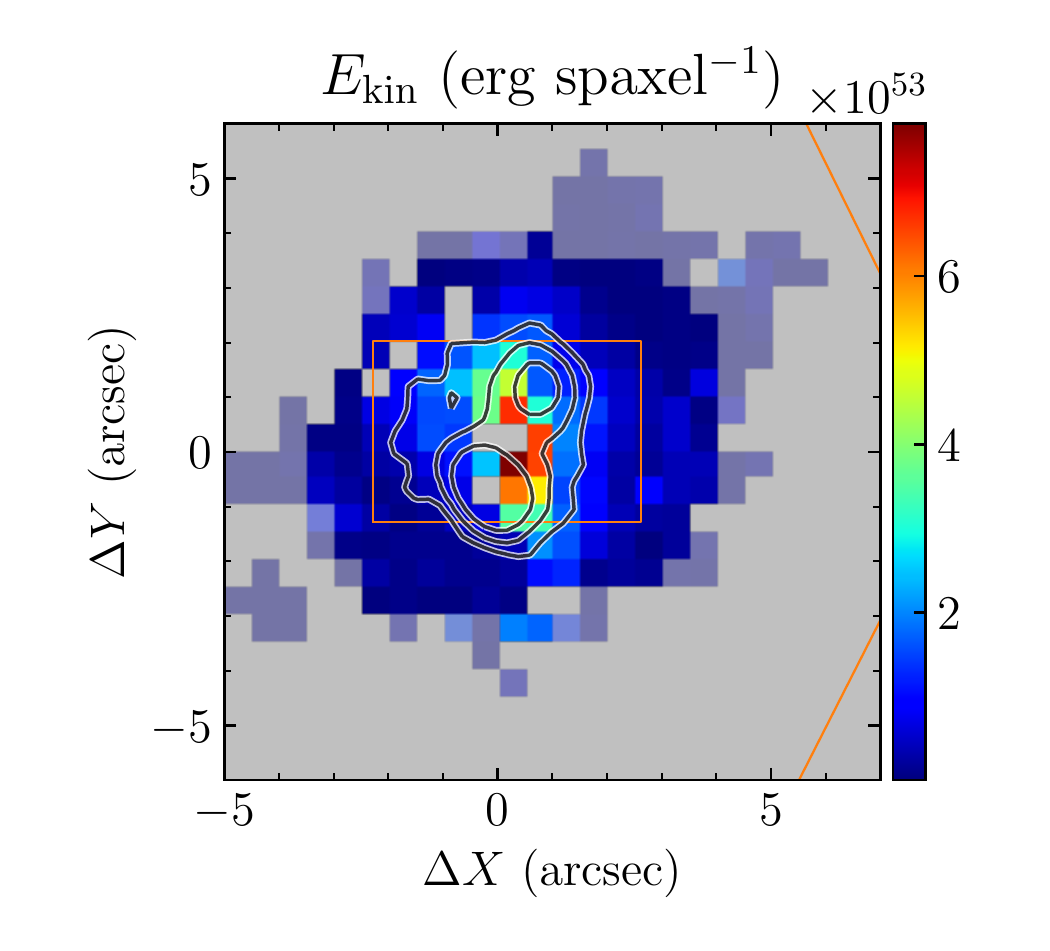}
    \includegraphics[trim=1.6cm 0cm 0 0cm,clip, width=0.375\textwidth]{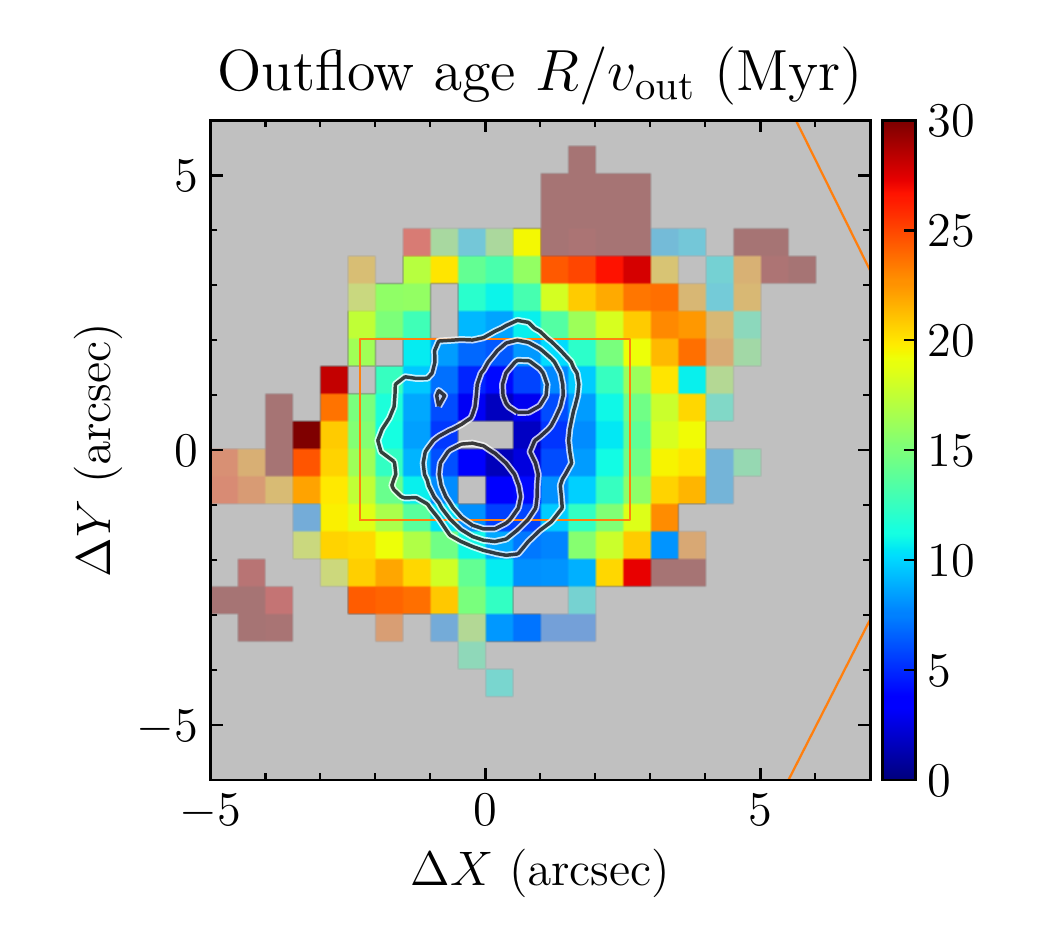}        
    \end{center}
    \vspace*{-0.75cm}
    \caption{Spatial distribution of the kinetic energy ({\it leftt}) age ({\it right}) of the ionized gas in the outflow as measured from the ``outflow'' component of the MaNGA emission line spectrum.  The contours indicate 3 GHz (S-band) emission  5, 10, 50$\times$ the rms of the image shown in Figure \ref{fig:wideband_maps}.} 
    \label{fig:outflow_age}
\end{figure}

With this velocity information, we can calculate the kinetic energy and ``age'' of the ionized gas in this outflow.  We determine the kinetic energy $K_{\rm ion}$ of the ionized gas in the MaNGA data by evaluating $K = \frac{1}{2}M_{\rm ion}v_{\rm out}^2$ in each spaxel (Figure \ref{fig:outflow_age}), and add together the values to measure a total $K_{\rm ion} = 2.4\times10^{55}~{\rm ergs}$ in the MaNGA data and $K_{\rm ion}=(4.6-8)\times10^{54}$~erg for the GMOS data.  These energies are comparable to the minimum energy estimated for the relativistic content of this outflow (Table~\ref{tab:rel_prop}, \S\ref{sec:outflow_cr}), suggesting these two components are in rough equipartition.  Furthermore, we estimate the age of the MaNGA outflow in each spaxel as:
\begin{eqnarray}
\label{eqn:age}
t_{\rm age} & = & \frac{R}{v_{\rm out}},
\end{eqnarray}
where $R$ is the projected physical separation between the spaxel and the center of the galaxy.  As shown in Figure \ref{fig:outflow_age}, this suggests the outflow is $\sim6$~Myr old.  For the GMOS data, the $\sim1.4$ kpc extent of the outflow coupled with the \Ha+[N\ii] flux weighted average outflow velocity $v_\mathrm{out}$ value of 222~\kms\ suggests $t_\mathrm{age}=6.2$~Myr -- consistent with the results derived from the MaNGA data.  As shown in Table~\ref{tab:outflow_ion}, this suggests a mass outflow rate of $\sim 4~{\rm M}_\odot\, {\rm yr}^{-1}$ and a kinetic power of $\dot{E}_{\rm kin} \sim (0.2-1)\times10^{41}~{\rm ergs~s}^{-1}$.

\subsection{Active Galactic Nucleus}
\label{sec:agn}

\begin{figure*}[tb]
\begin{center}
    \includegraphics[trim=0.5cm 1.6cm 0.5cm 0.4cm,clip,height=0.3\textheight]{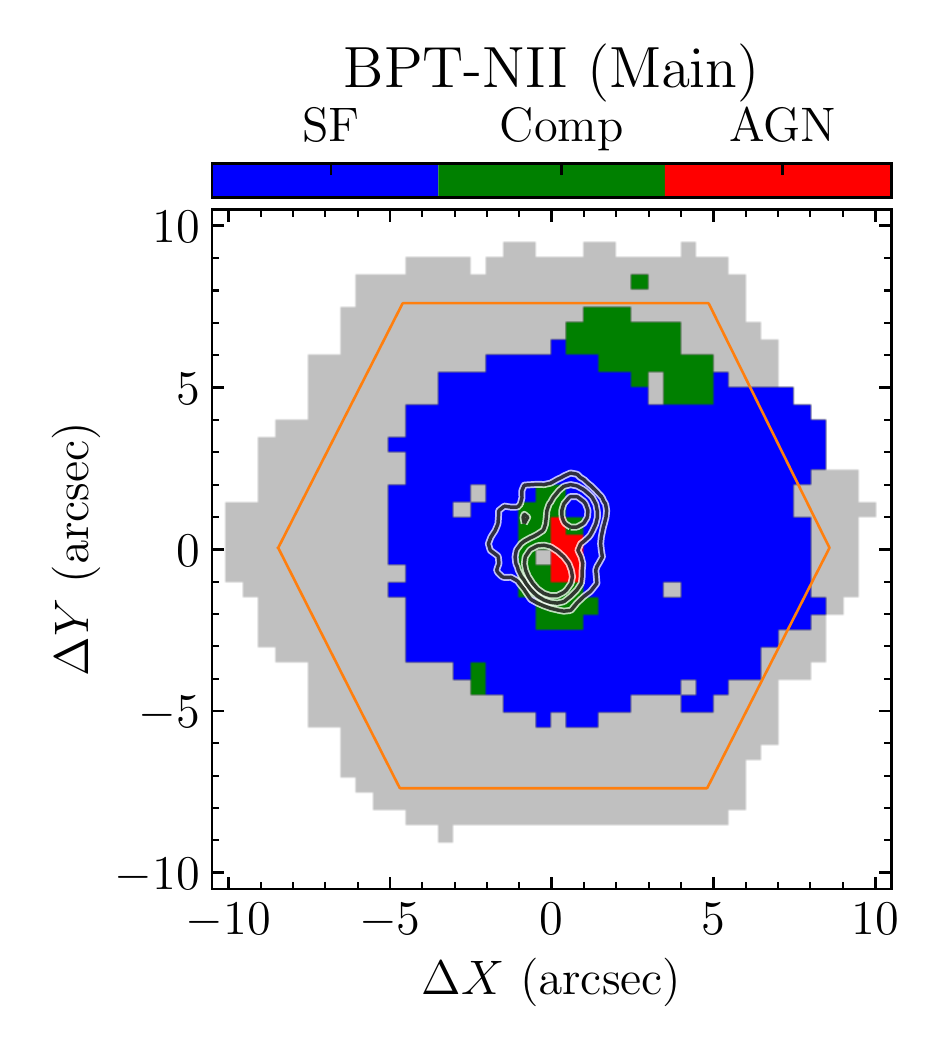}
\hspace*{1.5cm}
    \includegraphics[trim=2.0cm 1.6cm 0.5cm 0.4cm,clip,height=0.3\textheight]{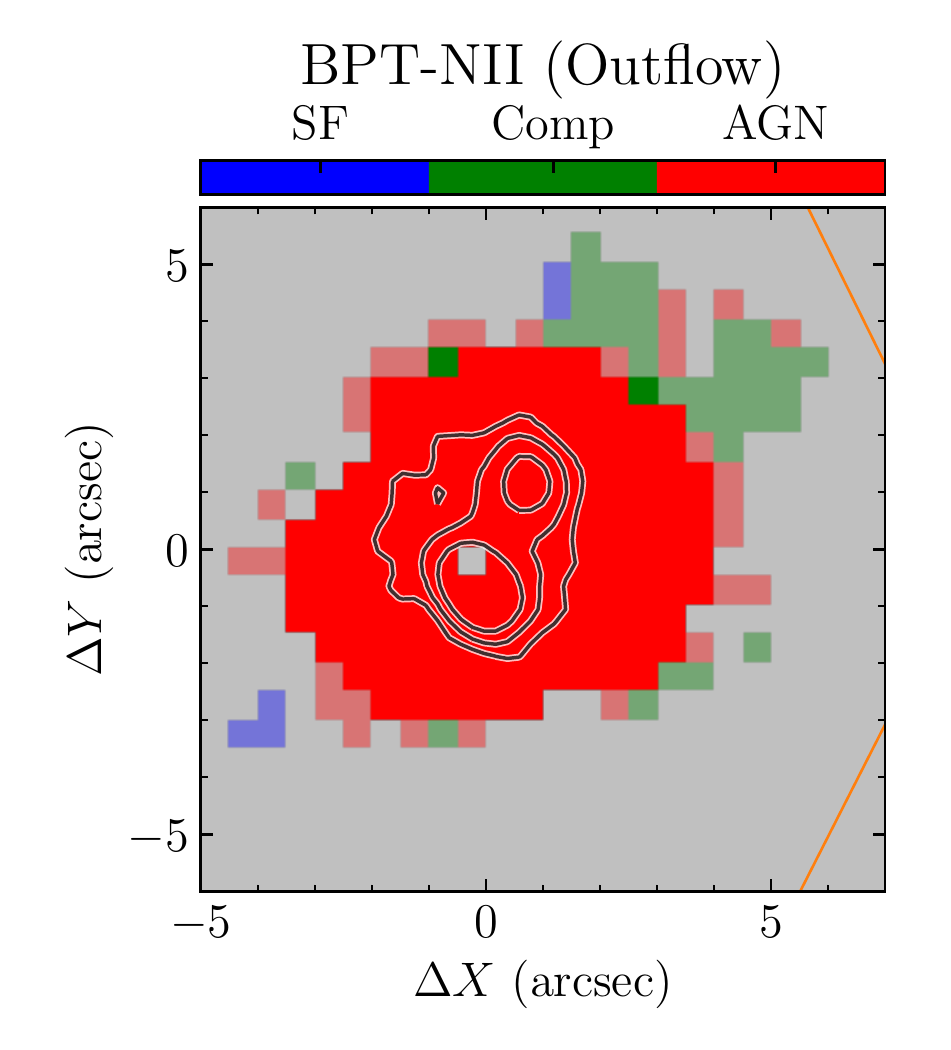}
\end{center}
\caption{Spatial distribution of the source of ionizing photons as inferred from the N{\sc ii} BPT diagram (Figure \ref{fig:manga_bpt}) of the ``main'' ({\it left}) and ``outflow'' ({\it right}) components in the emission line spectra of each spaxel.  In both figures the contours indicate 3 GHz (S-band) emission  $5, 10, 50\times \mathrm{RMS}$ of the image shown in Figure \ref{fig:wideband_maps}.}
\label{fig:manga_bptmap_nii}
\end{figure*}

The central location and morphology of this outflow is suggestive of an AGN origin.  While the emission line spectra observed by MaNGA (\S\ref{sec:emision_line_fitting}; Figures \ref{fig:manga_bpt}, \ref{fig:bpt_sii} \& \ref{fig:manga_bptmap_nii}) and the high velocity dispersion measured by GMOS (\S\ref{sec:gmos}; Figure \ref{fig:emission_gmos_WHAN}) of the central gas are all consistent with AGN activity -- its existence does not provide it is responsible for generating this outflow (e.g., \citealt{shimizu19}) nor explain how accretion onto the central SMBH results in the $\sim100 - 200~{\rm km~s}^{-1}$ shocks (\S\ref{sec:outflow_kin}) responsible for creating its observed relativistic (\S\ref{sec:outflow_cr}) and ionized components (\S\ref{sec:outflow_thermal}).

Determining if the AGN can power the observed outflow first requires estimating the AGN's bolometric luminosity $L_{\rm bol}$.  Current methods using the emission line spectra of the AGN assume that  the emitting material is photoionized by material accreting onto the SMBH (e.g., \citealt{heckman04}, \citealt{Netzer2009} and references therein).  As argued in \S\ref{sec:outflow_kin}, the ``outflow'' component of the emission line spectra is believed to be dominated by shock-heated material.  Therefore, the ``main'' component of the observed emission line spectrum should result in a more accurate estimate of $L_{\rm bol}$.  

\begin{figure*}[tb]
\begin{center}
    \includegraphics[trim=0.4cm 0 0.0cm 0,clip,height=0.24\textheight]{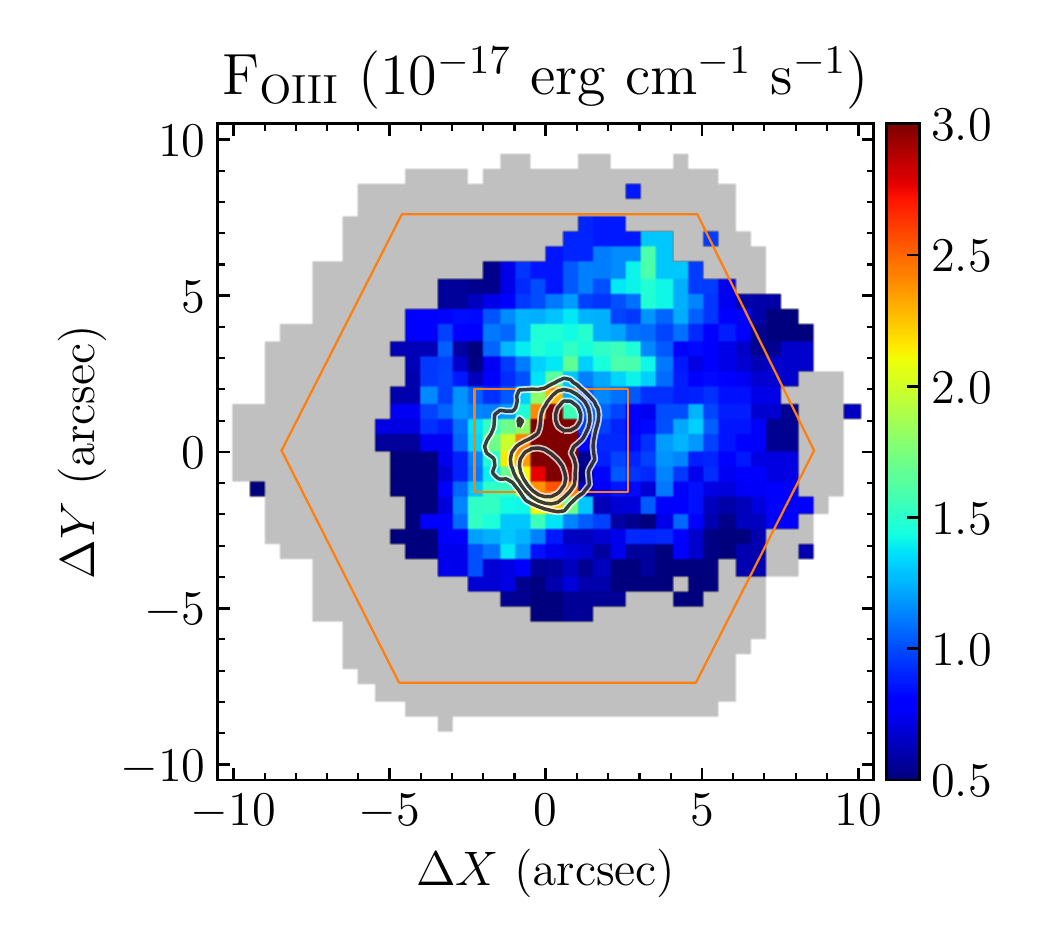}
    \includegraphics[trim=0.4cm 0 0.0cm 0,clip,height=0.24\textheight]{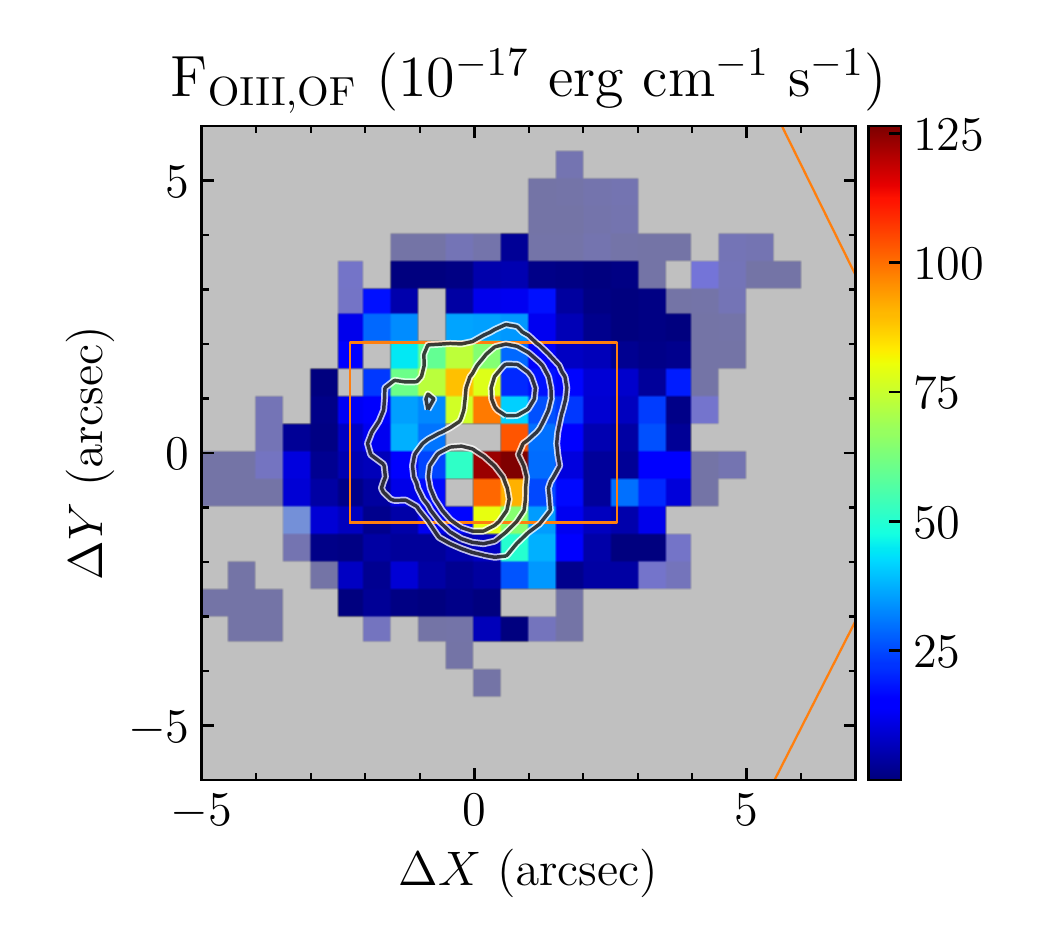}
    \includegraphics[trim=1.3cm 0 0.0cm 0,clip,width=0.24\textheight]{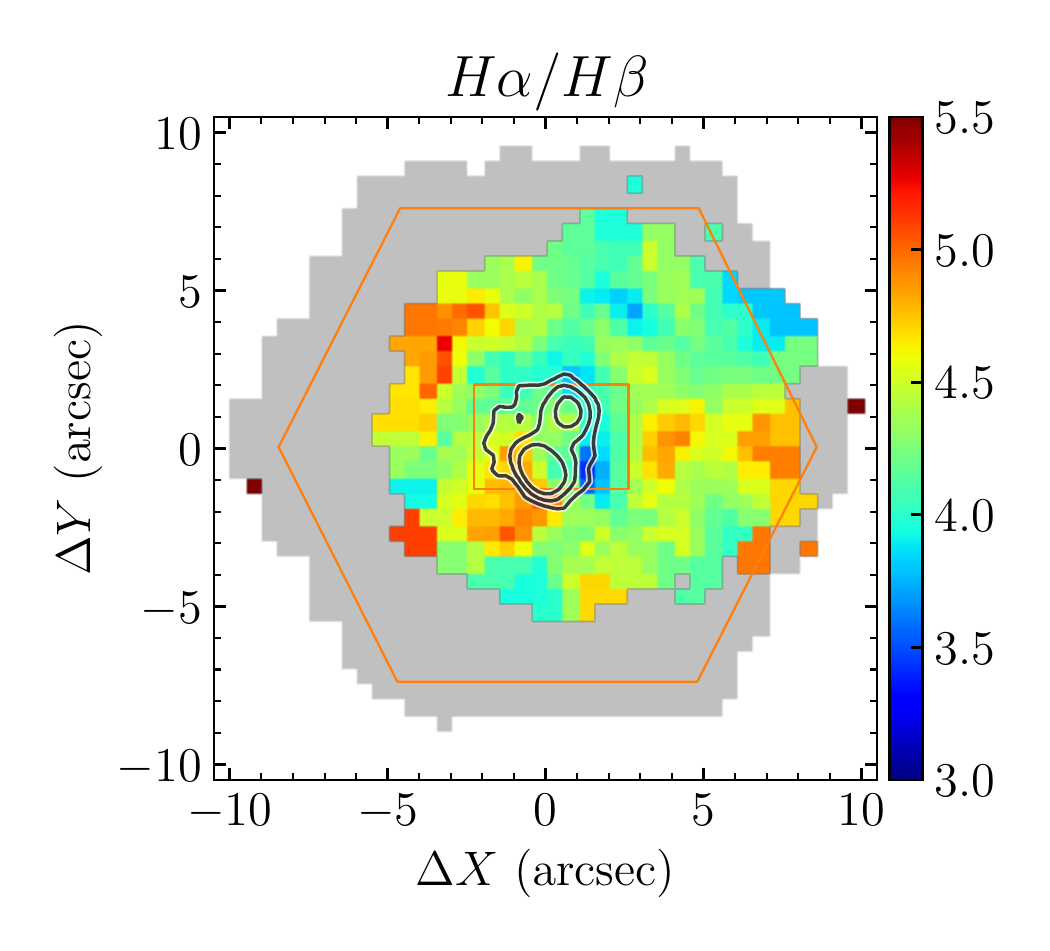}
\\
\end{center}
\vspace*{-0.5cm}
\caption{{\it Left}: [O{\sc iii}] flux of both the ``main'' and ``outflow'' component of the ionized gas as measured by MaNGA. {\it Right}: Balmer decrement H$\alpha$/H$\beta$ of the ``main'' component.  In all panels, the contours indicate 3 GHz (S-band) emission  $5, 10, 50\times \mathrm{RMS}$ of the image shown in Figure \ref{fig:wideband_maps}.}
\label{fig:manga_oiii_bd}
\end{figure*}

One of the most common techniques for determining $L_{\rm bol}$ uses the extinction-corrected luminosity of the [O\iii] line $L_{[{\rm O}\textsc{iii}]}^{\rm cor}$ (e.g., \citet{KauffmannHeckman2009}):
\begin{eqnarray}
\label{eqn:lbol_oiii}
L_{\rm bol} & \sim & (600-800) L_{[{\rm O}\textsc{iii}]}^{\rm cor},
\end{eqnarray}
where we used the observed Balmer decrements (Figure \ref{fig:manga_oiii_bd}) and \citet{Cardelli1989} attenuation law to calculate the value of $A_V$ along the line of sight.

This relation was derived by analyzing the SDSS spectrum of the central regions of high luminosity narrow-line AGN \citep{heckman04}, and did not attempt to separate between the [O\iii] emission from photoionized and shock heated material.  If large-scale outflows were rare in this AGN sample, then including the [O\iii] emission from the outflow in this calculation would significantly overestimate the true value of $L_{\rm bol}$.  However, since this differentiation was not made in we derive $L_{\rm bol}$ using both the total [O\iii] flux and the [O\iii] flux measured in just the ``main'' component.

Furthermore, as shown in Figures \ref{fig:bpt_sii} \& \ref{fig:manga_bptmap_nii}, only in the central $\sim1-2^{\prime \prime}$ of this galaxy are the line ratios of the ``main'' component consistent with photoionization by an AGN.  This is smaller than the $3^{\prime \prime}$ aperture used to derive Equation \ref{eqn:lbol_oiii} \citep{heckman04}.  To estimate the possible effect resulting from this discrepancy, we measured the [O\iii] flux in both regions.  The values of $L_{[\rm O\textsc{iii}]}$ and $L_{\rm bol}$ resulting from the different choices in region and components are given in Table \ref{tab:Lo3_EddingtonRatio}.

\begin{deluxetable}{lcc}[tb]
    \tablecaption{Bolometric luminosities and Eddington ratios. \label{tab:Lo3_EddingtonRatio}}
    \tablecolumns{3}
    \tablewidth{0pt}
    \tablehead{
       \colhead{Parameter} & \colhead{AGN region} & \colhead{3\arcsec}
    }
    \startdata
    \multicolumn{3}{c}{\citet{KauffmannHeckman2009} for main component}\\
     \hline
     $A_V$, mag & 1.22 & 1.25 \\
     $L_\mathrm{[OIII]}$, erg s$^{-1}$ & $3.95\times10^{{39}}$ & $8.79\times10^{{39}}$ \\
     $L_\mathrm{bol}$, erg s$^{-1}$ & $(2.4-3.2)\times10^{42}$ & $(5.3-7.0)\times10^{42}$ \\
     $\mathcal{R}_\mathrm{Edd.}$, \% & $0.025-0.033$ & $0.056-0.074$ \\
     \hline
     \multicolumn{3}{c}{\citet{KauffmannHeckman2009} for total flux}\\
     \hline
     $A_V$, mag & 1.99 & 2.01 \\
     $L_\mathrm{[OIII]}$, erg s$^{-1}$ & $2.21\times10^{{40}}$ & $6.50\times10^{{40}}$ \\
     $L_\mathrm{bol}$, erg s$^{-1}$ & $(1.3-1.8)\times10^{43}$ & $(3.9-5.2)\times10^{43}$ \\
     $\mathcal{R}_\mathrm{Edd.}$, \% & $0.140-0.186$ & $0.412-0.549$ \\
     \hline
     \hline
     \multicolumn{3}{c}{\citet{Netzer2009} + $\lambda^{-0.7}$ extinction law}\\
     \hline
     $L_\mathrm{bol}$, erg s$^{-1}$ & $2.98\times10^{{43}}$ & $8.88\times10^{{43}}$ \\
     $\mathcal{R}_\mathrm{Edd.}$, \% & 0.314 & 0.937 \\
     \hline
     \multicolumn{3}{c}{\citet{Netzer2009} + MW extinction law CCM89}\\
     \hline
     $L_\mathrm{bol}$, erg s$^{-1}$ & $2.62\times10^{{43}}$ & $7.67\times10^{{43}}$ \\
     $\mathcal{R}_\mathrm{Edd.}$, \% & 0.277 & 0.810 \\
    \enddata
\end{deluxetable}

Furthermore, the physical relationship between the bolometric luminosity of an AGN and the emission line spectrum of the photoionized gas depends on the spectrum produced by the material accreting onto the SMBH (e.g., \citealt{Netzer2009} and references therein).  The considerable diversity in the observed spectral energy distribution (SED) of AGN (e.g., \citealt{elvis94}) suggests different relations are appropriate for different types of AGN.  The line ratios of the ``main'' component in the central regions primarily fall within the LIER region of the [S\ii] BPT diagram (Figure \ref{fig:bpt_sii}).  Since the ``main'' component excludes primarily shock-heated material, this emission likely results from material photoionized by the AGN accretion disk.  However, LIER AGN were effectively excluded from the sample of \citet{heckman04} used to derive the $L_{\rm bol}-L_{[{\rm O}\textsc{iii}]}$ relationship given in Equation \ref{eqn:lbol_oiii} \citep{KauffmannHeckman2009}.  As a result, we also estimate $L_{\rm bol}$ using a relation involving the extinction-corrected $H\beta$ luminosity of the material photoionized by the AGN which is argued to be less sensitive to the SED of the accretion disk and therefore more appropriate for LI(N)ER AGN (Equation 1 in \citealt{Netzer2009}):
\begin{eqnarray}
\label{eqn:lbol_lier}
\log L_{\rm bol} & = & \log L_{\rm H\beta} + C +
\max \left[ 0.0, 0.31 \left(\frac{L_{[{\rm O}\textsc{iii}]}}{L_{\rm H\beta}} -0.6 \right) \right],
\end{eqnarray}
where $C$ depends on the extinction law.  To account for possible variations in the properties of dust along the line of sight, we repeat this analysis using the same two extinction laws discussed by \citet{Netzer2009}: optical depth $\tau_\lambda \propto \lambda^{-0.7}$ law originally derived for starburst galaxies (e.g., \citealt{wild07}; $C=3.48$), and the \citet{Cardelli1989} extinction law for Milky-Way type galaxies ($C=3.75$).  Again, we use the same two spatial regions used in the previous method.  As shown in Table \ref{tab:Lo3_EddingtonRatio}, the values of $L_{\rm bol}$ derived using this method are comparable to those derived using $L_{[{\rm O}\textsc{iii}]}$.

\begin{figure*}[tb]
    \begin{center}
    \includegraphics[width=0.475\textwidth]{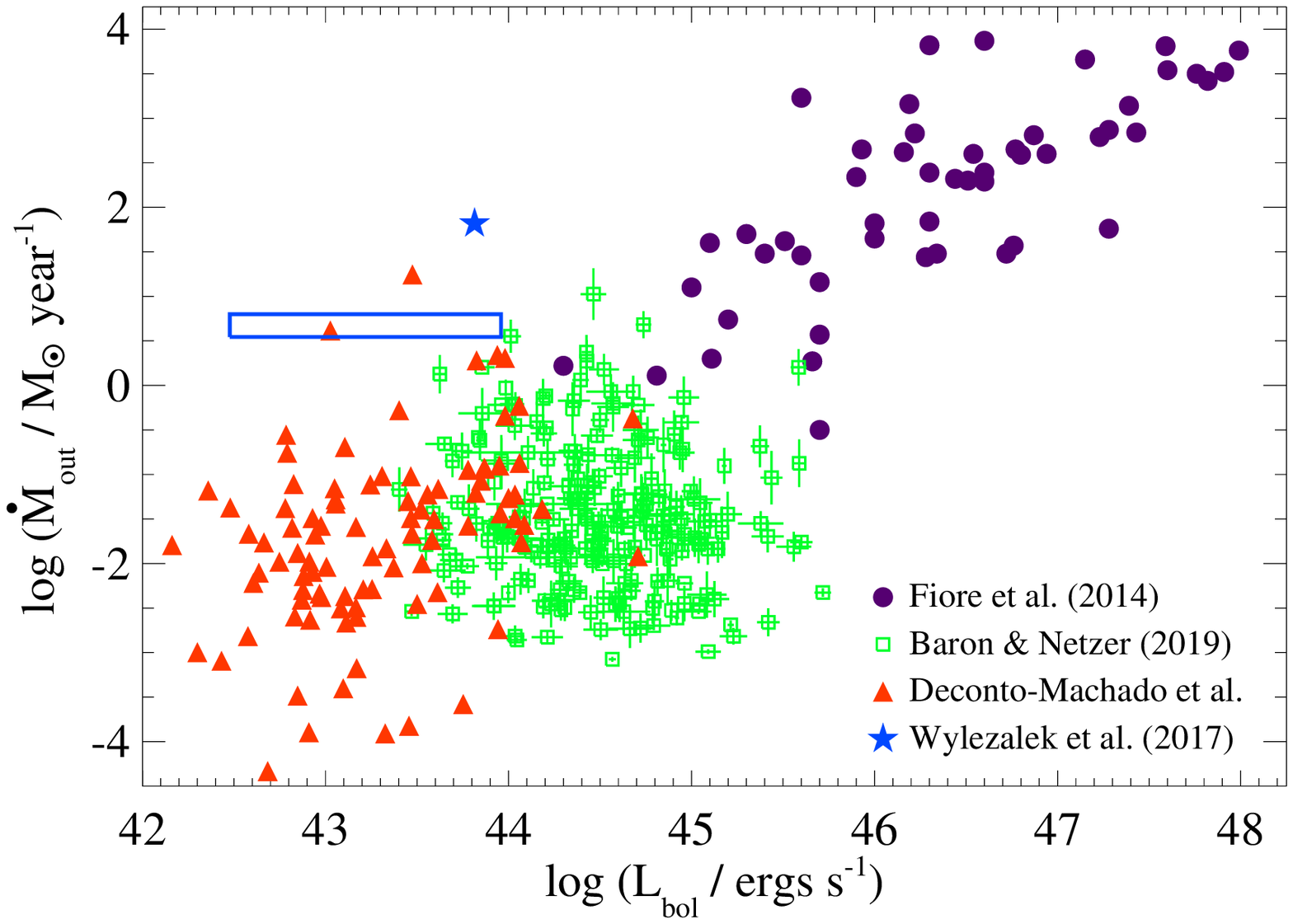}        
    \includegraphics[width=0.475\textwidth]{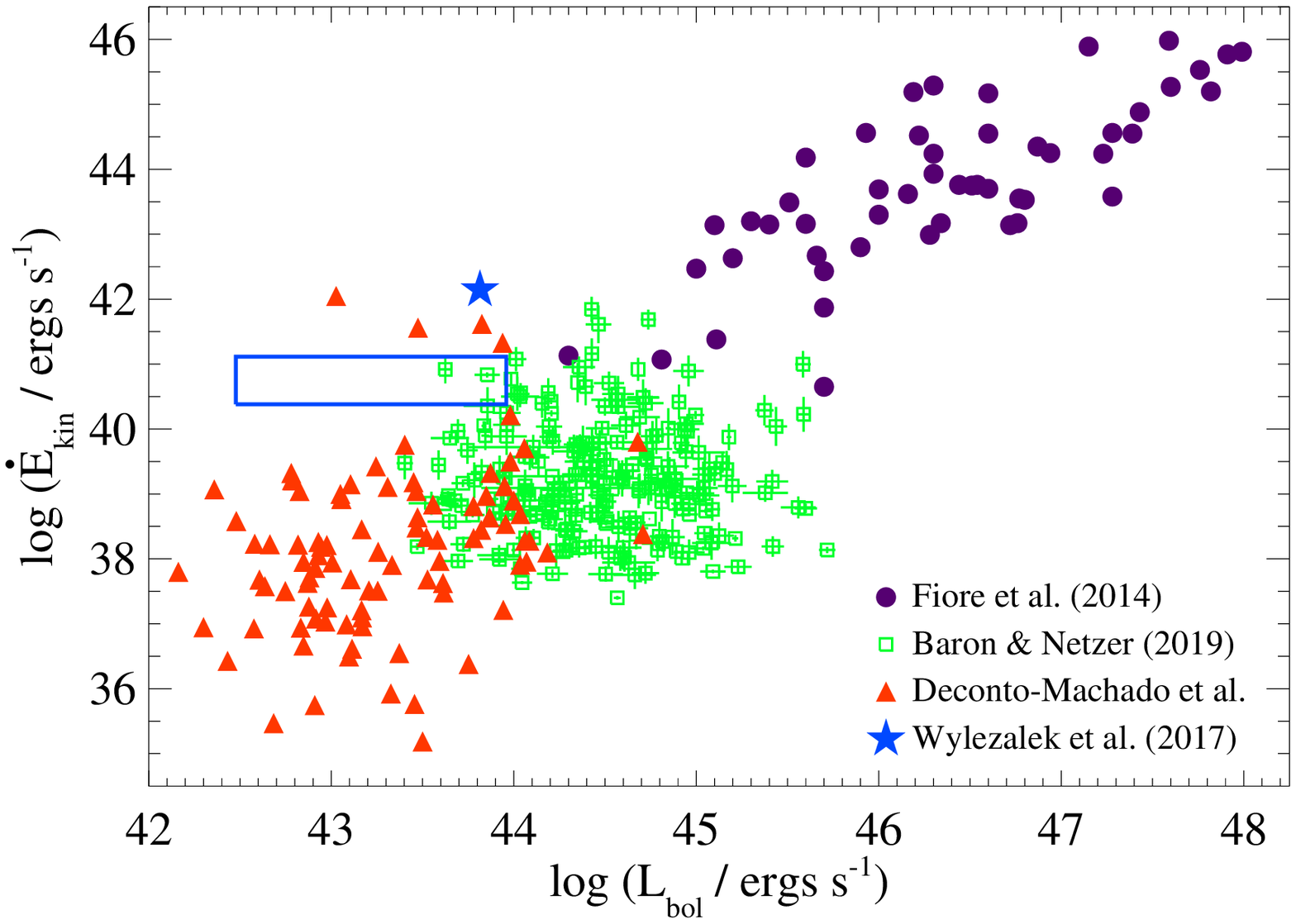}        
    \end{center}
    \vspace*{-0.5cm}
    \caption{Mass outflow rate $\dot{M}_{\rm out}$ ({\it left}) and kinetic power $\dot{E}_{\rm kin}$ ({\it right}) as a function of AGN bolometric luminosity $L_{\rm bol}$ for ionized gas outflows believed to result from AGN activity. \added{The blue rectangle shows the location of the galaxy MaNGA 1-166919 according to the range of our estimates of the outflow parameters.}}
    \label{fig:outflow_agn}
\end{figure*}

To determine if this AGN could power the outflow observed in MaNGA 1-166919, we compare its properties with those of ``known'' AGN driven outflow.  For example, \citet{kang18} found that:
\begin{eqnarray}
\label{eqn:r_out}
\log \left(\frac{R_{\rm out}}{\rm kpc}\right) & \sim & 0.028\log \left( \frac{L_{[{\rm O}\textsc{iii}]}}{\frac{\rm ergs}{\rm s}} \right) - 11.27
\end{eqnarray}
where $R_{\rm out}$ and $L_{[{\rm O}\textsc{iii}]}$ are, respectively, the radius and [O\iii] luminosity of the outflow.  The range of $L_{[{\rm O}\textsc{iii}]}$ specified in Table \ref{tab:Lo3_EddingtonRatio} suggests $R_{\rm out} \sim 1.1-1.5~{\rm kpc}$ -- in very good agreement with the  size of radio lobes detected in the wideband radio images (Table \ref{tab:12arcsec_wideband_imfit}; Figure \ref{fig:wideband_maps}) as well as the high $\sigma_{\rm gas}$ region inferred from the GMOS emission line spectra (Figure \ref{fig:gmos_gaskin}), which suggest $R_{\rm out} \sim 1.0-1.7~{\rm kpc}$.  Additional studies have found that the mass outflow rate $\dot{M}_{\rm out}$ of an ionized gas outflow is correlated with the bolometric luminosity $L_{\rm bol}$ of the AGN (e.g., \citealt{Fiore2017, Baron2019}, Deconto-Machado et al.\ in prep).
As shown in Figure \ref{fig:outflow_agn}, the mass outflow rate we estimate for this galaxy is significantly higher than the bulk of galaxies with a similar bolometric AGN luminosity.  A similar results is observed for the kinetic power $\dot{E}_{\rm kin}$ of this outflow, which again is higher than other AGN with similar $L_{\rm bol}.$

This suggests that the AGN activity in MaNGA 1$-$166919 results an outflow differently than most AGN with a similar bolometric luminosity.  If the generation of the outflow is physically connected to the accretion of material onto the SMBH, the accretion mode in 1$-$166919 is different than the others.  A key distinguishing parameter between ``radiative'' and ``jet'' mode accretion  onto a SMBH (as discussed in \S\ref{sec:intro}) is the Eddington Ratio ${\mathcal R}$, defined to be:
\begin{eqnarray}
\label{eqn:edd_ratio}
{\mathcal R} & \equiv & \frac{L_{\rm bol}}{L_{\rm Edd}},
\end{eqnarray}
where $L_{\rm Edd}$ is the Eddington luminosity of the central SMBH (e.g., \citealt{rybicki86,HeckmanBest2014} and references therein):
\begin{eqnarray}
\label{eqn:l_edd}
L_{\rm Edd} & \approx & 3.3\times10^{4} \left(\frac{M_{\rm BH}}{{\rm M}_\odot} \right)~L_\odot,
\end{eqnarray}
where $M_{\rm BH}$ is the mass of the SMBH.

To estimate $M_{\rm BH}$, we use the observed correlation between this quantity and the stellar velocity dispersion $\sigma_\star$ of the host galaxy's central bulge (see recent review by \citealt{kormendy13}).  A decomposition of the surface brightness of this galaxy into a bulge and disk component suggests its bulge has an effective radius $R_\mathrm{bulge}=3\farcs16$, ellipticity $e=0.16$, and positional angle PA$_\mathrm{bulge} = 110^\circ$ (Table 2 in \citealt{Simard2011ApJS..196...11S}).  We then estimate the central stellar velocity dispersion by calculating the light-weighted average of  $\sigma_\star^2 + v_\star^2$ (both shown in Figure \ref{fig:manga_stellarprop}) within $R_{\rm bulge}$ assuming the above geometry -- yielding $\sigma_\star = 165.2\pm0.5$ km s$^{-1}$.  The $M_{\rm BH} - \sigma_\star$ relationship derived by \citet{vdBosch2016}:
\begin{eqnarray}
\label{eqn:m-sigma}
\log \left(\frac{M_{\rm BH}}{\rm M_\odot} \right) & \approx & (-4.0\pm0.5) + (5.4\pm0.2) \log \left(\frac{\sigma_\star}{\frac{\rm km}{\rm s}} \right),
\end{eqnarray}
yields $M_{\rm BH} = 7.5^{+0.8}_{-0.7}\times10^{7}~{\rm M}_\odot$, which has an Eddington luminosity $L_{\rm Edd}$ (Equation \ref{eqn:l_edd}): 
\begin{eqnarray}
\label{eqn:ledd_calc}
L_{\rm Edd} & = & 2.5^{+0.3}_{-0.2}\times10^{12}~L_\odot \approx 9.5^{+1.0}_{-0.9}\times10^{45}~\frac{\rm ergs}{\rm s}.
\end{eqnarray}

For the range of $L_{\rm bol}$ calculated above (Table \ref{tab:Lo3_EddingtonRatio}), this implies ${\mathcal R}\lesssim 1\%$ -- suggestive of ``jet-mode'' accretion onto the SMBH (e.g. \citealt{best12}).  Recent theoretical work suggests that, for a given AGN luminosity, ``jet-mode'' accretion results in a more massive and energetic outflow that ``radiative-mode'' accretion (e.g., \citealt{cielo18}), consistent with the comparison described above (Figure \ref{fig:outflow_agn}).

\begin{table*}[tb]
    \caption{Comparison of radio and host galaxy properties of HERAGN and LERAGN}
    \begin{center}
    \begin{tabular}{ccccc}
    \hline
    \hline
    Property & MaNGA $1-166919$ & LERAGN & HERAGN & Citation \\ 
    \hline
    $L_{\rm 1.4~GHz}$ & $3.4\times10^{22}~\frac{\rm W}{\rm Hz}$ & $\lesssim 10^{26}~\frac{\rm W}{\rm Hz}$ & $\gtrsim10^{26}~\frac{\rm W}{\rm Hz}$ & \citealt{best12} \\
    $g-r$ & 0.66 & $\sim 0.8-1.0$ & $\sim0.45-0.85$ & \citealt{smolcic09} \\
    $M_\star$ & $6\times10^{10}~{\rm M}_\odot$ & $\gtrsim5\times10^{10}~{\rm M}_\odot$ & $\sim(2-15)\times10^{10}~{\rm M_\odot}$ & \citealt{smolcic09} \\
    $\log\left[\left(\frac{\sigma_\star}{\rm km~s^{-1}}\right)^4\right]$ & 8.8 & $\sim8.8-10$ & $\sim8-9.2$ & \citealt{smolcic09} \\
    \hline
    \hline
    \end{tabular}
    \end{center}
    \label{tab:heragn_leragn}
\end{table*}
The different AGN accretion modes are believed to occur in different radio AGN and host galaxies  (e.g., \citealt{HeckmanBest2014, smolcic16} and references therein), with radiative mode accretion typically associated with High Excitation Radio AGN (HERAGN) while ``jet-mode'' accretion is believed to occur in Low Excitation Radio AGN (LERAGN).  As shown in Table \ref{tab:heragn_leragn}, the radio luminosity of this AGN is consistent with a LERAGN (though there are radio quiet HERAGN; e.g., \citealt{best12}) but the properties of the host galaxy -- especially its color -- are reminiscent of HERAGN.  This suggests that AGN activity in MaNGA 1-166919 is currently driving the transition of the host galaxy from HERAGN-like to LERAGN-like properties.  This requires understanding how the AGN affects the surrounding ISM, which we discuss in \S\ref{sec:hostgal}.

\subsection{Outflow / Host Galaxy Interaction}
\label{sec:hostgal}

The appearance of this outflow on the kpc-scales probed by the observations described in \S\ref{analysis_radio} and \S\ref{analysis_optical} is less dependent on its initial geometry and content and more sensitive to the structure of the surrounding ISM (e.g., \citealt{wagner12, wagner13, wagner16}) and the relative orientation of the outflow to the galactic disk. The offset between the central axis of the outflow (as determined by its radio morphology) and the polar axis of the regularly rotating gas disk \added{projected onto the minor axis of the galaxy} (Figure \ref{fig:main_rot}) suggests a significant inclination between the two -- expected to increase the impact of the jet on the surrounding ISM (e.g., \citealt{cielo18, mukherjee18, murthy19}).    This interaction  is expected to suppress (``negative feedback'') star formation in some regions and enhance (``positive feedback'') star formation in the other parts of the host galaxy (e.g., \citealt{AYWagner2015, dugan17, cielo18, mukherjee18}).

\begin{figure*}[tbh]
    \centering
    \includegraphics[height=0.255\textheight]{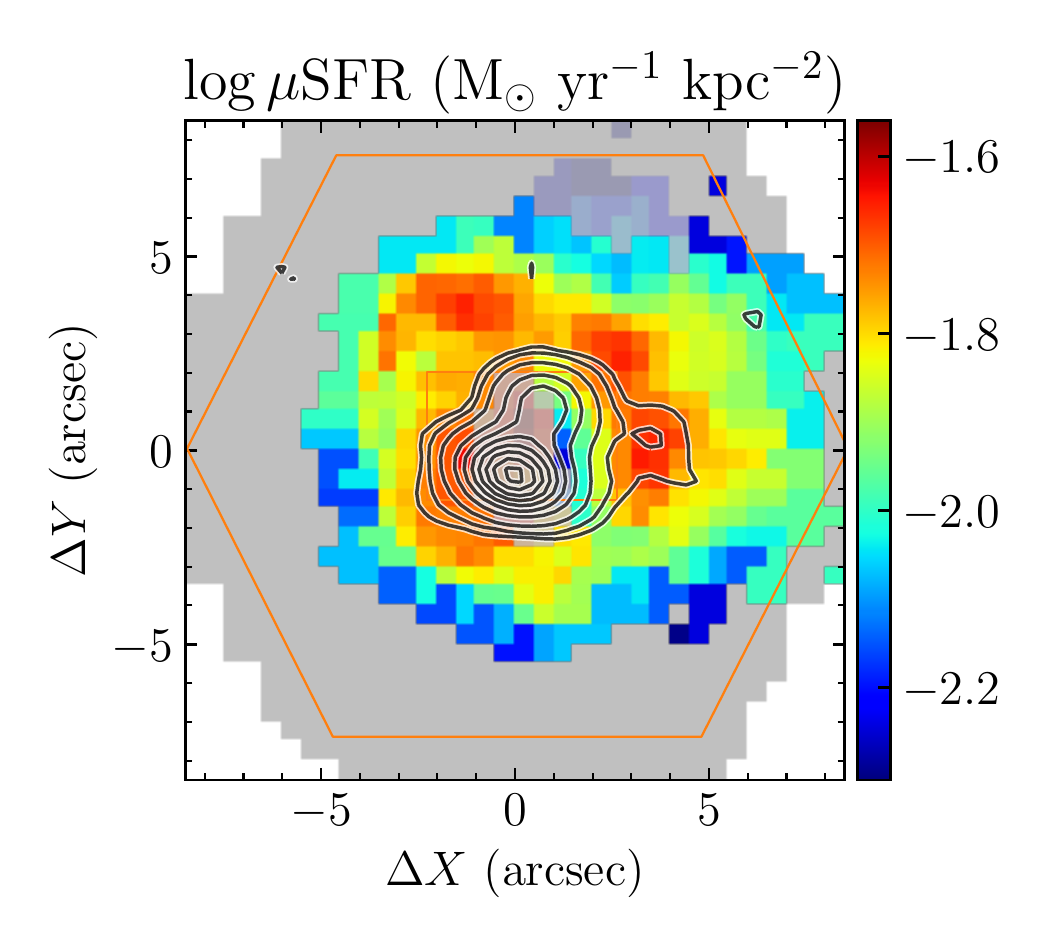}
    \includegraphics[trim=1.3cm 0 0.0cm 0,clip,height=0.255\textheight]{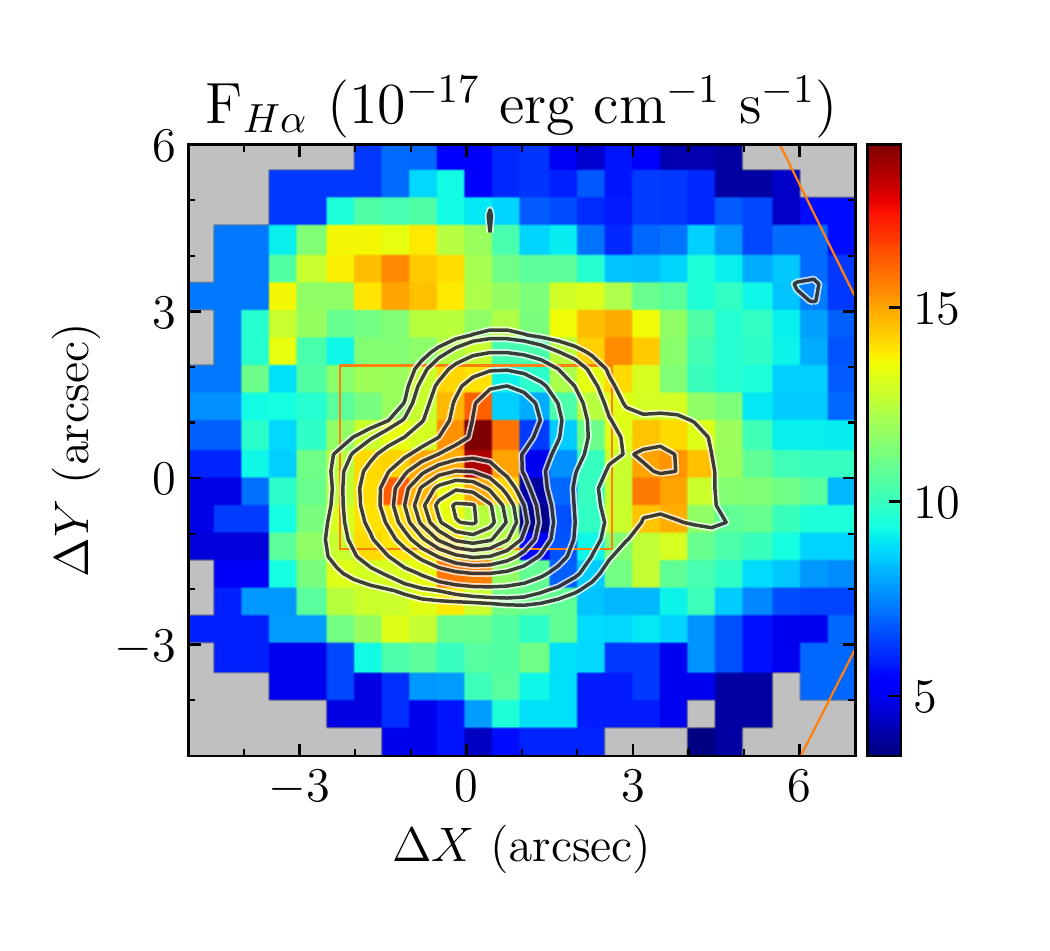}
    \includegraphics[trim=1.3cm 0 0.7cm 0,clip,height=0.255\textheight]{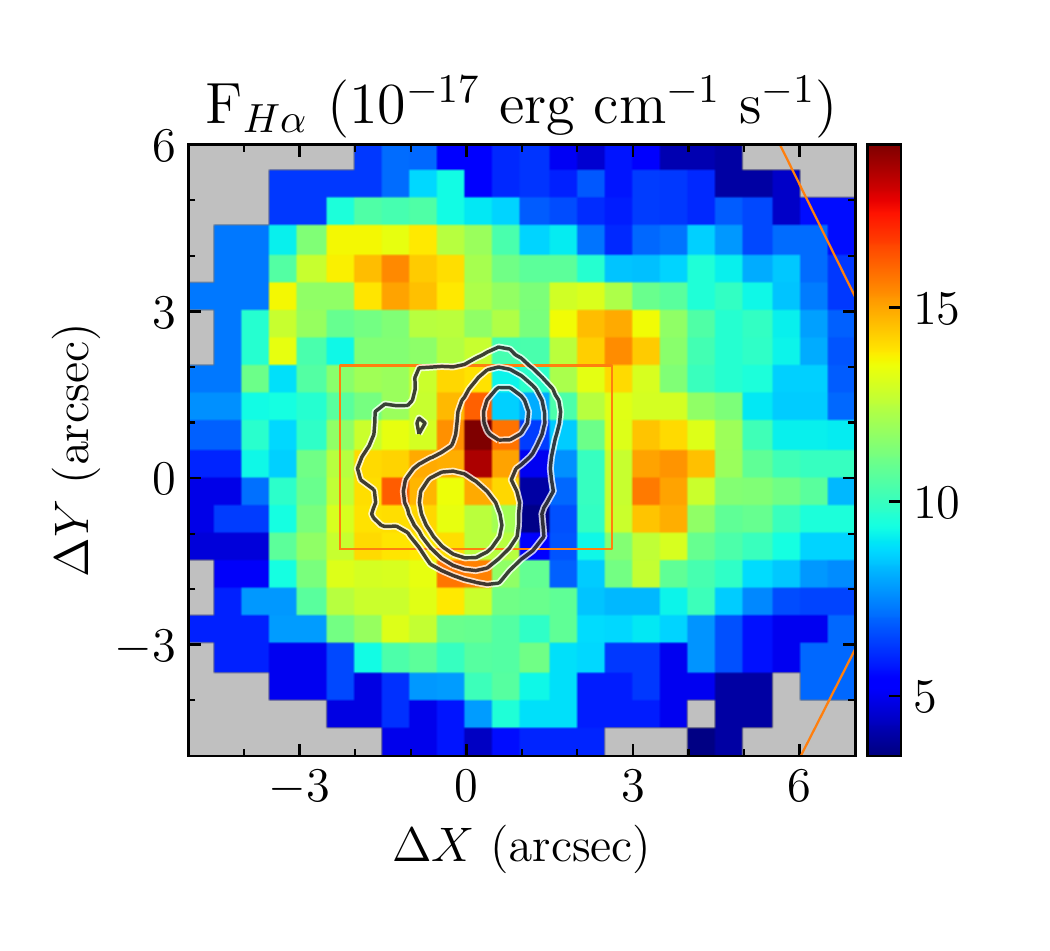}
    \vspace*{-0.25cm}
    \caption{{\it Left}: Map of SFR surface density calculated using the extinction corrected \Ha\ luminosity of the main ionized gas component in the MaNGA dataset. Spaxels showing a non-thermal gas excitation have been dimmed.  The contours indicate 3 GHz (S-band) emission  5, 10, 50$\times$ the rms of the image shown in Figure \ref{fig:wideband_maps}. {\it Center \& Right}: Extinction corrected \Ha\ luminosity of the main ionized gas component in the MaNGA dataset overlaid with contours indicating the 6 GHz (C-band) flux density ({\it center}; $3,5,10,20,30,40,50,60,70,80 \times \mathrm{RMS}$) and 3 GHz (S-band) flux density (same as {\it left} panel).}
    \label{fig:muSFR}
\end{figure*}

To understand how this outflow affects its host galaxy, we need to measure the amount and distribution of star formation.  This is best done using tracers for the star formation rate (SFR), such as the H$\alpha$ and 1.4 GHz luminosity (e.g., \citealt{Kennicutt1998,Kennicutt2012} and references therein).  As shown in Figure \ref{fig:wideband_maps}, there is little 1.4 GHz emission detected outside the outflow region.  Therefore, we can use the non-detection of diffuse 1.4 GHz emission to determine an upper-limit on the SFR.  The conversion between 1.4 GHz luminosity and SFR is believed to (e.g., \citealt{murphy11, Kennicutt2012}):
\begin{eqnarray}
\label{eqn:mu_sfr_radio}
\log\left(\frac{\rm SFR}{\rm M_\odot~yr^{-1}} \right) & \approx & \log\left( \frac{L_{\rm 1.4~GHz}}{\rm ergs~s^{-1}~Hz^{-1}} \right) - 28.20.
\end{eqnarray}
For the beam size and rms of the 1.4 GHz image (Table \ref{tab:radio_img_prop}), a $<3\sigma$ detection of diffuse radio emission in this galaxy corresponds to an upper-limit of the SFR surface density $\mu{\rm SFR} \lesssim 0.2~M_{\odot}~{\rm year}^{-1}~{\rm kpc}^{-2}$.  A more sensitive measure of $\mu{\rm SFR}$ is the H$\alpha$ emission of the ``main'' component to the ionized gas in this galaxy.  We first correct the observed H$\alpha$ flux for extinction using the spaxel-by-spaxel method described in \S\ref{sec:outflow_thermal} \& \ref{sec:agn} for the Balmer decrements shown in Figure \ref{fig:manga_oiii_bd}.  To convert the extinction-corrected \Ha\ luminosity of each spaxel into  SFR, we use the relation (e.g, \citealt{hao11,murphy11,Kennicutt2012}):
\begin{eqnarray}
\label{eqn:mu_sfr_ha}
\log\left(\frac{\rm SFR}{\rm M_\odot~yr^{-1}} \right) & \approx & \log\left( \frac{L_{\rm H\alpha}}{\rm ergs~s^{-1}} \right) - 41.27.
\end{eqnarray}
We then divide the SFR by the projected area of each spaxel to calculate $\mu$SFR. \added{As shown in Figure \ref{fig:muSFR}, the most intense regions of star formation in this galaxy have $\mu{\rm SFR} \lesssim 0.025~{\rm M}_\odot~{\rm year}^{-1}$ -- significantly lower than the upper-limit derived aboved from the non-detection of radio emission outside the outflow region.}

\added{Furthermore, } as shown in Figure \ref{fig:muSFR}, the regions with highest SFR at located near the edge of the outflow. 
Enhanced star formation near the outflow's boundary is observed in numerical simulations of such systems, typically concentrated along the jet axis and/or a ``ring'' around the outflow (e.g., \citealt{dugan17, mukherjee18}) -- similar to what is observed here (Figure \ref{fig:muSFR}).  The region of high SFR $\sim3\arcsec$ W of the center of the galaxy is coincident with $\sim5\sigma$ emission present in the 6.0 GHz (C-band) image of this source (Figure \ref{fig:muSFR}).  The flat radio spectrum ($\alpha \sim 0$; \S\ref{sec:radio_spec}, Figure \ref{fig:alpha_map}) detected in this region is suggestive of thermal bremmstrahlung radiation from a H\ii\ region, consistent with the significant SF detected in this region.  There is a region of low H$\alpha$ emission (Figure \ref{fig:muSFR}) and low Balmer Decrement (Figure \ref{fig:manga_oiii_bd}) located just beyond the W border of the outflow.  As mentioned in \S\ref{sec:outflow_kin}, high energy photons and particles produced ``downstream'' of the shock can heat and ionize the pre-shock material -- potentially destroying dust molecules (decreasing the Balmer Decrement) and fully ionizing the surrounding medium (resulting in a low H$\alpha$ luminosity if the gas is too hot to recombine).

To assess the global impact of this outflow on star formation in the host galaxy, we compare its total SFR  of $\mathrm{SFR}\approx3$~M$_\odot\, \mathrm{yr}^{-1}$ -- consistent with the $\mathrm{SFR}_\mathrm{SED}=3.7$~~M$_\odot\, \mathrm{yr}^{-1}$  derived from an independent analysis of its spectral energy distribution (SED)  (\href{https://salims.pages.iu.edu/gswlc/}{GSWLC-2} catalog; \citealt{Salim2018ApJ...859...11S}).  For star forming galaxies, the SFR is thought to be strongly dependent on the galaxy's stellar mass $M_\star$, with an analysis of star-forming galaxies observed by the SDSS suggesting that (e.g., \citealt{elbaz07} and references therein):
\begin{eqnarray}
\label{eqn:sfr_sdss}
{\rm SFR} & \sim & (5-16)\left[\frac{M_\star}{10^{11}~{\rm M}_\odot} \right]^{0.77} ~\frac{\rm M_\odot}{\rm year}.
\end{eqnarray}
For the measured $M_\star \approx 6\times10^{10}~{\rm M}_\odot$ of MaNGA 1-166919, this relation suggests ${\rm SFR} \sim 3.5-11~{\rm M}_\odot~{\rm year}^{-1}$ -- the lower range of which is consistent with the value derived above.  This suggests that the radio quiet AGN activity at the center of this galaxy has not (yet) quenched star formation, as observed in other such galaxies (e.g., \citealt{comerford20}), and the location of this galaxy in the ``Green Valley'' of the color-magnitude diagram is due in part to extinction.

\section{Summary and Conclusions}
\label{sec:summary}

In this paper, we present a detailed analysis of the radio (\S\ref{analysis_radio}) and optical (\S\ref{analysis_optical}) properties of MaNGA 1-166919 to determine the origin, content, and impact of its kpc-scale outflow.  Together, this data allows us to measure the properties of the central AGN (\S\ref{sec:agn}), the kinematics of this outflow (\S\ref{sec:outflow_kin}), the energetics of its relativistic (\S\ref{sec:outflow_cr}) and ionized gas components (\S\ref{sec:outflow_thermal}), as well as its impact on its host galaxy.  Such information is needed develop a complete model for how the AGN affects its host galaxy.

\begin{figure}[tbh]
    \begin{center}
    \includegraphics[width=0.475\textwidth]{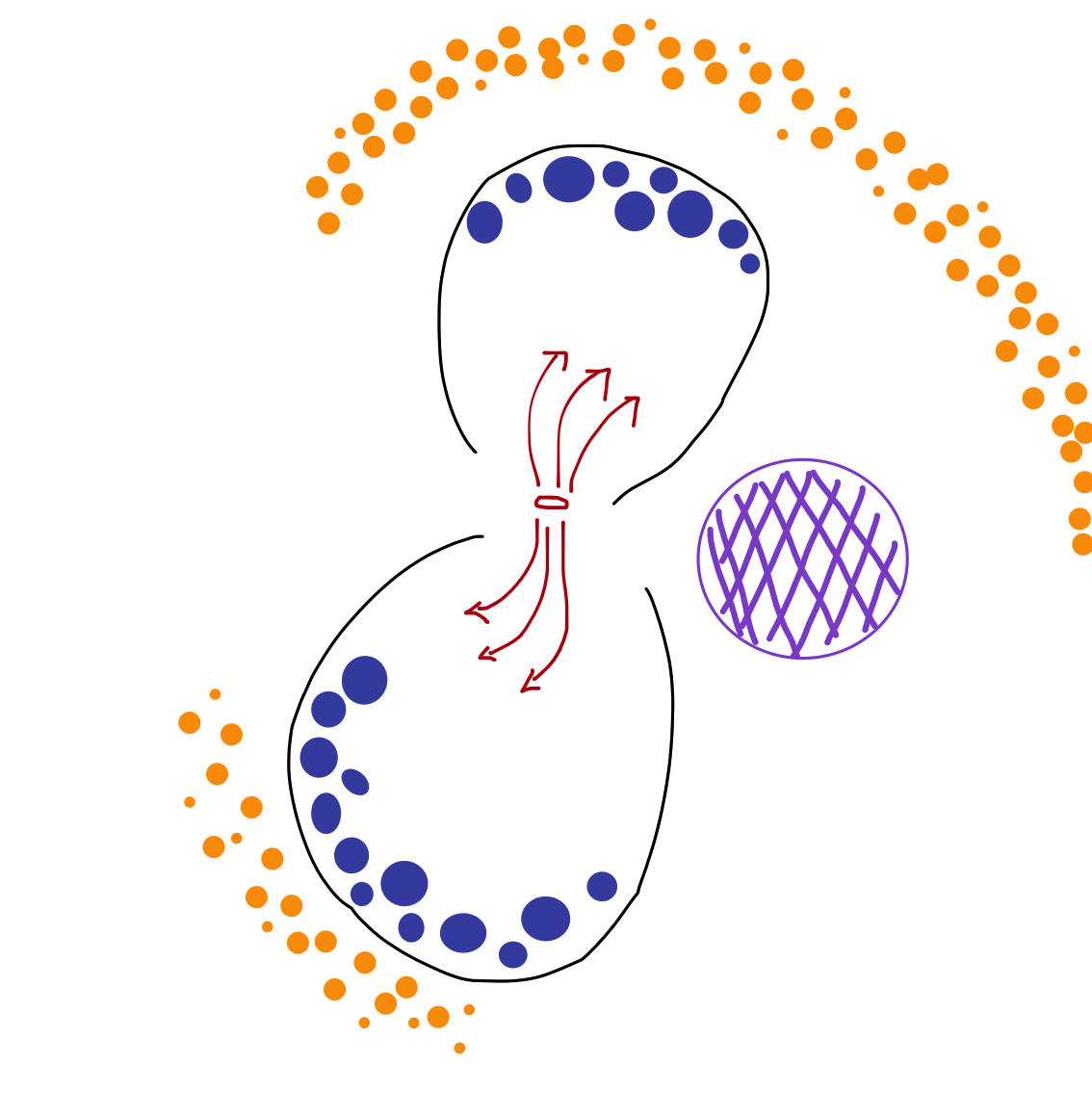}
    \end{center}
\vspace*{-0.5cm}
\caption{Schematic diagram of MaNGA 1-166919. In the center of the galaxy is low-luminosity, low-Eddington AGN injecting fast moving material into its surrounding (red circle and arrows). This material drives shocks into the ISM (black lines) with the shocked material responsible for the observed relativistic electrons and hot ionized gas (blue circles). Around this shock we detect regions of enhanced star formation (orange circles) as well as a region of low star formation (purple hashed region).}
\label{fig:blob_schematic_figure}
\end{figure}

 As shown in Figure \ref{fig:blob_schematic_figure}, our results indicate the center of this galaxy hosts a low-luminosity AGN powered by low level ($L_{\rm bol} \lesssim 0.01 L_{\rm Edd}$) accretion onto the SMBH at its center (\S\ref{sec:agn}).  The material ejected during this accretion drives ``bi-conical'', $\gtrsim100-200~{\rm km~s}^{-1}$ shocks (\S\ref{sec:outflow_kin}) into the surrounding medium responsible for producing the observed relativistic electrons (\S\ref{sec:radio_spec}) and ionized gas (\S\ref{sec:outflow_kin}), which have comparable energies ($\sim10^{54} - 10^{55}~{\rm ergs}$; Tables \ref{tab:rel_prop} \& \ref{tab:outflow_ion}).  Furthermore, the kinetic power and mass outflow rate of the ionized gas is observed to be higher than that of other AGN with comparable $L_{\rm bol}$ (Figure \ref{fig:outflow_agn}), suggesting that low-Eddington accretion may be more efficient in producing outflows than their high-Eddington counterparts.  Lastly, we detect regions of both enhanced and diminished star formation around the outflow (Figure \ref{fig:muSFR}), suggesting it results in ``positive'' and ``negative'' feedback in the host.  However, currently the global SFR of this galaxy is consistent with the SFR of star-forming galaxies with similar stellar masses (\S\ref{sec:hostgal}) -- though the relatively small size and young age ($\sim {\rm 6~Myr}$) of the outflow suggests it may, in the future, more profoundly impact star-formation in its host.

Such a complete picture of the outflow-mediated interaction between the AGN and its surroundings in MaNGA 1-166919 is only possible by analyzing spatially-resolved, multi-wavelength data.  This is now possible for large samples of outflow galaxies, and similar analyses will allow one determine how the properties and impact of such outflows are related to the properties of the central AGN, host galaxy, and age of the systems -- critical for developing a complete model for the role outflows play in galaxy evolution.

\acknowledgments
The research reported in this publication was supported by Mohammed Bin Rashid Space Centre (MBRSC), Dubai, UAE, under Grant ID number 201701.SS.NYUAD. The contributions of AAY, IK, and JDG are further supported by NYU Abu Dhabi research grant AD022.
IK also acknowledges the support from the Russian Scientific Foundation grant 17-72-20119 and the Interdisciplinary Scientific and Educational School of Moscow University ``Fundamental and Applied Space Research''.

Funding for the Sloan Digital Sky Survey IV has been provided by the Alfred P. Sloan Foundation, the U.S. Department of Energy Office of Science, and the Participating Institutions. SDSS-IV acknowledges support and resources from the Center for High-Performance Computing at the University of Utah. The SDSS web site is www.sdss.org.

SDSS-IV is managed by the Astrophysical Research Consortium for the Participating Institutions of the SDSS Collaboration including the Brazilian Participation Group, the Carnegie Institution for Science, Carnegie Mellon University, the Chilean Participation Group, the French Participation Group, Harvard-Smithsonian Center for Astrophysics, Instituto de Astrof\'isica de Canarias, The Johns Hopkins University, Kavli Institute for the Physics and Mathematics of the Universe (IPMU) / University of Tokyo, the Korean Participation Group, Lawrence Berkeley National Laboratory, Leibniz Institut f\"ur Astrophysik Potsdam (AIP), Max-Planck-Institut f\"ur Astronomie (MPIA Heidelberg), Max-Planck-Institut f\"ur Astrophysik (MPA Garching), Max-Planck-Institut f\"ur Extraterrestrische Physik (MPE), National Astronomical Observatories of China, New Mexico State University, New York University, University of Notre Dame, Observat\'ario Nacional / MCTI, The Ohio State University, Pennsylvania State University, Shanghai Astronomical Observatory, United Kingdom Participation Group, Universidad Nacional Aut\'onoma de M\'exico, University of Arizona, University of Colorado Boulder, University of Oxford, University of Portsmouth, University of Utah, University of Virginia, University of Washington, University of Wisconsin, Vanderbilt University, and Yale University.

This research has made use of the NASA/IPAC Extragalactic Database (NED), which is funded by the National Aeronautics and Space Administration and operated by the California Institute of Technology.    This research has made use of NASA's Astrophysics Data System Bibliographic Services.  The National Radio Astronomy Observatory is a facility of the National Science Foundation operated under cooperative agreement by Associated Universities, Inc. Based on observations obtained at the international Gemini Observatory, a program of NSF’s NOIRLab, which is managed by the Association of Universities for Research in Astronomy (AURA) under a cooperative agreement with the National Science Foundation. on behalf of the Gemini Observatory partnership: the National Science Foundation (United States), National Research Council (Canada), Agencia Nacional de Investigaci\'{o}n y Desarrollo (Chile), Ministerio de Ciencia, Tecnolog\'{i}a e Innovaci\'{o}n (Argentina), Minist\'{e}rio da Ci\^{e}ncia, Tecnologia, Inova\c{c}\~{o}es e Comunica\c{c}\~{o}es (Brazil), and Korea Astronomy and Space Science Institute (Republic of Korea).

\facilities{VLA, Sloan, Gemini}
\software{CASA \citep{casa}, LMFIT \citep{lmfit}, AstroPy \citep{Astropy2013A&A...558A..33A,Astropy2018AJ....156..123A}, MIRIAD \citep{miriad}}

\bibliographystyle{aasjournal}
\bibliography{Blob_revision}
\end{document}